\documentclass[11pt]{article}
\pdfoutput=1
\usepackage{jcapmod}

\usepackage{booktabs}
\usepackage[english]{babel}
\usepackage{amsmath,amssymb,amsbsy,amstext, amsthm, simplewick}
\usepackage{hyperref}
\usepackage{graphicx}
\usepackage{amsfonts}
\usepackage{amssymb}
\usepackage[small]{caption}
\usepackage{upgreek}
  \usepackage{framed}
\usepackage{enumitem}
\usepackage{footmisc}

\usepackage{booktabs}
\usepackage[english]{babel}
\usepackage{amsmath,amssymb,amsbsy,amstext, amsthm, simplewick}
\usepackage{wrapfig}
\usepackage{hyperref}
\usepackage{graphicx}
\usepackage{amsfonts}
\usepackage{amssymb}
\usepackage{subfig}
\usepackage{upgreek}
 \usepackage{exscale,relsize}
 \usepackage[makeroom]{cancel}
\usepackage{soul}
\usepackage{bbold}
\usepackage{lipsum}
\usepackage{mdframed}
\usepackage{mathtools}
\usepackage[dvipsnames]{xcolor}
\usepackage{tikz}
\usepackage{tikz-cd}
\usepackage[export]{adjustbox}

 \newcommand{\circled}[2][]{%
  \tikz[baseline=(char.base)]{%
    \node[shape = circle, draw, inner sep = 1pt]
    (char) {\phantom{\ifblank{#1}{#2}{#1}}};%
    \node at (char.center) {\makebox[0pt][c]{#2}};}}
\robustify{\circled}

\usepackage{colortbl}
\definecolor{lightgreen}{cmyk}{0.2, 0, 0.2, 0.2}
\definecolor{lightgray}{cmyk}{0.1,0.2,0,0.1}
\definecolor{lightgray2}{cmyk}{0.1,0.1,0,0.1}

\makeatletter
\newlength{\apb@width}
\newcommand{\autoparbox}[2][c]{\settowidth{\apb@width}{#2}\parbox[#1]{\apb@width}{#2}}

\newcommand{\Cen}[2]{%
  \ifmeasuring@
    #2%
  \else
    \makebox[\ifcase\expandafter #1\maxcolumn@widths\fi]{$\displaystyle#2$}%
  \fi
}
\makeatother

\newcommand{\beq}{\begin{equation}\begin{aligned}}
\newcommand{\eeq}{\end{aligned}\end{equation}}

\numberwithin{equation}{section}

\def\beq{\begin{equation}}
\def\eeq{\end{equation}}

\def\Beq{\begin{equation}\begin{aligned}}
\def\Eeq{\end{aligned}\end{equation}}

\def\bea{\begin{eqnarray}}
\def\eea{\end{eqnarray}}

\def\beq{\begin{equation}}
\def\eeq{\end{equation}}
\def\bea{\begin{eqnarray}}
\def\eea{\end{eqnarray}}

\def\bp{{\bf p}}
\def\bk{{\bf k}}
\def\bx{{\bf x}}

\def\bq{{\bf q}}

\def\S{\mathcal{N}_s(\sigma/H)^2}

\DeclareRobustCommand{\SkipTocEntry}[4]{}
\DeclareSymbolFont{extraup}{U}{zavm}{m}{n}
\DeclareMathSymbol{\varheart}{\mathalpha}{extraup}{86}
\DeclareMathSymbol{\vardiamond}{\mathalpha}{extraup}{87}

\begin{document}

\hypersetup{pageanchor=false}

\begin{titlepage}

\setcounter{page}{1} \baselineskip=15.5pt \thispagestyle{empty}

\begin{flushright}
{\tt IFT-UAM/CSIC-20-14}
\end{flushright}


\vspace{0.1cm}
\begin{center}

{\fontsize{20.74}{24}\selectfont  \sffamily \bfseries  Curvature Perturbations From Stochastic\\[10pt] Particle Production During Inflation}

\end{center}

\vspace{0.2cm}


\begin{center}
{\fontsize{12}{30}\selectfont  Marcos A.~G.~Garcia$^{\spadesuit,\clubsuit}$\footnote{marcosa.garcia@uam.es}, Mustafa A. Amin$^{\clubsuit}$\footnote{mustafa.a.amin@gmail.com}, Daniel Green$^{\vardiamond}$\footnote{drgreen@physics.ucsd.edu}}
\end{center}

\begin{center}

\vskip 7pt

\textsl{$^{\clubsuit}$ Department of Physics \& Astronomy, Rice University, Houston, Texas 77005, U.S.A.}\\
\textsl{$^{\spadesuit}$ Instituto de F\'isica Te\'orica (IFT) UAM-CSIC, Campus de Cantoblanco, 28049, Madrid, Spain}\\
\textsl{$^{\vardiamond}$ Department of Physics, University of California, San Diego, La Jolla, CA 92093, U.S.A.}

\vskip 7pt

\end{center}

\vspace{0.5cm}
\hrule \vspace{0.3cm}
\noindent {\sffamily \bfseries Abstract} \\[0.1cm]
We  calculate the curvature power spectrum sourced by spectator fields that are excited repeatedly and non-adiabatically during inflation. In the absence of detailed information of the nature of spectator field interactions, we consider an ensemble of models with
intervals between the repeated interactions and interaction strengths drawn from simple probabilistic distributions. We show that the curvature power spectra of each member of the ensemble shows rich structure with many features, and there is a large variability between different realizations of the same ensemble. Such features can be probed by the cosmic microwave background (CMB) and  large scale structure observations. They can also have implications for  primordial black hole formation and CMB spectral distortions.

The geometric random walk behavior of the spectator field allows us to calculate the ensemble-averaged power spectrum of curvature perturbations semi-analytically. For sufficiently large stochastic sourcing, the ensemble averaged power spectrum shows a scale dependence arising from the time spent by modes outside the horizon during the period of particle production, in spite of there being no preferred scale in the underlying model. We find that the magnitude of the ensemble-averaged power spectrum overestimates the typical power spectra in the ensemble because the ensemble distribution of the power spectra is highly non-Gaussian with fat tails. 

\vskip 10pt
\hrule
\vskip 10pt

\vspace{0.6cm}
 \end{titlepage}

\hypersetup{pageanchor=true}

\tableofcontents

\newpage

\section{Introduction}
Cosmic inflation provides a causal mechanism for generating the seemingly acausal initial conditions for structure formation. However, the detailed physics driving inflation, and its connection to well tested microphysics of the Standard Model is not well understood. Phenomenologically, a slowly rolling single scalar field driving inflation is sufficient to explain most current observations (roughly scale-invariant, and Gaussian primordial fluctuations \cite{Akrami:2018odb,Beutler:2019ojk}), though puzzles remain \cite{Chowdhury:2019otk}, and new opportunities for probing the detailed physics of inflation are on the horizon \cite{Abazajian:2016yjj, Shandera:2019ufi,Meerburg:2019qqi,Chluba:2019kpb,Slosar:2019gvt}.
\\ \\
\noindent{\bf Motivation}: The field content and their dynamics during inflation might be simple. However, most high-energy completions of the Standard Model include many interacting fields at inflationary energies \cite{Baumann:2014nda}, whose dynamics need not {\it a priori} be simple. It is still plausible that the field content during inflation is rich, and their dynamics are complex, leading to features in the power spectrum and higher-point correlators (see, for example, \cite{Slosar:2019gvt,Meerburg:2019qqi,Durakovic:2019kqq}) which might have been missed in the data so far. Important consequences might also arise from features in the power spectrum that lie beyond the scales relevant for the cosmic microwave background (CMB) and large-scale structure (LSS), but could be relevant to, for example, primordial black hole formation \cite{Khlopov:2008qy,Panagopoulos:2019ail}, reheating \cite{Amin:2014eta,Lozanov:2019jxc}, and CMB spectral distortions \cite{Chluba:2011hw,Chluba:2019kpb}. In this paper, we provide some relatively model-independent results regarding the form of curvature perturbations in such ``non-trivial" inflationary scenarios.
\\ \\
\noindent{\bf Modeling Complex Scenarios}: In the presence of many fields and complex dynamics, the perturbations of spectator fields can experience a time-dependent effective mass that varies in a non-trivial manner. Without committing to any particular ultraviolet-completed scenario, we model this time-dependence as a series of localized (in time), non-adiabatic changes in the effective mass of the spectator fields, with their locations and strengths derived from simple ensemble distributions \cite{Amin:2015ftc,Amin:2017wvc,StochasticSpectator}.  This modeling generates an ensemble of realizations of the effective mass -- a proxy for our limited understanding of the actual microphysics of inflation.\footnote{In earlier works \cite{Green:2014xqa,Amin:2015ftc,Amin:2017wvc,StochasticSpectator}, some of us were motivated by the connection between particle production in cosmology and current conduction in disordered wires (also see \cite{Zanchin:1997gf,Bassett:1997gb}). Apart from the elegant mathematical correspondence, the primary drive there was that certain universal features, such as Anderson Localization \cite{anderson1958absence,mello2004quantum} in one dimension, arise independent of the details of the systems -- motivating a search for similar universality in particle production. For different approaches in the context of the early universe where simple behavior arises in spite of underlying complexity of the models, also see \cite{Cheung:2007st,Weinberg:2008hq,Tye:2008ef,LopezNacir:2011kk,Dias:2012nf,McAllister:2012am,Marsh:2013qca,Easther:2013rva,Price:2014xpa,Price:2015qqb, Dias:2016slx,Giblin:2017qjp,Bjorkmo:2017nzd,Guo:2017vxp,Dias:2017gva,Dias:2018koa}.} Explicit inflationary models, with the phenomenology dictated by time-varying effective masses of spectator fields, have been investigated in, for example, \cite{Berera:1998px, Kofman:2004yc, Green:2009ds,Cook:2011hg,Barnaby:2012xt,Sorbo:2011rz,Mirbabayi:2014jqa,Mukohyama:2014gba,Abolhasani:2019wri}. 
\\ \\
\noindent{\bf Excited Spectator Fields}: The perturbations of the spectator fields experience a stochastic, approximately exponential growth, driven by the repeated non-adiabatic changes in their effective mass. This behavior was discussed in detail in~\cite{Amin:2015ftc,Amin:2017wvc} for spectator fields in Minkowski spacetime and \cite{StochasticSpectator} for those in de Sitter spacetime. In these works, the behavior of fields as a function parameters related to the effective mass was carried out. This was done for an individual realization within an ensemble, as well as in an ensemble averaged sense. In many cases, the non-trivial evolution of the spectator fields is most prominent on superhorizon scales.
\begin{figure}[t!] 
   \centering
   \includegraphics[width=\textwidth]{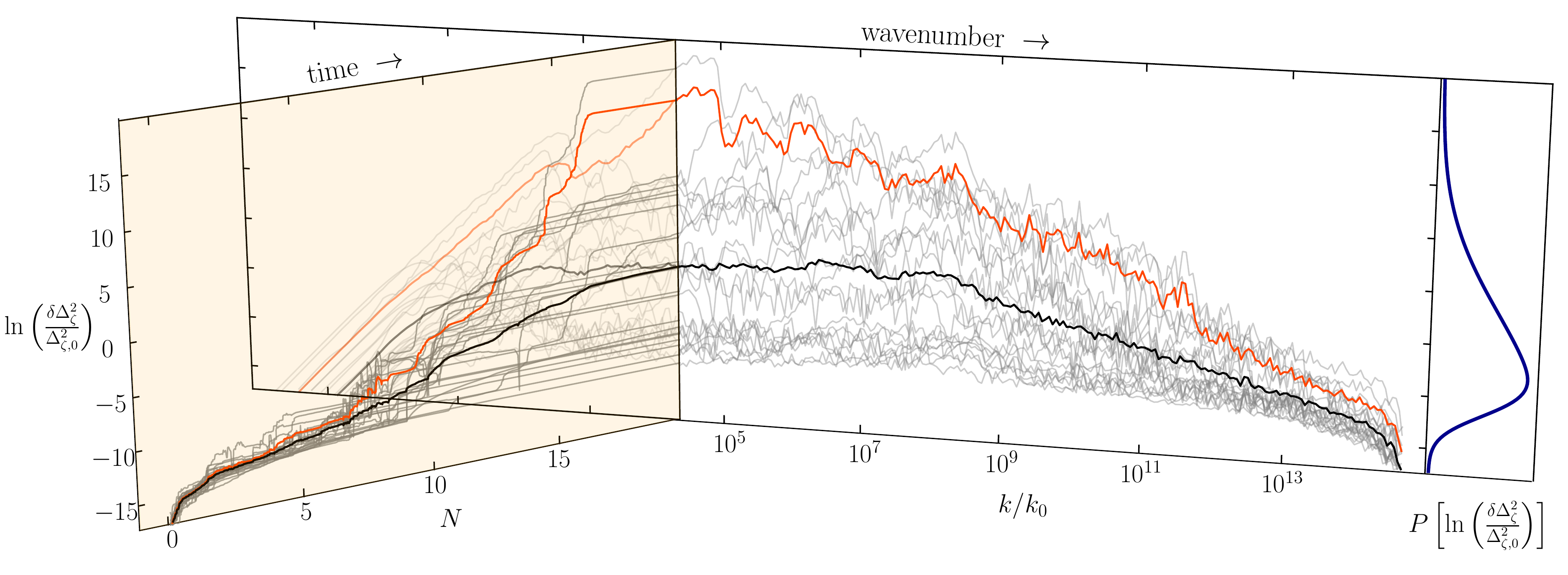} 
   \caption{Ratio of the component of the power spectrum sourced by stochastic particle production, $\delta\Delta_{\zeta}^2$, to the component of the power spectrum sourced solely by the vacuum fluctuation, $\Delta_{\zeta,0}^2$, as a function of the number of $e$-folds $N$ and wavenumber $k$. Here the characteristic disorder strength is given by $\S=25$, stochastic particle production is assumed to be effective for $N_{\rm tot}=20$ $e$-folds, and we have assumed that $\Delta_{\zeta,0}^2=\Delta_{\zeta,{\rm Planck}}^2\simeq 2.1\times10^{-9}$. The wavenumber $k_0$ is that of the curvature mode that leaves the horizon at $N=0$. Each gray curve corresponds to a particular realization of disorder, for a total of 20 unique realizations. The red (black) curve shows the arithmetic (geometric) sample mean. The blue curve shown in the rightmost panel shows the reconstructed probability density function for $\ln(\delta\Delta_{\zeta}^2/\Delta_{\zeta,0}^2)$ at $N=20$, $k/k_0=e^{10}$. 
}
   \label{fig:3D}
\end{figure}
\\ \\
\noindent{\bf Sourced Curvature Perturbations}: Curvature perturbations are sourced by the excited spectator field perturbations -- calculating this sourced curvature spectrum is the main goal of this paper.  We summarize the main results here for convenience.
\begin{itemize}
\item We find that the curvature perturbations sourced by the spectator field can exceed the usual vacuum contribution, {\it without} the spectator field dominating the background energy density of the universe. 
\item The curvature power spectra generated (via the excited spectator fields) by each realization of the effective-mass ensemble can be highly non-trivial. For a finite duration of the epoch during which repeated non-adiabatic particle production in the spectator field takes place, the sourced component of the curvature power spectrum has a shape resembling a ``tilted plateau" with additional small-scale features on top in any given realization. At very low wavenumbers, the sourced part of the spectrum rises with a slope determined by causality, while at very high wavenumbers the spectrum decays due to the lack of excitation of deep sub-horizon modes. 
\item In the ensemble averaged sense, we calculate the shape and amplitude of the curvature power spectrum semi-analytically (see Fig.~\ref{fig:kplots1}) in terms of (i) $\S$, where $\sigma^2$ is the variance of the strength of the effective mass, $\mathcal{N}_{\rm s}\gg 1$ is the mean number of non-adiabatic changes per $e$-fold of expansion, and (ii) the total number of $e$-folds ($N_{\rm tot}$) during with repeated, non-adiabatic particle production takes place. Although in an ensemble sense, the effective mass realizations do not break scale invariance, the resulting sourced power spectra can do so. There are features related to the beginning and end of the non-adiabatic period, as well as a $\S$ dependent tilt.
\item For an ensemble with the same $\S$ and $N_{\rm tot}$, realization to realization, we numerically find that the  sourced spectrum can vary by many orders of magnitude and have many additional features (compared the ensemble averaged one). This large variation within the same ensemble, and the highly non-Gaussian distribution of the power-spectra amplitudes, are a consequence of the stochastic, exponential behavior of the spectator fields.

For a given sample of realizations, the geometric mean  provides a far better estimate of the typical realization, than the arithmetic mean. From the semi-analytic calculation, the arithmetic ensemble mean over-estimates the typical realization, whereas a ``variance-suppressed" arithmetic ensemble mean, underestimates the typical realization. 
\end{itemize}
\begin{figure}[!t] 
   \centering
   \includegraphics[width=0.9\textwidth]{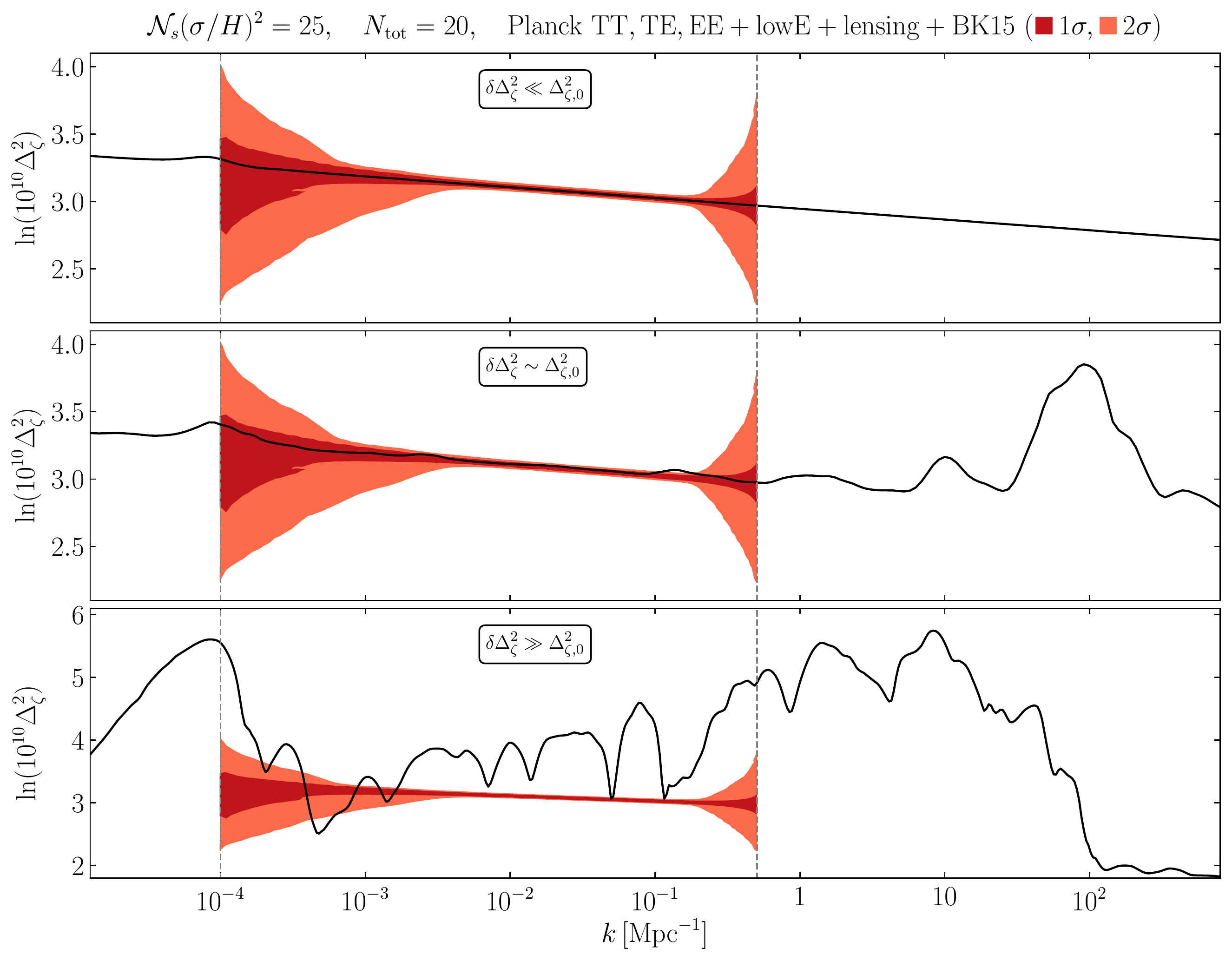} 
   \caption{The total power spectrum $\Delta_{\zeta}^2$ obtained as the sum of the component sourced by the vacuum fluctuation $\Delta_{\zeta,0}^2$ and the component sourced by stochastic particle production $\delta \Delta_{\zeta}^2$, for three different realizations of the disorder. All three are drawn from the same ensemble, characterized by the interactions with strength $\S=25$ and duration $N_{\rm tot}=20$ in $e$-folds. Shown in the background are the 1 and $2\sigma$ contours of the Planck 2018 reconstructed power spectrum. Top: a realization dominated by the vacuum fluctuation, which we fit to the Planck data. Middle: realization for which the vacuum and the stochastic components are comparable in the experimentally constrained window. Bottom: a realization dominated by the stochastic source. In all three cases the amplitude is constrained to coincide with the Planck value at the pivot scale $k_{\star} = 0.05\,{\rm Mpc}^{-1}$. For the above choice of parameters, the curvature power spectra shown in the top two panels are more likely than the bottom one. Heuristically, this can be seen from the probability distribution drawn in Fig.~\ref{fig:3D}.}
   \label{fig:planck}
\end{figure}
We note that some of the conclusions of this paper are qualitatively different from related earlier works where the scalar metric fluctuations are also sourced by particle production.  The most noteworthy difference is that, despite choosing a statistically uniform temporal density of scattering events, the power spectrum is not scale invariant (even in an ensemble averaged sense).  In contrast, disorder in single field inflation~\cite{Green:2014xqa}, dissipative modes~\cite{Berera:1995ie,Green:2009ds,LopezNacir:2011kk} or resonant particle production~\cite{Flauger:2009ab,Flauger:2016idt} lead to nearly scale invariant spectra (up to small period features) after ensemble averaging.  In these examples, the fluctuations freeze outside the horizon and therefore a large deviation from scale invariance would signal a preferred time during the evolution.  The effect of stochastic particle production described in this paper accumulates outside the horizon and is sensitive to the number of $e$-folds during which the curvature mode is super-horizon.  As a result, we find strongly scale-dependent behavior in the power spectrum despite having a model where there is no preferred time other than the end of inflation/particle production.  Aspects of the dynamics explored in this paper are reminiscent of {\it intermittent non-Gaussianity}~\cite{Bond:2009xx, bond} as both show sensitivity to super-horizon physics. 
\\ \\
\noindent{\bf Observational Implications}: A stochastically sourced power spectrum ($\delta\Delta_{\zeta}^2$) adds a scale-de\-pen\-dent contribution to the roughly scale invariant one from vacuum fluctuations ($\Delta_{\zeta,0}^2$) over a finite interval in $k$ if the duration of non-adiabatic particle production is finite. This sourced interval in $k$ may lie to the left or right of, or overlap with, the observational window accessible to CMB and large-scale structure measurements. Moreover, $\delta\Delta_\zeta^2$ can be above or below $\Delta_{\zeta,0}^2$ depending on underlying parameters of the model and the particular realization of the effective mass that is used from within a given ensemble.

As an example of the large variation of the curvature spectrum from realization to realization, in Fig.~\ref{fig:planck}, we present three results for the curvature power spectrum (continuous black curves), composed of a sum of the vacuum, adiabatic part  and a stochastically sourced part. The three realizations are drawn from the {\em same} statistical ensemble, characterized by the strength of the stochastic interactions ($\S=25$) and the duration of the non-adiabatic epoch in $e$-folds, $N_{\rm tot}=20$. All three results are normalized to match the Planck 2018 reconstructed power spectrum (shown as the shaded regions) at the pivot scale $k_{\star} = 0.05\,{\rm Mpc}^{-1}$~\cite{Aghanim:2018eyx}. 

The top panel depicts a realization for which the stochastic component of the spectrum is sub-dominant relative to the adiabatic one in the observable window, $\delta\Delta^2_{\zeta}\ll \Delta_{\zeta,0}^2$. The latter is chosen to fit the experimental data. In such a case, the corrections are unobservably small almost everywhere (except for a small bump at large scales barely outside the observational window). It is worth noting that even in this case, there remains a possibility that observationally relevant higher point correlation functions might be generated.

The middle panel shows a realization for which the adiabatic and the stochastic components of $\Delta_{\zeta}^2$ are comparable in the Planck domain. While mostly scale-invariant, some departures from scale-invariance are expected here, which makes this case observational interesting. Note that in this case, even if the spectrum is roughly featureless in the experimentally constrained region, it can hide large bumps at lower or higher scales, with a potential for rich phenomenology. 

Finally, the lower panel shows the case in which $\delta\Delta^2_{\zeta}\gg \Delta_{\zeta,0}^2$. In this scenario, the dominant stochastic component of the spectrum is too noisy to account for the observed curvature fluctuation. Note that the amplitude of the spectrum can vary by several orders of magnitude not only between realizations but also within the same realization. 

As can be roughly seen from the probability distribution in Fig.~\ref{fig:3D}, the curvature power spectra shown in the top and middle panel of Fig.~\ref{fig:planck} are more likely than the bottom one (for the chosen set of parameters).

\par\bigskip

\noindent{\bf Simplifying Assumptions:} In order to keep the analysis manageable, we restrict ourselves to the study of the power spectrum sourced by a single excited spectator field. Generalizations might be possible based on \cite{Amin:2017wvc}. Although our mathematical framework is in principle valid for any mass of this spectator field, we limit our discussion to a conformally massive ($M^2=2H^2$) spectator field. This choice is technically convenient, as the vacuum mode functions of the field have a simple, free-field form when written as functions of conformal time. Moreover, since $M>H$, the isocurvature fluctuations are suppressed compared to the curvature perturbations; they would decay exponentially outside the horizon (in contrast to the curvature perturbations) after the particle production ends during inflation. 

Phenomenologically, using $M\sim H$ is of interest, given that in many supergravity models the large vacuum density during inflation $V\sim H^2 M_P^2$, where $M_P$ denotes the Planck mass, typically leads to induced masses for scalar fields of the order of the expansion rate $H$~\cite{Goncharov:1984qm,Evans:2013nka}. Moreover, the stochastic excitation of conformally massive spectators was explored in detail in our previous work~\cite{StochasticSpectator}, both in the single disorder realization sense and in the statistical sense -- we take advantage of these results in this work. 

An additional simplification for our study corresponds to the modeling of the non-adiabatic changes of the effective mass of the field (events) as Dirac-delta functions in physical time. When needed, we regulate the temporal width of the delta-function with the help of a momentum cutoff. Moreover, to regulate the usual UV divergence of momentum integrals, we use an adiabatic subtraction scheme \cite{Parker:1974qw,Birrell:1977,Anderson:1987yt,Fulling:1989nb,parker2009quantum,Kohri:2017iyl}.

Finally, for this study, we limit ourselves to a regime where the energy density of the spectators remains small compared to the background energy density, and we also ignore the effect of curvature perturbations on the production of the spectator fields.
\\ \\
\bigskip
\noindent Our paper is organized as follows:\par\medskip

\noindent {\bf Section~\ref{sec:PS}} develops the formalism that is necessary to compute the curvature power spectrum sourced by a non-adiabatically excited spectator field. In Section~\ref{sec:description} we determine the coupling of a generic spectator field to the quasi-de Sitter Goldstone mode (related to the curvature perturbation), and use it to calculate the form of the power spectrum enhancement for an arbitrary effective mass. We also include some useful mathematical results related to the Green's function of the Goldstone mode. In Section~\ref{sec:spectfield} we briefly summarize the main results of our previous work~\cite{StochasticSpectator} concerning the dynamics of a stochastically excited, conformally massive spectator field in de Sitter space.\par\medskip

\noindent {\bf Section~\ref{sec:stochform}} describes the approximations made to compute the sourced power spectrum, and contains our numerical results. Section~\ref{sec:ddscatterers} is devoted to the calculation of the spectrum in the limit of very short-time non-adiabatic interactions, modeled by Dirac-delta functions. Section~\ref{sec:numres} contains our numerical results for individual realizations of the disorder, for weak, moderate and strong scattering strengths. The one-point statistical properties of the power spectrum are discussed based on numerical results in Section~\ref{sec:dist}.\par\medskip

\noindent {\bf Section~\ref{sec:averageps}} presents our analytical results coming from the computation of the mean power spectrum. \par\medskip

\noindent{\bf Section~\ref{sec:backreact}} presents the domain of validity of our approximations in the light of perturbativity and backreaction constraints. \par\medskip

\noindent{\bf Section~\ref{sec:cmb}} includes our discussion of the phenomenological and observational consequences of our results. We mostly focus on the dissection of our results in the light of CMB and matter power spectrum observations.\par\medskip

\noindent {\bf Section~\ref{sec:conc}} contains a summary of our results and our conclusions. \par\medskip

\noindent In Appendix~\ref{app:diracdelta} we provide a step-by-step calculation of the non-adiabatically sourced power spectrum in the case of Dirac-delta scatterers. Appendix~\ref{sec:cutoff} contains a detailed account of the ultraviolet dependence of the sourced power spectrum, arising from excited sub-horizon modes. There we also determine the limit in which our results can be regarded as universal, as in that they depend only on the strength of the stochasticity and the duration of the particle production epoch. In Appendix~\ref{app:numerics} we describe the nuances and approximations behind our numerical results. Finally, in Appendix~\ref{sec:intdom} we present the detailed analytical computation of the ensemble averaged power spectrum.

\section{Sourced Curvature Power Spectrum: Formalism}\label{sec:PS}

In this section, we develop the formalism that is necessary to compute the curvature power spectrum sourced by a non-adiabatically excited spectator field. We will arrive at an expression, \eqref{eq:deltazeta}, for the power spectrum of curvature perturbations in terms of time integrals over the Green's functions of the curvature perturbation, \eqref{eq:green1}, and a time-dependent momentum integral over non-adiabatically excited modes of the spectator field.

\subsection{Curvature Perturbations from Spectator Fields}\label{sec:description}

Consider a spectator field\footnote{$\chi$ is assumed to not dominate the total energy density of the universe.} $\chi(t,\bx)$ of mass $M$ in a homogeneous and isotropic quasi-de Sitter spacetime. We assume that the coupling of this field with a time-dependent background can be parametrized by a time-dependent effective mass $m(t)$. As discussed in the Introduction, this effective mass can arise due to random non-adiabatic events derived from complicated interactions with other fields at the background level. 

Let us now write the effective action for the Goldstone boson, $\pi(\bx,t)$, associated with the time-translation invariance of the quasi-de Sitter background~\cite{Creminelli:2006xe,Cheung:2007st}. Due to the time-dependence of the effective mass of $\chi$, a coupling between the Goldstone mode $\pi$ and the spectator field is induced,\footnote{We use the ``mostly minus" sign convention for the metric.}
\begin{align} \notag
\mathcal{S}\;&=\; \frac{1}{2}\int \sqrt{-g}\, d^4x\Big[c(t+\pi) \partial_{\mu}\pi\partial^{\mu}\pi + \partial_{\mu}\chi\partial^{\mu}\chi - \left( M^2 + m^2(t+\pi) \right)\chi^2 \Big]\\ \label{eq:spi}
&=\; \frac{1}{2}\int \sqrt{-g}\, d^4x \left[c(t) \partial_{\mu}\pi\partial^{\mu}\pi + \partial_{\mu}\chi\partial^{\mu}\chi - \left( M^2 + m^2(t) \right)\chi^2 - \frac{dm^2(t)}{dt}\chi^2\pi + \cdots\right]\,,
\end{align}
where 
\beq
c(t) \;\equiv\;  2M_P^2 |\dot{H}| \;\simeq\; {\rm const.}\qquad\textrm{and}\qquad \zeta\simeq H\pi\,.
\eeq
In the above line, $\zeta$ is the usual curvature perturbation and the connection to the Goldstone mode is valid on superhorizon scales. In going from the first to the second line in Eq.~\eqref{eq:spi}, we have disregarded $\pi$-$\partial \pi$ couplings (the so-called decoupling limit~\cite{Baumann:2011su}), as well as couplings of the form $(d^nm^2(t)/dt^n)\chi^2\pi^{n}$ for $n\geq 2$. Note that the expansion performed to arrive at (\ref{eq:spi}) is controlled by the smallness of  $\zeta\simeq H\pi$ (also see Section~\ref{sec:ddscatterers}). 

Denoting the scale factor by $a$, and promoting the Goldstone and spectator fields to operators, we can rewrite the action of the above system as
\begin{align}\notag
\mathcal{S}\;\simeq\; \frac{1}{2}\int d^3\bx d\tau\, a^2(\tau) \bigg[c(\tau) \Big((\partial_{\tau}\hat{\pi})^2 &- (\nabla\hat{\pi})^2\Big) + (\partial_{\tau}\hat{\chi})^2 - (\nabla\hat{\chi})^2\\
& - a^2(\tau)\left( M^2 + m^2(\tau) \right)\hat{\chi}^2 - a(\tau)\frac{dm^2(\tau)}{d\tau}\hat{\chi}^2\hat{\pi} \bigg]\,,
\end{align}
where $\tau$ is the conformal time, related to cosmic time $t$ via $dt/d\tau=a$. The equation of motion for $\hat{\pi}$ is then given by
\beq
\hat{\pi}''(\bx,\tau) + 2\mathcal{H}\hat{\pi}'(\bx,\tau) - \nabla^2\hat{\pi}(\bx,\tau) \;=\; -\frac{a(\tau)}{2c(\tau)}\frac{dm^2(\tau)}{d\tau}\hat{\chi}^2(\bx,\tau)\,,
\eeq
where a prime denotes differentiation with respect to $\tau$, and $\mathcal{H}=a'/a$. Fourier transformation\footnote{We use the same Fourier convention here as in~\cite{StochasticSpectator}, namely $\hat{\pi}(\bx,\tau)= (2\pi)^{-3/2}\int d^3\bk\,e^{-i\bk\cdot\bx} \hat{\pi}_k(\tau)$.} equivalently leads to the equation of motion satisfied by each mode of $\pi$, 
\beq\label{eq:eompik}
\hat{\pi}''_{\textbf{k}}(\tau) + 2\mathcal{H}\hat{\pi}'_{\textbf{k}}(\tau) + k^2\hat{\pi}_{\textbf{k}}(\tau) \;=\;  -\frac{a(\tau)}{2c(\tau)}\frac{dm^2(\tau)}{d\tau}\int \frac{d^3\bp}{(2\pi)^{3/2}} \hat{\chi}_{\bp}(\tau)\hat{\chi}_{\bp-\bk}(\tau)\,.
\eeq
Denoting by $G_k(\tau,\tau')$ the Green's function of $\pi$, the formal solution of the previous equation is given by 
\beq
\hat{\pi}_{\textbf{k}}(\tau) \;=\; \hat{\pi}^{(0)}_{\textbf{k}}(\tau) - \int d\tau'\,G_k(\tau,\tau')\, \frac{a(\tau')}{2c(\tau')}\frac{dm^2(\tau')}{d\tau'}\int \frac{d^3\bp}{(2\pi)^{3/2}} \hat{\chi}_{\bp}(\tau')\hat{\chi}_{\bp-\bk}(\tau')\,,
\eeq
where $\hat{\pi}^{(0)}_{\textbf{k}}(\tau)$ denotes the homogeneous solution. 

We now proceed to compute the two-point function of $\pi$ sourced by the spectator field $\chi$. It can be readily verified that the cross term involving $\hat{\pi}^{(0)}_\textbf{k}$ and the $\hat{\chi}$-dependent part of $\hat{\pi}$ vanishes. Therefore, the inhomogeneous component of the two-point function takes the form
\begin{align}\notag
\langle 0| \hat{\pi}_{\textbf{k}}(\tau)\hat{\pi}_{\textbf{k}'}(\tau) |0\rangle \;=\; \frac{1}{4}\int d\tau'&\,d\tau''\, G_k(\tau,\tau')G_k(\tau,\tau'') \frac{a(\tau')}{c(\tau')}\frac{a(\tau'')}{c(\tau'')}\frac{dm^2(\tau')}{d\tau'}\frac{dm^2(\tau'')}{d\tau''}\\ \label{eq:pipi1}
&\times \int \frac{d^3\bp}{(2\pi)^{3/2}}\frac{d^3\bq}{(2\pi)^{3/2}}\, \left\langle 0\left|\hat{\chi}_{\bp}(\tau')\hat{\chi}_{\bp-\bk}(\tau')\hat{\chi}_{\bq}(\tau'')\hat{\chi}_{\bq-\bk'}(\tau'') \right|0\right\rangle \,.
\end{align}
We use an explicit bra-ket notation for (quantum) vacuum expectation values, which account for the effect of zero-point fluctuations. These quantum expectation values should be distinguished from statistical averages computed over the ensemble of possible realizations of the disorder $m^2(t)$. We will introduce these statistical averages in the upcoming sections.\\

The six-dimensional momentum integral over the unequal time correlator of the four $\chi$s appearing in equation \eqref{eq:pipi1} can be unpacked further. Introducing the canonically normalized field
\beq\label{eq:Xdef}
\hat{X}\equiv a \hat{\chi}\,,
\eeq
the mode expansion of the spectator can be written as 
\beq\label{eq:Xmodeexp}
\hat{X}_\textbf{k}(\tau) \;=\; X_k(\tau) \hat{a}_{\bk} + X_k^{*}(\tau) \hat{a}^{\dagger}_{-\bk}\,,
\eeq
where $[\hat{a}_{\bk},\hat{a}^{\dagger}_{\bk'}] = \delta^{(3)}(\bk-\bk')$, $[\hat{a}_{\bk},\hat{a}_{\bk'}] = [\hat{a}^{\dagger}_{\bk},\hat{a}^{\dagger}_{\bk'}]=0$. This expansion leads to the consistent quantization of $\hat{X}$ provided that its mode functions satisfy the Wronskian condition $X_k(\tau)X_k^{*\prime}(\tau) - X_k'(\tau)X_k^{*}(\tau)=i$, and reduce to Bunch-Davies mode functions in the infinite past. Substitution of (\ref{eq:Xdef}) and (\ref{eq:Xmodeexp}) into (\ref{eq:pipi1}) leads to the following expression after the Wick decomposition of the $\hat{\chi}$ four-point function,
\begin{align}\notag
\langle 0| \hat{\pi}_{\textbf{k}}(\tau)\hat{\pi}_{\textbf{k}'}(\tau) |0\rangle \;=\; &\frac{1}{4}\int d\tau'\,d\tau''\, \frac{G_k(\tau,\tau')}{a(\tau')c(\tau')}\frac{G_k(\tau,\tau'')}{a(\tau'')c(\tau'')} \frac{dm^2(\tau')}{d\tau'}\frac{dm^2(\tau'')}{d\tau''}\\ \notag
&\times \int \frac{d^3\bp}{(2\pi)^{3/2}}\frac{d^3\bq}{(2\pi)^{3/2}}\, \Big\{ \langle 0|\hat{X}_{\bp}(\tau')\hat{X}_{\bp-\bk}(\tau') |0\rangle \langle 0| \hat{X}_{\bq}(\tau'')\hat{X}_{\bq-\bk'}(\tau'') |0\rangle \\ \notag
&\hspace{94pt} + \langle 0|\hat{X}_{\bp}(\tau')\hat{X}_{\bq}(\tau'')|0\rangle \langle 0| \hat{X}_{\bp-\bk}(\tau') \hat{X}_{\bq-\bk'}(\tau'') |0\rangle\\ \label{eq:pipi2}
&\hspace{94pt} + \langle 0|\hat{X}_{\bp}(\tau')\hat{X}_{\bq-\bk'}(\tau'')|0\rangle \langle 0| \hat{X}_{\bp-\bk}(\tau') \hat{X}_{\bq}(\tau'') |0\rangle \Big\}  \,.
\end{align}
The first term inside the momentum integral in the above equation corresponds to a pure $\mathbf{k}=0$ contribution. We can verify this using commutation relations for the creation and annihilation operators. Such a zero-mode contribution is irrelevant for our computation and can be promptly discarded. The remaining terms must be kept, but lead to divergences in the ultraviolet. We regularize the field correlators by means of the {\em adiabatic subtraction} (AS) scheme~\cite{Parker:1974qw,Birrell:1977,Anderson:1987yt,Fulling:1989nb,parker2009quantum,Kohri:2017iyl}. As a concrete example,  
\begin{align} \notag
\langle 0|\hat{X}_{\bp}(\tau')\hat{X}_{\bq}(\tau'')|0\rangle_{\rm AS} \;&\equiv\; \langle 0|\hat{X}_{\bp}(\tau')\hat{X}_{\bq}(\tau'')|0\rangle - \langle 0|\hat{X}_{\bp}^{0}(\tau')\hat{X}_{\bq}^{0}(\tau'')|0\rangle\\ \notag
&=\; \left[  X_p(\tau')X_q^*(\tau'') - X_p^0(\tau')X_q^{0*}(\tau'') \right]\,\delta^{(3)}(\bp+\bq)\\
&\equiv\; \left[  X_p(\tau')X_q^*(\tau'') \right]_{\rm AS}\,\delta^{(3)}(\bp+\bq)\,.
\end{align}
Here $\hat{X}_\bp^0(\tau)$ denote the adiabatic (unexcited) modes of the spectator field. In the absence of a non-adiabatic sourcing for $\chi$, the contribution to the $\pi$ two-point function (\ref{eq:pipi2}) vanishes identically. 

Substitution of the adiabatically subtracted correlators into (\ref{eq:pipi2}) and an integration with respect to $\bq$ leads to the  following final form for the $\pi$ two-point function:
\begin{align}\notag
\langle 0| \hat{\pi}_{\textbf{k}}(\tau)\hat{\pi}_{\textbf{k}'}(\tau) |0\rangle \;=\; &\frac{1}{2}\int d\tau'\,d\tau''\, \frac{G_k(\tau,\tau')}{a(\tau')c(\tau')}\frac{G_k(\tau,\tau'')}{a(\tau'')c(\tau'')} \frac{dm^2(\tau')}{d\tau'}\frac{dm^2(\tau'')}{d\tau''}\\ \label{eq:pipi3}
&\times \int \frac{d^3\bp}{(2\pi)^{3}}\,  \left[  X_p(\tau')X_p^*(\tau'') \right]_{\rm AS}  \big[  X_{|\bp-\bk|}(\tau')X_{|\bp-\bk|}^*(\tau'') \big]_{\rm AS} \, \delta^{(3)}(\bk+\bk') \,. 
\end{align}

\subsubsection*{The Curvature Power Spectrum}
Recall that the comoving curvature perturbation $\zeta$ is related to the Goldstone boson $\pi$ by the rescaling
\beq
\zeta \;\simeq\; H\pi\,,
\eeq
and its dimensionless power spectrum is in turn defined by the relation
\beq
\langle0| \zeta(\bk)\zeta(\bk') |0\rangle \;\equiv\;\frac{2\pi^2}{k^3} \Delta_{\zeta}^2(k)\, \delta^{(3)}(\bk+\bk')\,.
\eeq
We now recall that we have assumed that the background dynamics are independent of the spectator $\chi$, and correspond to those of an expanding inflationary spacetime. Thus, the component of the curvature power spectrum that is not sourced by $\chi$ is given by (assuming slow-roll):
\beq\label{eq:zetasr}
\Delta_{\zeta,0}^2(k) \;\simeq\; \frac{H^2}{8M_P^2 \pi^2\epsilon} \;=\; \frac{H^4}{4\pi^2c}\,,
\eeq
where $\epsilon = -\dot{H}/{H^2}$ denotes the first Hubble flow function~\cite{Liddle:2003as,Leach:2002ar}, which we assume to be approximately constant and small, $\epsilon\ll 1$, as is required to support typical slow-roll inflation. When needed, we will use the 2018 Planck value for the amplitude of the curvature spectrum
\beq
\Delta_{\zeta}^2(k_{\star})_{\rm Planck} \;=\; 2.1\times 10^{-9} \, ,
\eeq
at the pivot scale $k_{\star}=0.05\,{\rm Mpc}^{-1}$~\cite{Aghanim:2018eyx}. Note that the measured spectrum of course includes the sum of the sourced and unsourced spectra; the above value for the unsourced spectra will nevertheless serve as a useful benchmark. From here onwards we will denote by $\Delta_{\zeta,0}^2$ the vacuum (or adiabatic) contribution to the power spectrum, and by $\delta \Delta_{\zeta}^2$ the non-adiabatically sourced term, so that 
\Beq
\Delta_{\zeta}^2 = \Delta_{\zeta,0}^2 + \delta \Delta_{\zeta}^2\,.
\Eeq
For observationally constrained scales we must therefore have $\Delta_{\zeta}^2 = \Delta_{\zeta,\,{\rm Planck}}^2$.

Combining (\ref{eq:pipi3})-(\ref{eq:zetasr}) we arrive to the following expression for the component of the power spectrum sourced by $\chi$:
\begin{align}\notag
\delta\Delta_{\zeta}^2(k) \;=\; 4\pi^2 (\Delta_{\zeta,0}^2)^2 \frac{k^3}{H^6} \int d\tau'&\,d\tau''\,  \frac{G_k(\tau,\tau')}{a(\tau')}\frac{G_k(\tau,\tau'')}{a(\tau'')} \frac{dm^2(\tau')}{d\tau'}\frac{dm^2(\tau'')}{d\tau''}\\ \label{eq:deltazeta}
&\times \int \frac{d^3\bp}{(2\pi)^{3}}\, \left[  X_p(\tau')X_p^*(\tau'') \right]_{\rm AS}  \big[  X_{|\bp-\bk|}(\tau')X_{|\bp-\bk|}^*(\tau'') \big]_{\rm AS} \,.
\end{align}
Note that the right-hand side of (\ref{eq:deltazeta}) is multiplied by two powers of the unperturbed power spectrum, which implies that the stochastic excitation of the curvature mode must overcome a suppression of at least $\mathcal{O}(10^{-9})$ relative to the background value. This suppression will of course be larger if we expect $\delta \Delta_{\zeta}^2$ to dominate over the vacuum contribution to the power spectrum, $\Delta_{\zeta}^2 \simeq \delta \Delta_{\zeta}^2$. 

To explicitly calculate $\delta \Delta_\zeta^2$, we need (i) the quasi-de Sitter Green's function for $\pi$, (ii) the form of the effective mass $m^2(t)$ and the corresponding time evolution of the modes of the spectator field $\chi=X/a$. 

\subsection*{Quasi-de Sitter Green's Functions}\label{sec:greens}

In order to evaluate the power spectrum correction (\ref{eq:deltazeta}), it would be useful to have a compact, closed-form solution for the Green's function of the Goldstone $\pi$. To this end, first let $\pi_k^{(0)}$ be a solution to equation of motion Eq.~\eqref{eq:eompik} without a source term. The equation satisfied by $\pi_k^{(0)}$ can can be rewritten in a more familiar form as
\beq
v_{k}'' + \left(k^2-\frac{a''}{a}\right)v_k \;=\; 0\,,
\eeq
where $ v_k(\tau) \equiv \pi^{(0)}_{k}(\tau)\,a(\tau)$. In the slow-roll approximation, $\epsilon\ll 1$, this equation reduces to
\beq\label{eq:eomvsr}
 v_k'' + \left(k^2 - \frac{\nu^2-\frac{1}{4}}{\tau^2}\right) v_k \;\simeq\; 0\,,
\eeq
where $\nu = \frac{3}{2}+\epsilon$. The solution of (\ref{eq:eomvsr}) that satisfies the Wronskian condition $v_k(\tau)v_k^{* \prime}(\tau) - v_k'(\tau)v_k^{*}(\tau)=i/c$, and which reduces to the Bunch-Davies vacuum initial condition for the canonically normalized field $\sqrt{c}v_k$, is given by
\beq
v_k(\tau) \;=\; \sqrt{\frac{\pi}{4c}}(-\tau)^{1/2}H_{\nu}^{(1)}(-k\tau)\,,
\eeq
where $H^{(1)}_{\nu}$ denotes the Hankel function of the first kind. Therefore, the corresponding mode function for the Goldstone field can be written as
\beq
\pi_k^{(0)}(\tau) \;=\; H\sqrt{\frac{\pi}{4c}}(-\tau)^{3/2}H_{\nu}^{(1)}(-k\tau)\,. \label{eqn:Pi0}
\eeq
Using this solution, we arrive at the Green's function for equation \eqref{eq:eompik}:
\begin{align} \displaybreak[0] \notag
G_k(\tau,\tau')\;&=\; \frac{\pi_k^{(0)}(\tau)\pi_k^{(0)*}(\tau')-\pi_k^{(0)}(\tau')\pi_k^{(0)*}(\tau)}{\partial_{\tau} \pi_k^{(0)}(\tau') \,\pi_k^{(0)*}(\tau')-\pi_k^{(0)}(\tau')\,\partial_{\tau} \pi_k^{(0)*}(\tau')} \theta(\tau-\tau')\\ \displaybreak[0] \label{eq:green1}
&= \frac{\pi (-k\tau)^{3/2}}{2k\sqrt{-k\tau'}}\left[ J_{\nu}(-k\tau)Y_{\nu}(-k\tau')-Y_{\nu}(-k\tau)J_{\nu}(-k\tau') \right] \theta(\tau-\tau')\\\displaybreak[0]
&= \frac{[k\tau'-k\tau]\cos(k\tau - k\tau')+[1+(k\tau)(k\tau')]\sin(k\tau-k\tau')]}{k^3\tau'^2} \theta(\tau-\tau') + \mathcal{O}(\epsilon)\notag\,,
\end{align}
where, in the second line, $J_\nu$ and $Y_\nu$ are Bessel functions of the first and second kind, and we have expanded in the slow roll parameter in the third line. Note that $\tau$ is the conformal time where we wish to evaluate the final curvature perturbation, whereas $\tau'$ is the integration variable over which the source term (multiplied by this Green's function) will be integrated (see  equation \eqref{eq:deltazeta}). For the curvature perturbations from inflation, we are typically interested in power spectra (and hence the Green's function) on superhorizon scales with $|k\tau|\ll1$. However, $|k\tau'|$ is not restricted a-priori apart from $|k\tau'|>|k\tau|$.

For future convenience, we note that
\beq\label{eq:notcurlyg}
\frac{G_k(\tau,\tau')}{\tau'}\;\simeq\; 
\begin{cases}
\dfrac{\cos(k\tau')}{(k\tau')^2}\,, &|k\tau|\ll 1\ll |k\tau'|\,,\\[10pt]
-\dfrac{1}{3}\,, &|k\tau|\ll |k\tau'|\ll1\,.
\end{cases}
\eeq
It will also turn out to be useful to define the following derived quantity, 
\beq\label{eq:curlyg}
\mathcal{G}_k(\tau,\tau')\;\equiv\; \frac{dG_k(\tau,\tau')}{d\tau'} + \frac{G_k(\tau,\tau')}{\tau'}\,,
\eeq
which yields
\beq\label{eq:curlygapp}
\mathcal{G}_k(\tau,\tau')\;\simeq\; 
\begin{cases}
-\dfrac{\sin(k\tau')}{k\tau'}\,, &|k\tau|\ll 1\ll |k\tau'|\,,\\[10pt]
-\dfrac{2}{3}\,, &|k\tau|\ll |k\tau'|\ll1\,,
\end{cases}
\eeq
to lowest order in slow-roll.

\subsection{Non-adiabatically Excited Spectator Fields in de Sitter Space}\label{sec:spectfield}

The final and most important ingredients that go into the evaluation of the power spectrum correction (\ref{eq:deltazeta}) correspond to the effective mass $m^2(t)$ and the spectator field $\chi$.  The equation of motion for the $k$-mode of the $\chi$ field is given by
\beq\label{eq:eomchi}
\left(\frac{d^2}{dt^2} + 3H\frac{d}{dt} + \frac{k^2}{a^2} + M^2 + m^2(t)\right) \chi_k(t) \;=\;0\,,
\eeq
where we have ignored the sourcing of $\chi$ by $\pi$ perturbations.  Discarding such ``sourcing" amounts to ignoring $\pi dm^2(t)/dt$  compared to $m^2(t)$ in the equation of motion for $\chi$. As a heuristic justification, note that for a typical non-adiabatic variation in $m^2(t)$ with a temporal width $w$, we have $\pi dm^2(t)/dt\sim \pi m^2(t)/w\sim (\zeta/Hw) m^2(t)$. Hence, we can potentially ignore the impact of $\pi$ on $\chi$ if $Hw\gg \zeta$ (see also the discussion at the end of this section, and see footnote \ref{fn:Hw}).

We assume that the effective mass consists of localized, non-adiabatic events (scatterings), which have random strengths and which are randomly located in cosmic time. Further assuming that the temporal width $w$ of the scatterers is much smaller than the characteristic period of the mode-functions $\chi_k$, we approximate the stochasticity of the background dynamics by a sum of Dirac-delta scatterers, 
\begin{align} \notag
m^2(t) \;&=\; \sum_j m_j\, \delta(t-t_j)\\ \label{eq:mdelta}
&=\; \sum_j \frac{m_j}{a(\tau_j)}\, \delta(\tau-\tau_j) \;=\; m^2(\tau)\,.
\end{align}
Note that when needed, the delta function can be ``fattened", so that $\delta(t-t_i)\sim 1/w$ over a time-interval $w$ around $t_i$ and zero otherwise. For the scaled mode functions $X_k=a\chi_k$, the equation of motion is given by
\beq\label{eq:eomXk}
X_k''(\tau) + \left[ k^2 - \frac{a''}{a} + a^2M^2 + \sum_i m_i a(\tau_i) \delta (\tau-\tau_i) \right] X_k(\tau)\;=\;0\,.
\eeq
In earlier work ~\cite{StochasticSpectator}, some of us investigated the dynamics of the ensemble of solutions (realizations) for the mode functions $X_k(\tau)$ for two particular cases: for {\em conformally massive fields} ($M^2=2H^2$) and for {\em massless fields} ($M^2=0$) under the condition that the scatterer amplitudes have zero-mean and are independent at different times,
\beq\label{eq:mstats}
\langle m_j\rangle = 0\,,\qquad \langle m_j m_i\rangle = \sigma^2 \delta_{ij}\,.
\eeq
In the above expressions, $\langle \cdots \rangle$ denotes taking the ensemble average with respect to a distribution,
and that the scatterer locations are distributed uniformly over cosmic time. 

Below, we summarize the results derived in  \cite{StochasticSpectator} related to the stochastic excitation of the $X_k$ which are relevant for the present paper (i.e.~for the conformal mass scenario). 

\begin{figure}[!t] 
   \centering
      \includegraphics[width=0.7\textwidth]{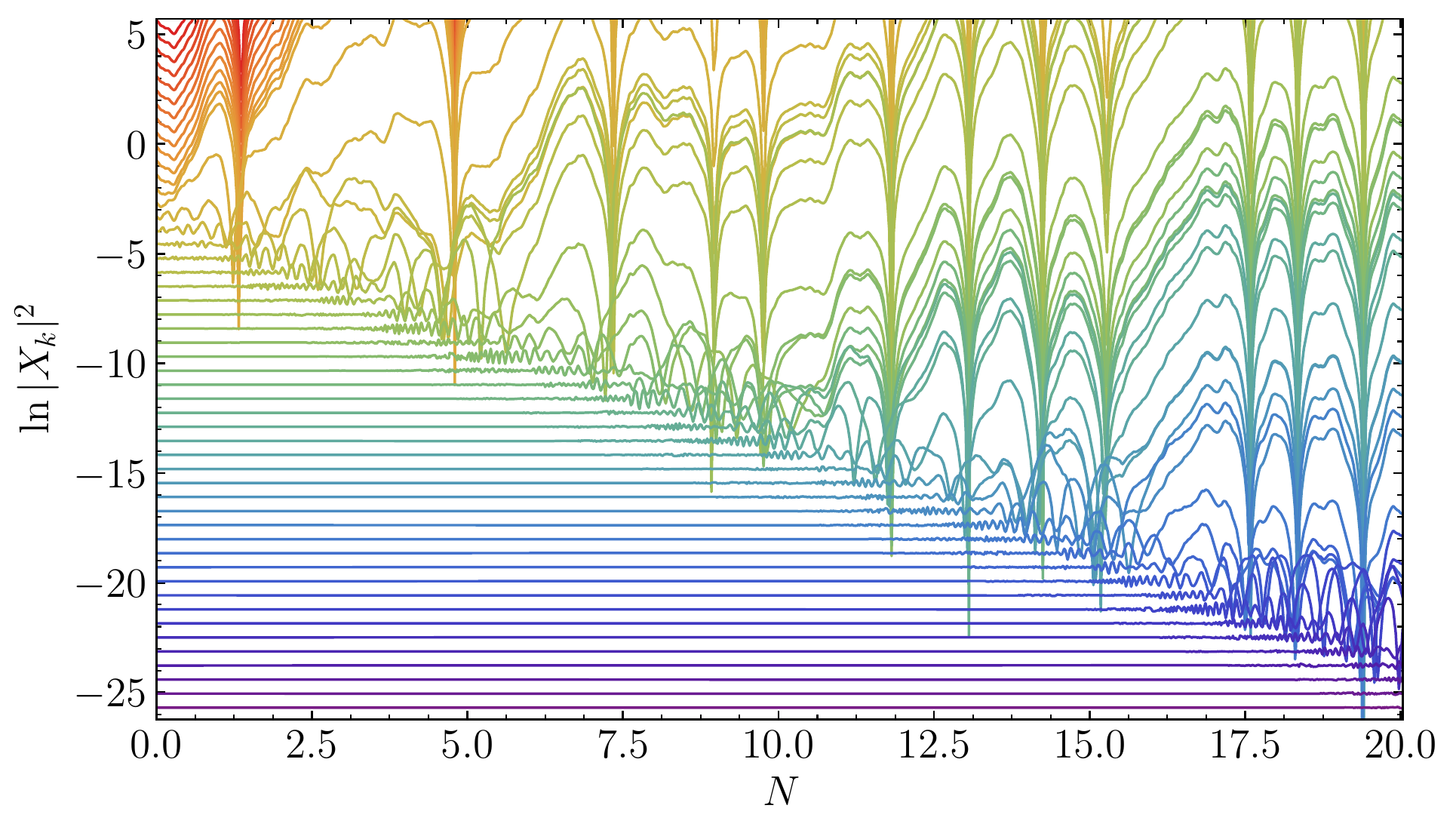} 
   \includegraphics[width=0.7\textwidth]{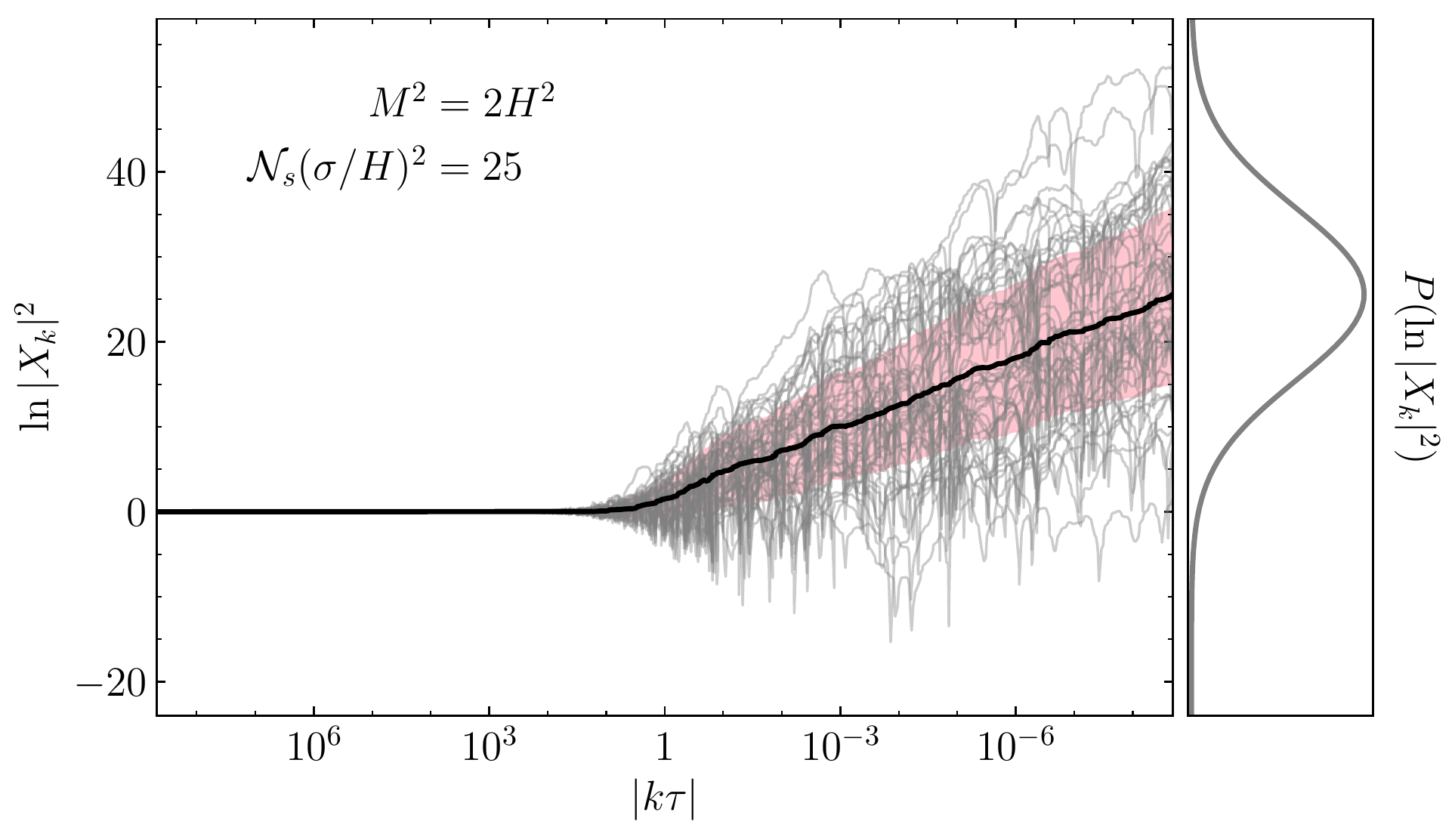} 
   \caption{Top Panel: The behavior of $\ln |X_k|^2$ for $M^2=2H^2$, for different $k$ as a function of $N=\ln a$ for a fixed realization of $m^2(t)$. The colors represent different $k$ modes, with purple having the largest $k$, to red having the smallest. The non-trivial evolution of $|X_k|$ is due to repeated non-adiabatic particle  production. The different $k$ modes approximately follow each other deep inside the horizon (straight lines). The relative amplitude of different $k$ modes typically changes stochastically close to horizon crossing (and can even reverse order) and, this relative amplitude is thereafter maintained for superhorizon evolution. Bottom Panel: The behavior of $\ln |X_k|^2$ normalized with respect to $|X_k^0|^2$, for different realizations of $m^2(t)$, as a function of conformal time. The grey curves are $\ln|X_k|^2$ corresponding to different realizations of $m^2(t)$, the black curve is the ensemble mean of $\ln |X_k|^2$ and the pink region represents trajectories within one standard deviation of the mean. The instantaneous probability distribution of $\ln |X_k|^2$ is shown in the right panel. The trajectories of $\ln |X_k|^2$ undergo a ``random walk" like behavior, and  have a Gaussian distribution (over the ensemble) at all times, i.e.~$|X_k|$ is log-normally distributed.}
  
   \label{fig:trajectories_chi}
\end{figure}
\begin{enumerate}
\item In absence of scattering, $m^2(t)=0$, the vacuum mode function takes a particularly simple form given by 
\Beq
X^0_k(\tau)=a(\tau){\chi^0_k(\tau)}=\frac{1}{\sqrt{2k}}e^{-ik\tau}\,. \label{eqn:modevac}
\Eeq
Under the influence of repeated non-adiabatic changes in $m^2(t)$, the mode function $X_k(\tau)$ deviates from the vacuum solution (\ref{eqn:modevac}). The evolution of $X_k$ becomes stochastic, with the largest deviations from the vacuum mode functions typically seen once the $X_k$ mode is outside the horizon. See top panel in Fig.~\ref{fig:trajectories_chi}. Note that despite the seemingly similar behavior for different $k$ modes inside and outside the horizon, there is stochasticity close to horizon crossing. While the behavior of $X$ modes seen in Fig.~\ref{fig:trajectories_chi} is useful for developing intuition, the translation from the behavior of $X$ modes to the power spectrum which we will compute in the next section is not immediate, since it involves non-trivial integrals over many momentum modes of $X$.
\item In the limit of large scatterer density in a given interval $\Delta t$ in cosmic time, the cha\-rac\-te\-ris\-tic ``strength'' of the non-adiabatic events can be uniquely determined by the so-called scattering strength parameter
\begin{equation}\label{eq:paramdef}
\mathcal{N}_s \left(\frac{\sigma}{H}\right)^2 \;\equiv\; \frac{N_s}{H\Delta t} \left(\frac{\sigma}{H}\right)^2\,,
\end{equation}
where $N_s$ denotes the number of scatterings, and $\mathcal{N}_s$ denotes the (dimensionless) density of scatterers (number of scatterers per $e$-fold). This result is independent of the distribution of the scatterer amplitudes $m_i$, provided that it satisfies (\ref{eq:mstats}), and that the locations $t_i$ are roughly uniformly distributed.\footnote{E.g.~uniformly distributed $t_i$ over a given time interval, or normally distributed $t_i$ centered on an equispaced grid in cosmic time. The result is also valid for a non-random, equispaced grid of $t_i$.}

\item The random variable $\ln |X_k(t)|^2$ is normally distributed (as an ensemble over realizations of $m^2(t)$) on super and sub-horizon scales, for any scattering strength and for any $k$. Equivalently, $|X_k(t)|^2$ is log-normally distributed.

\begin{figure}[!t]
\centering
    \includegraphics[width=0.75\textwidth]{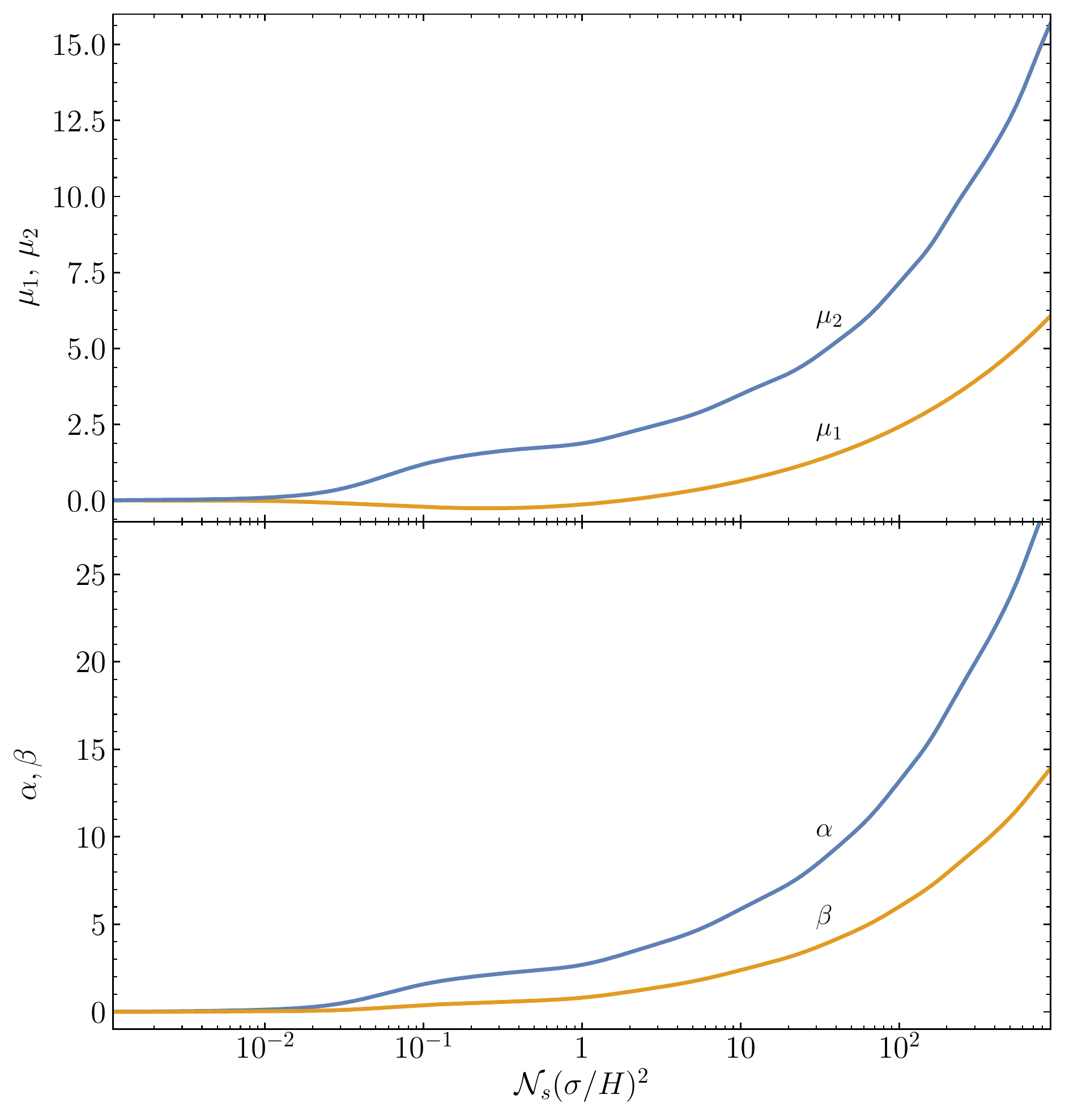}
    \caption{Numerically evaluated, ensemble averaged growth rates for super-horizon evolution of the spectator field magnitude, $|X_k|$, as functions of $\mathcal{N}_s (\sigma/H)^2$ in the conformal case. Upper panel: The mean and variance rates: $\mu_1=\partial_{Ht}\langle \ln |X_k|^2\rangle$ and $\mu_2=\partial_{Ht}\,{\rm Var}[ \ln |X_k|^2]$ are shown (adapted from \cite{StochasticSpectator}). Lower panel: $\alpha=\mu_1+(3/2)\mu_2$ and $\beta=\mu_1+(1/2)\mu_2$ which are are linear combinations of the growth rates are shown. These combinations be useful when discussing the ensemble averaged curvature perturbation power spectra.}
    \label{fig:f1f2}
\end{figure}

\item Sufficiently deep inside the horizon, the scalar field is approximately in its vacuum state,
\begin{flalign} \label{eq:lnchisubh}
& \text{($|k\tau|\gg 1$)} & \Cen{3}{
\begin{aligned}
\langle \ln |X_k|^2\rangle \;&\simeq\; - \ln(2k)\,,\\
{\rm Var}\left[ \ln |X_k|^2\right] \;&=\; \frac{1}{4}\mathcal{N}_s\left(\frac{\sigma}{H}\right)^2\left(k\tau\right)^{-2} \;\ll\; 1\,.
\end{aligned}}      &&  
\end{flalign}
Note that the vacuum approximation strictly coincides with the sub-horizon regime only for $\S \sim \mathcal{O}(1)$.\footnote{For $\S\gg 1$, the vacuum approximation is broken somewhat inside the horizon, whereas for $\S\ll 1$ it continues to be valid even outside the horizon.} Outside the horizon, $\ln|X_k|$ evolves linearly with cosmic time (in an ensemble averaged sense), with
\begin{flalign} \label{eq:lnchirateconf}
& \text{($|k\tau|\ll 1$)} & \Cen{3}{
\begin{aligned}
\partial_{Ht} \langle \ln |X_k|^2\rangle \;&=\; \mu_1\,,\\
\partial_{Ht} {\rm Var}\left[ \ln |X_k|^2\right] \;&=\; \mu_2\,, 
\end{aligned}}      &&  
\end{flalign}
where the rates $(\mu_1,\mu_2)$ are functions of $\S$. The values of $\mu_1$ and $\mu_2$ as a function of $\S$ are shown in the upper panel of Fig.~\ref{fig:f1f2}. For future convenience we define the following functions of $\S$,
\Beq \label{eq:alphabeta}
&\alpha=\mu_1+\frac{3}{2}\mu_2 \,,\\
&\beta=\mu_1+\frac{1}{2}\mu_2  \,,
\Eeq
which are shown in the bottom panel of Fig.~\ref{fig:f1f2}.
\item On super-horizon scales, $\ln |X_k|^2$ satisfies the properties of a drifted (Brownian) random walk. In particular, as mentioned above, the mean and variance of $\ln |X_k|^2$ grow linearly with time, and for the drift-less variable 
\beq
Z_k(t) \;\equiv\; \ln |X_k|^2-\langle \ln|X_k|^2\rangle\,,
\eeq
we find 
\begin{align}\notag
\langle Z_{k}(t) Z_{k'}(t') \rangle \;\simeq \; \mu_2 H &\min \left[t-t_k,t'-t_k,t-t_{k'},t'-t_{k'}\right]\\ \label{eq:twopointconf}
&\times\theta(t-t_k)\theta(t-t_{k'})\theta(t'-t_k)\theta(t'-t_{k'}) \,,
\end{align}
where $t_k, t_{k'}$ indicate the time of horizon crossing for $k$ and $k'$ modes. This is equivalent to $|X_k|^2$ performing a geometric random walk. All $n$-point correlation functions for the field magnitude can then be computed in terms of the $Z_k$ two point functions, 
\beq\label{eq:Xnpoint}
\langle |X_{k_1}(t_1)|^2\cdots |X_{k_n}(t_n)|^2 \rangle \;=\; \exp\left[\sum_{i=1}^n \langle \ln|X_{k_i}(t_i)|^2\rangle + \frac{1}{2}\sum_{i,j=1}^n \langle Z_{k_i}(t_i)Z_{k_j}(t_j) \rangle \right]\,.
\eeq
\end{enumerate}
We encourage the interested reader to consult the source of these results in our earlier paper \cite{StochasticSpectator}. These results were derived by a combination of analytical and numerical methods using the Transfer Matrix and Fokker-Planck formalisms. It is worth noting that the preceding results technically apply only for a pure de Sitter background, but we do not expect them to be significantly altered by slow-roll corrections.

\subsubsection*{Ignoring Dissipation}\label{footnote:dissipation}
Before ending this section, we briefly comment on the issue of dissipation of curvature perturbations (see the general discussion in \cite{LopezNacir:2011kk}). So far, we have set up the formalism for calculating the sourcing of curvature fluctuations ($\zeta$) from the excited spectator fields ($X$), but have ignored the backreaction of $\zeta$ on $X$. Formally, the source term in the equation of motion for the curvature perturbations (cf.~\eqref{eq:eompik}) includes a term independent of $\zeta$, and a {\it linear response} (dissipation) term that has correlations with curvature perturbations.\footnote{Both contributions are included in the source in the sense that the $\chi$ has a component due to the adiabatic particle production, and due to fluctuations in the curvature perturbations.} Because of such correlations, it is a-priori not possible to ignore the contribution of the linear response when calculating the sourced curvature power spectrum based just on the smallness of $\zeta$. 

However, it is still plausible to ignore the linear response contribution to $\Delta_\zeta^2$. Schematically, we note that the linear response contribution to $\Delta_\zeta^2$ involves unequal time {\it commutators} (Green's function) associated with the spectator field. In contrast, the contribution to $\Delta_\zeta^2$ in absence of the linear response term involves unequal time {\it correlators} of the spectator fields. Since the commutator only cares about relative differences between $X$ at different times, its growth ends up being suppressed compared to the correlators (at least in the regime of exponential $X$ growth outside the horizon which we are most interested in). 

We leave a more detailed consideration of these dissipation effects to future work.

\section{Sourced Curvature Power Spectrum: Sample Computation} \label{sec:stochform}

In the previous section, we presented the formalism and the tools required to compute the curvature power spectrum, when it is sourced by a stochastically excited spectator field. In what follows, we will perform this computation explicitly, assuming that the spectator field evolves in a way that is described in Section \ref{sec:spectfield}. A number of details related to the justification of our assumptions, as well as techniques for the numerical calculations are relegated to the Appendices.

\subsection{Dirac-delta Scatterers in the Large $\mathcal{N}_s$ Limit} \label{sec:ddscatterers}
In the following, we arrive at a simplified expression for the sourced curvature power spectrum by considering a sum of Dirac-delta scatterers for characterizing the effective mass of the spectator field. We carefully include a ultraviolet cutoff in momentum integrals  (related to the more realistic finite width for the scatterers) to regulate ultraviolet divergences that arise as a result of the Dirac-Delta scatterer assumption. By considering a large number density of scatterers, we argue that the contributions to the power spectrum which depend on the momentum cutoff do not qualitatively change our results.

Substituting the effective mass (\ref{eq:mdelta}) into (\ref{eq:deltazeta}), we find that the correction to the power spectrum sourced by $X$ is
\begin{align}\notag
\delta\Delta_{\zeta}^2(k) \;& =\; 4\pi^2 (\Delta_{\zeta,0}^2)^2 \frac{k^3}{H^6} \sum_{i,j} m_i m_j \int d\tau' \,d\tau''\, \frac{\delta(\tau'-\tau_i)}{a(\tau')} \frac{\delta(\tau''-\tau_j)}{a(\tau'')}  \frac{d}{d\tau'} \frac{d}{d\tau''} \bigg\{ \frac{G_k(\tau,\tau')}{a(\tau')}\frac{G_k(\tau,\tau'')}{a(\tau'')} \\ \label{eq:deltadeltazeta}
&\hspace{100pt} \times \int \frac{d^3\bp}{(2\pi)^{3}}\, \left[  X_p(\tau')X_p^*(\tau'') \right]_{\rm AS}  \big[  X_{|\bp-\bk|}(\tau')X_{|\bp-\bk|}^*(\tau'') \big]_{\rm AS} \bigg\} \,.
\end{align}
where we performed integration by parts in $\tau'$ and $\tau''$, and discarded the corresponding boundary terms, as we assume that the stochastic excitations only occur over a finite period of time. Distributing the conformal time derivatives, and taking into account the discontinuous nature of $X'_k(\tau)\equiv \partial_{\tau}X_k(\tau)$ at each Dirac-delta scatterer location, we may rewrite (\ref{eq:deltadeltazeta}) schematically as
\beq\label{eq:deltazetaK}
\delta\Delta_{\zeta}^2(k) \;=\; 4\pi^2 (\Delta_{\zeta,0}^2)^2 \sum_{i,j} \frac{m_i m_j}{H^2}(k\tau_i)^2(k\tau_j)^2 \left( \mathcal{K}^{\rm I}_{ij} + \mathcal{K}^{\rm II}_{ij} + \mathcal{K}^{\rm III}_{ij} \right)\,,
\eeq
where the time integrals are evaluated over the delta functions, and the momentum integral is now represented by the summation of the quantities $\mathcal{K}_{ij}$, defined as follows:
\begin{align} \label{eq:K1} \displaybreak[0] 
\mathcal{K}_{ij}^{\rm I} \;&=\;  \mathcal{G}_k(\tau,\tau_i) \mathcal{G}_k(\tau,\tau_j) \int \frac{d^3\bp}{(2\pi)^{3}k}\, \left[  X_p(\tau_i)X_p^*(\tau_j) \right]_{\rm AS}  \big[  X_{|\bp-\bk|}(\tau_i)X_{|\bp-\bk|}^*(\tau_j) \big]_{\rm AS}\,,\\[10pt] \notag
\mathcal{K}_{ij}^{\rm II} \;&=\; \mathcal{G}_k(\tau,\tau_i) G_k(\tau,\tau_j)       \int \frac{d^3\bp}{(2\pi)^{3}k}\, \bigg\{ \big[  X_p(\tau_i)X_p^{*\prime}(\tau_j^{-}) \big]_{\rm AS}  \big[  X_{|\bp-\bk|}(\tau_i)X_{|\bp-\bk|}^{*}(\tau_j) \big]_{\rm AS} \\ \notag \displaybreak[0]
&\hspace{100pt} + \big[  X_p(\tau_i)X_p^{*}(\tau_j) \big]_{\rm AS}  \big[  X_{|\bp-\bk|}(\tau_i)X_{|\bp-\bk|}^{*\prime}(\tau_j^{-}) \big]_{\rm AS} + {\rm h.c.}\bigg\}\\ \notag 
&\hspace{15pt} + \frac{1}{2}\left(\frac{m_i}{H\tau_i} \mathcal{G}_k(\tau,\tau_j) G_k(\tau,\tau_i)  + \frac{m_j}{H\tau_j} \mathcal{G}_k(\tau,\tau_i) G_k(\tau,\tau_j) + \frac{m_i m_j}{2H^2\tau_i\tau_j} G_k(\tau,\tau_i) G_k(\tau,\tau_j) \right) \\ \notag
&  \hspace{35pt}  \times \int \frac{d^3\bp}{(2\pi)^{3}k}\,   \bigg\{ X_p(\tau_i)X_p^{*}(\tau_j) \big[  X_{|\bp-\bk|}(\tau_i)X_{|\bp-\bk|}^*(\tau_j) \big]_{\rm AS}\\ \label{eq:K2}
&\hspace{108pt} + \big[  X_p(\tau_i)X_p^{*}(\tau_j) \big]_{\rm AS}  X_{|\bp-\bk|}(\tau_i)X_{|\bp-\bk|}^*(\tau_j)   \Big) \bigg\}\,, \\[10pt] \notag \displaybreak[0]
\mathcal{K}_{ij}^{\rm III} \;&=\; G_k(\tau,\tau_i) G_k(\tau,\tau_j)   \int \frac{d^3\bp}{(2\pi)^{3}k}\, \bigg\{ \big[  X_p^{\prime}(\tau_i^{-})X_p^{*\prime}(\tau_j^{-}) \big]_{\rm AS}  \big[  X_{|\bp-\bk|}(\tau_i)X_{|\bp-\bk|}^*(\tau_j) \big]_{\rm AS} \\ \notag
&\hspace{100pt} + \big[  X_p^{\prime}(\tau_i^{-})X_p^{*}(\tau_j) \big]_{\rm AS}  \big[  X_{|\bp-\bk|}(\tau_i)X_{|\bp-\bk|}^{*\prime}(\tau_j^{-}) \big]_{\rm AS}\\ \notag
&\hspace{100pt} + \big[  X_p(\tau_i)X_p^{*\prime}(\tau_j^{-}) \big]_{\rm AS}  \big[  X_{|\bp-\bk|}^{\prime}(\tau_i^{-})X_{|\bp-\bk|}^*(\tau_j) \big]_{\rm AS}\\ \notag
&\hspace{100pt} + \big[  X_p(\tau_i)X_p^{*}(\tau_j) \big]_{\rm AS}  \big[  X_{|\bp-\bk|}^{\prime}(\tau_i^{-})X_{|\bp-\bk|}^{*\prime}(\tau_j^{-}) \big]_{\rm AS}\\ \notag
&\hspace{100pt} + \frac{m_j}{2H\tau_j} \Big( X_p^{\prime}(\tau_i^{-})X_p^{*}(\tau_j) \big[  X_{|\bp-\bk|}(\tau_i)X_{|\bp-\bk|}^*(\tau_j) \big]_{\rm AS}\\ \notag
&\hspace{150pt}  + X_p(\tau_i)X_p^{*}(\tau_j)   \big[  X_{|\bp-\bk|}^{\prime}(\tau_i^{-})X_{|\bp-\bk|}^*(\tau_j) \big]_{\rm AS}\\ \notag
&\hspace{150pt} + \big[  X_p^{\prime}(\tau_i^{-})X_p^{*}(\tau_j) \big]_{\rm AS}  X_{|\bp-\bk|}(\tau_i)X_{|\bp-\bk|}^*(\tau_j) \\ \notag
&\hspace{150pt} + \big[  X_p(\tau_i)X_p^{*}(\tau_j) \big]_{\rm AS}  X_{|\bp-\bk|}^{\prime}(\tau_i^{-})X_{|\bp-\bk|}^*(\tau_j) + {\rm h.c.}\Big)\\ \label{eq:K3}
&\hspace{100pt} + \frac{m_i m_j}{2H^2\tau_i\tau_j} X_p(\tau_i)X_p^{*}(\tau_j)  X_{|\bp-\bk|}(\tau_i)X_{|\bp-\bk|}^*(\tau_j)  \bigg\} \,.
\end{align}
Here $\tau_i^{-}$ denotes the conformal time immediately before the discontinuity at the $i$-th scatterer location for the derivative of the corresponding mode function. Appendix~\ref{app:diracdelta} is devoted to the detailed derivation of Eqs.~(\ref{eq:deltazetaK})-(\ref{eq:K3}).

Our separation of the momentum integral in (\ref{eq:deltadeltazeta}) into three different types of $\mathcal{K}_{ij}$ is motivated by their distinct behaviors at large momentum $p \to \infty$. As we elaborate in Appendix~\ref{sec:cutoff}, $\mathcal{K}^{\rm I}_{ij}$ is convergent at large momenta, as this integral $\propto \Lambda^{-1}$, where $\Lambda$ is a suitably chosen ultraviolet (UV) momentum cutoff. On the other hand, both $\mathcal{K}^{\rm II}_{ij}$ and $\mathcal{K}^{\rm III}_{ij}$ can be decomposed into two components: (i) a convergent piece that is sourced by $X$ at super-horizon scales and is $\mathcal{O}(\mathcal{K}^{\rm I})$, and (ii) a divergent piece that scales logarithmically and linearly with $\Lambda$, respectively.

The apparent failure of the AS scheme, which was employed to ameliorate UV divergences in the momentum integrals involving $X$, arises from the singular nature of the Dirac-delta approximation for the scattering duration for the effective mass $m^2(\tau)$ (see Appendix~\ref{app:cutoff1} for a detailed discussion). In reality, any non-adiabatic event has finite temporal width, $w$, whose size is determined by the precise microphysics, which naturally provides an additional source of a comoving momentum cutoff. More precisely, this cutoff can be identified with the inverse scattering width, 
\beq\label{eq:cutoff}
\Lambda_i \;=\; (Hw\tau_i)^{-1}\,.
\eeq
The correction to the power spectrum (\ref{eq:deltadeltazeta}) should therefore be interpreted as integrating over momentum up to the cutoff (\ref{eq:cutoff}). This naively suggests that our results will depend our choice of $w$ and $\tau_i$. However, in Appendix~\ref{app:cutoff2} we show that in the limit where the scattering density is large, $\mathcal{N}_s\gg 1$, which is our regime of interest, this cutoff-dependence is subdominant relative to the superhorizon contribution of $X$ to the momentum integral, and will henceforth be neglected.\par\bigskip

We will now evaluate (\ref{eq:deltazetaK}) in the limit where the scattering density is large, $\mathcal{N}_s\gg 1$. This condition guarantees that the superhorizon evolution of $X$ is controlled solely by the scattering strength parameter $\S$~\cite{StochasticSpectator}. Moreover, as we describe in detail in Appendix~\ref{app:cutoff2}, this limit simplifies (\ref{eq:deltazetaK}) to 
\begin{align} \notag
 \delta\Delta_{\zeta}^2(k)  \;&\stackrel{\mathcal{N}_s\gg 1}{\approx}\; 4\pi^2 (\Delta_{\zeta,0}^2)^2 \sum_{i,j}  \frac{m_i m_j}{H^2}(k\tau_i)^2(k\tau_j)^2 \mathcal{G}_k(\tau,\tau_i) \mathcal{G}_k(\tau,\tau_j) \\ \label{eq:deltaDeltafinal}
&\hspace{90pt}\times \int \frac{d^3\bp}{(2\pi)^{3}k}\, \left[  X_p(\tau_i)X_p^*(\tau_j) \right]_{\rm AS}  \big[  X_{|\bp-\bk|}(\tau_i)X_{|\bp-\bk|}^*(\tau_j) \big]_{\rm AS}\,,
\end{align} 
which is correct up to $\mathcal{O}(1)$ factors. That is, to a good approximation, we can disregard the explicit calculation (though not the contributions) of the $\mathcal{K}^{\rm II}$ and $\mathcal{K}^{\rm III}$ terms. The convenience of this result relies on the fact that this expression is convergent even when $\Lambda \to \infty$, with the ultraviolet contributions of the deep subhorizon $X$-modes suppressed by AS regularization. The result (\ref{eq:deltaDeltafinal}) will henceforth serve as the basis of our analytical and numerical analyses. \par\medskip

 We now proceed to dissect this result, to discuss the regimes of interest for our purposes. We expect the integral to be dominated by the superhorizon modes of $X$, because $X_k$ will deviate exponentially from its vacuum form, $X_k^0$, primarily outside the horizon. The domination of the integral in (\ref{eq:deltaDeltafinal}) by super-horizon $X$-modes makes a closed-form analytical expression for $\Delta_{\zeta}^2$ difficult, if not impossible. Nevertheless, we will be content (for now) with crude but useful estimates, that will turn out to provide a good qualitative description of the numerical solutions. For the reader who is fine trusting the numerical results, Section~\ref{sec:anesti} can be skipped.

\subsection{Analytical Estimates}\label{sec:anesti}

\begin{figure}[t!]
\centering
\includegraphics[width=0.8\textwidth]{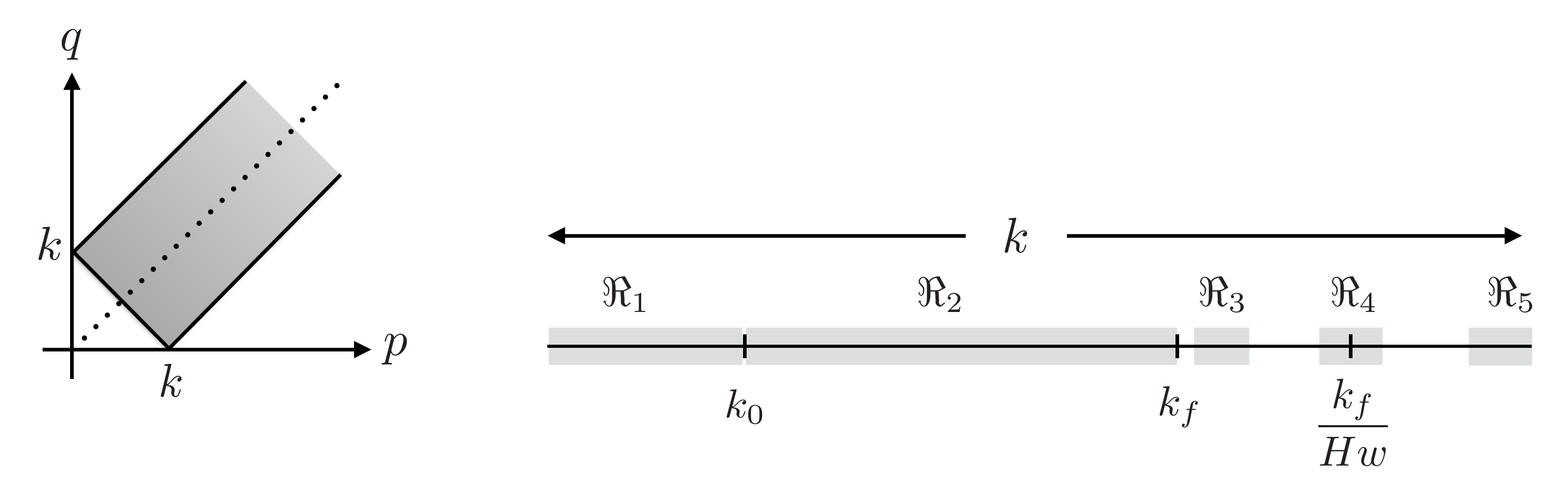}
\caption{The left panel shows the two dimensional region in momentum space over which the spectator field modes are integrated in (\ref{eq:K1cont0}).
The right panel delineates the different regimes in $k$-space of the curvature spectrum, which will be useful for understanding the features in the power spectrum. We define $k_0^{-1}$ as the size of the horizon when non-adiabatic excitation of spectator fields begins, and $k_f^{-1}$ for that when it ends. The wavenumber $k_f/Hw$ is the largest momentum mode of the spectator field (and hence the curvature perturbation) that is excited, see (\ref{eq:cutoff}).}
\label{fig:Integrations}
\end{figure}

Deriving analytical estimates of (\ref{eq:deltaDeltafinal}) is non-trivial; we will focus on simply getting a rough $k-$ dependence of $\delta\Delta_\zeta^2(k)$.
We use the equal-time approximation, $\tau_i = \tau_j$, where (\ref{eq:deltaDeltafinal}) further simplifies to\footnote{Only the diagonal $i=j$ terms in the sum (\ref{eq:deltaDeltafinal}) are positive definite. Off-diagonal contributions alternate signs stochastically, and for $\mathcal{N}_s\gg1$ we expect them to approximately cancel each other due to the non-correlation of the scatterer amplitudes $m_i$, cf.~(\ref{eq:mstats}). This is proven to be the case for the ensemble-averaged $\delta\Delta_{\zeta}^2$ in Section~\ref{sec:averageps}.  The validity of this approximation for each realization of the ensemble is discussed from the numerical perspective in Appendix~\ref{app:numerics} (see also Footnote~\ref{foot:eqtime}). 
}
\begin{align}\notag
\delta\Delta_{\zeta}^2(k) \;&\simeq\; 4\pi^2 (\Delta_{\zeta,0}^2)^2 \sum_{i} \frac{m_i^2}{H^2}(k\tau_i)^4 \mathcal{G}_k^2(\tau,\tau_i) \int \frac{d^3\bp}{(2\pi)^{3}k}\, |X_p(\tau_i)|^2_{\rm AS}  |X_{|\bp-\bk|}(\tau_i)|^2_{\rm AS}\\ \label{eq:K1cont0}
&=\; (\Delta_{\zeta,0}^2)^2 \sum_{i} \frac{m_i^2}{H^2}(k\tau_i)^4 \mathcal{G}_k^2(\tau,\tau_i) k^{-2}\int_0^{\infty} p\,dp\int_{|p-k|}^{p+k}q\,dq\, |X_p(\tau_i)|^2_{\rm AS}  |X_{q}(\tau_i)|^2_{\rm AS}\,,
\end{align}
where in going from the first to the second line, we took advantage of the azimuthal symmetry of the integrand, and then traded the angular integration for another momentum integral (see Eq.~(\ref{eq:pqint})). Since there are multiple scales in our problem, it will be useful to summarize all of them in the following list:
\Beq \label{eq:definitions}
k & \;=\; \textrm{wavenumber of the curvature perturbation $\pi$}, \\
p, q & \;=\; \textrm{wavenumber of $X$ modes (to be integrated over) }, \\
\tau_0 &\;=\; \textrm{conformal time when scatterings begin},\\
\tau_f &\;=\; \textrm{conformal time when scatterings end},\\
k_0 &\;=\; \textrm{momentum at horizon crossing when scatterings begin, i.e.} \,k_0\equiv |\tau_0|^{-1}\,,\\
k_f &\;=\; \textrm{momentum at horizon crossing when scatterings end, i.e.} \,k_f\equiv |\tau_f|^{-1}\,,\\
\Lambda_f  &\;=\; \textrm{the largest momentum excited for $X$, i.e. } \, \Lambda_f \equiv k_f / H w, \textrm{ cf. (\ref{eq:cutoff})} \, .
\Eeq
The relevant hierarchies between these scales are summarized in Fig.~\ref{fig:Integrations}. There we also define the momentum domains $\Re_1$ - $\Re_5$ that separate these scales. We discuss the behavior of the curvature spectrum in each of these different regions  below. We emphasize that the following discussion about the behavior of the curvature perturbation is heuristic, and is mean to serve as a rough guide to understanding the many qualitative (and not quantitative) features at various momentum scales. The reader should refer to Appendices~\ref{sec:cutoff} and \ref{sec:intdom} for more detailed steps. We start with the behavior of $\delta\Delta_\zeta^2$ in $\Re_5$ since it is easiest to understand. The most relevant aspects for the scale dependence of $\delta\Delta_\zeta^2$ arise in the regions $\Re_2$ and $\Re_1$, for which we also provide an intuitive understanding. 

\begin{enumerate}
\item[$\Re_5$:] If  $k\gg \Lambda_f = k_f/Hw$, then the Goldstone mode ($\pi_k$) in question remains inside the horizon during the whole duration of scatterings. Importantly, for $k\gg \Lambda_f$, the $X_p$ modes that lie within the domain of integration in (\ref{eq:K1cont0}) shown in the left panel of Fig.~\ref{fig:Integrations} will not be excited away from the vacuum. Based on our AS scheme, these $X_p$ modes do not contribute to the power spectrum, and therefore $\delta\Delta_{\zeta}^2\simeq 0$.
\item[$\Re_4$:] Modes with $ k_f \ll k\lesssim k_f/Hw$ can be affected by the scattering events at subhorizon scales. In this case the momentum integral will be sourced only by those $X$ modes that are in the allowed integration region in Fig.~\ref{fig:Integrations}. For example, for a given $\tau_i$, this domain corresponds to the triangular region $p,q<\Lambda_i$, $p+q>k$, where $\Lambda_i\equiv (Hw|\tau_i|)^{-1}$ is the instantaneous cutoff scale. The integrated modes ($p, q\gtrsim k\gg k_f$) will be sub-horizon. For such modes a last scattering approximation can be used. Physically, this approximation is related to the fact that the largest deviation form the vacuum mode function $X_p^0$ due to a single scattering is obtained at the latest times, and inside the horizon, we can ignore accumulation over many scatterers.  Namely, $X_p\simeq X_p^0 + \delta X_p$, with $\delta X_p(\tau_i)\propto(m_i a(\tau_i)/p)X_p^0(\tau_i)$ (see Appendix~\ref{app:cutoff1} for details). We then have 
\begin{align}\notag
\delta\Delta_{\zeta}^2(k\lesssim k_f/Hw) \;&\sim\; (\Delta_{\zeta,0}^2)^2 \left(\frac{\sigma}{H}\right)^2\sum_{i} (k\tau_i)^2 k^{-2}\int_{k-\Lambda_i}^{\Lambda_i} p\,dp\int_{|p-k|}^{\Lambda_i}q\,dq\, |X_p^0(\tau_i) \delta X_p(\tau_i) |^2\\ \notag
&\sim\; (\Delta_{\zeta,0}^2)^2 \left(\frac{\sigma}{H}\right)^4\sum_{i} (k\tau_i)^2 k^{-2}\frac{k^2}{(k\tau_i)^2}\\ \label{eq:logdie}
&\sim\; (\Delta_{\zeta,0}^2)^2  \mathcal{N}_s\left(\frac{\sigma}{H}\right)^4 \ln\left(\frac{k_f}{kHw}\right)\,.
\end{align}
In the first line we replaced $m_i^2$ by $\sigma^2$ and used $\mathcal{G}_k^2(\tau,\tau_i)\sim (k\tau_i)^{-2}$ (since the $|k\tau_i|\gg 1$, see \eqref{eq:curlygapp}). We also used the fact that the integrand can be approximated by setting $p=q$. In the second line, the area of integration in the $p-q$ plane is $\sim k^2$, and the integral is well approximated by setting $p=q=k/2$. 
The logarithm in the third line is related the total number of scatterings from $\tau_{\rm in}\sim k^{-1}$ to $\tau_{\rm end}=\Lambda_f^{-1}$; the sum over scatterers is just proportional to the time the mode spends in the domain that allows for
growth.  We note that in this case only sub-horizon modes are excited, and therefore our result depends on the cutoff and on $\mathcal{N}_s$ and $\sigma$ independently. Nevertheless, in the saturation limit $\mathcal{N}_sHw\sim 1$, the power spectrum may be written in terms of the square of the scattering parameter $\S$ and the scatterer width. The interested reader can find further details in Appendix~\ref{app:cutoff2}. 

The key takeaway is that there is a red tilted spectrum with a $\log(1/k)$ scaling, which eventually vanishes for $k\gtrsim k_f/Hw$.

\item[$\Re_3$:] If now the Goldstone mode has $k\gtrsim k_f$, the cutoff dependence can be neglected (see Appendix \ref{app:cutoff2}). The main contribution comes from super-horizon $X$-modes at $p,q\ll k$ since they grow exponentially. Given this exponential excitation of these modes, we schematically write
\begin{align}\label{eq:K1kf}
\delta\Delta_{\zeta}^2(k\gtrsim k_f) \;&\sim\; (\Delta_{\zeta,0}^2)^2 \left[ \mathcal{N}_s\left(\frac{\sigma}{H}\right)^2 \right]^2 e^{\gamma_1 N_{\rm tot}} \left(\frac{k}{k_0}\right)^{-\gamma_2}\,,
\end{align}
where $N_{\rm tot}$ denotes the number of $e$-folds of expansion during which scatterings are active, and where $\gamma_{1,2}>0$ are functions of the scattering strength parameter $\S$. The quadratic factor in the scattering strength is expected from the excitation of sub-horizon modes (see previous case and the discussion following Eq.~(\ref{eq:K3cont})). Note that we do not expect a scale invariant curvature spectrum, but a red one, since the curvature modes are all sub-horizon ($k\gtrsim k_f$), and must die off logarithmically for $k_f\ll k\lesssim k_f/(Hw)$ as discussed in $\Re_4$. This previous result, and those which follow below, will be analytically derived in the ensemble average sense in Section~\ref{sec:averageps}. 

\item[$\Re_2$:] We now turn to the more interesting regime $k_f>k>k_0$. Here we expect the bulk of the contribution to come from those $X$ modes that leave the horizon while scatterings are active. Note that for these modes, the number of $e$-folds of exponential growth are given by $N_{\rm tot}-N_*(k)=N_{\rm tot}-\ln(k/k_0)$. In this case the upper limit of the momentum integral in (\ref{eq:K1cont0}) for the modes in question is $|\tau_i|^{-1}\gg k$.
Recalling from (\ref{eq:curlygapp})  that $\mathcal{G}_k(\tau,\tau_i)\simeq {\rm const.}$ outside the horizon, we can then write
\begin{align}\notag
\delta\Delta_{\zeta}^2(k_f>k>k_0)  \;&\sim \; (\Delta_{\zeta,0}^2)^2 \left(\frac{\sigma}{H}\right)^2 \sum_{i} (k\tau_i)^4  k^{-2}\int_0^{|\tau_i|^{-1}} p\,dp\int_{|p-k|}^{p+k}q\,dq\, |X_p(\tau_i)|^2_{\rm AS}|X_q(\tau_i)|^2_{\rm AS} \\ \notag
&\sim\; (\Delta_{\zeta,0}^2)^2 \left(\frac{\sigma}{H}\right)^2 \sum_{|k\tau_i|<1}  (k\tau_i)^4e^{(4-\gamma)(N(\tau_i)-\ln(k/k_0))}\\ \notag
&\sim\; (\Delta_{\zeta,0}^2)^2 \left(\frac{\sigma}{H}\right)^2 \sum_{|k\tau_i|<1}  (k\tau_i)^{-\gamma}\\ \label{eq:K1cont}
&\sim\; (\Delta_{\zeta,0}^2)^2 \mathcal{N}_s\left(\frac{\sigma}{H}\right)^2 e^{\gamma N_{\rm tot}} \left(\frac{k}{k_0}\right)^{-\gamma}\,.
\end{align}
In going from the first line to the second, we used $|X_p(\tau_i)|^2\sim p^{-1}e^{(4-\gamma)(N(\tau_i)-N_*(p)/2)}$, and used $k$ as the relevant scale for the domain of integration. In going from the second to the third, we used $N(\tau_i)=\ln(\tau_0/\tau_i)$ and $k_0=|\tau_0|^{-1}$. In the following line, we carried out the summation over $i$  by using $\tau_i=\tau_0e^{-iH\delta t}$ where $\delta t$ is the typical interval between scatterings (the detailed steps for a similar summation are provided in equation \eqref{eq:howtosum} of Appendix~\ref{app:cutoff2}). In (\ref{eq:K1cont}), $\gamma$ denotes a function of $\S$.  With $\S\gg 1$, the exponential enhancement of the modes is significant and $\gamma$ is positive. In addition to a potentially sizable exponential enhancement of the result, we expect to observe a non-scale invariant, red-tilted spectrum.\footnote{In the weak scattering limit $\S\le 1$, we can take $\gamma\approx 0$.}

To understand the scale dependence intuitively, note that the enhancement of the spectrum arises primarily from the exponential growth during the time that modes spend outside the horizon during scattering. This time depends on the wavenumber, since the wavenumber determines the moment of horizon crossing. In the strong scattering case, the more time a mode spends outside the horizon, the more it is enhanced. Hence, we should expect lower $k$ modes to be more enhanced compared to the higher $k$ ones.

\item[$\Re_1$:] Finally, for modes that are already at superhorizon scales before scatterings begin $(k< k_0)$, the dominant contribution to (\ref{eq:K1cont0}) is provided by $X$ modes that are at superhorizon scales throughout the entire scattering duration, $p,q\lesssim k_0$. The power spectrum correction then takes the form
\begin{align}\notag
\delta\Delta_{\zeta}^2(k < k_0)  \;&\sim\; (\Delta_{\zeta,0}^2)^2 \mathcal{N}_s\left(\frac{\sigma}{H}\right)^2 k_0^{-2}\cdot k^3k_0 \frac{e^{\gamma N_{\rm tot}}}{k_0^2}\\ \label{eq:K1k0}
&\sim\; (\Delta_{\zeta,0}^2)^2 \mathcal{N}_s\left(\frac{\sigma}{H}\right)^2 e^{\gamma N_{\rm tot}} \left(\frac{k}{k_0}\right)^{3}\,.
\end{align}
Note again, the expression above implies a non-scale invariant curvature power spectrum. 

To understand the $k^3$ behavior intuitively, note that the $\zeta$ modes were already outside the horizon when scatterings began. While these modes are excited by the stochastic source of $X$ modes, causality forbids correlations from being established in these super-horizon scales, and the resulting spectrum is that of a white noise process, $\langle 0|\zeta(\bk)\zeta(\bk')|0\rangle = {\rm const.}$ This in turn leads to the $k^3$ scaling in $\Delta^2_\zeta$.
\end{enumerate}
With these estimates at hand, we now proceed to evaluate the approximation (\ref{eq:deltaDeltafinal}) numerically in the next section, and analytically, in the ensemble average sense, in Section~\ref{sec:averageps}.

\subsection{Numerical Results}\label{sec:numres}

In this section we evaluate (\ref{eq:deltaDeltafinal}) numerically. Although this expression has already been simplified through various approximations, its numerical evaluation is still challenging for several reasons. Even for a modest, but non-negligible duration of scatterings, in the large $\mathcal{N}_s$ limit, the number of terms in the sum will scale as $N_s^2\gg 1$ (recall that $N_{\rm s}$ is the total number of scatterers, whereas $\mathcal{N}_{\rm s}$ is the number of scatterers per $e$-fold of expansion). Each term requires the numerical evaluation of a three-dimensional momentum integral, for which the integrand must in turn be evaluated numerically using the transfer matrix formalism, as $\chi$ experiences multiple scattering events over the course of inflation \cite{StochasticSpectator}. We have nevertheless found an approximate way to evaluate $\delta\Delta_{\zeta}^2$ in a reasonable computational time-scale. This approximation is discussed in detail in Appendix~\ref{app:numerics}. \par\medskip

In the present  section we will present and discuss the form of the curvature spectra for different realizations of the disorder, with different scattering strengths and different durations $N_{\rm tot}$. Due to the difficulty in evaluating (\ref{eq:deltaDeltafinal}) for very large number densities of scatterings, we have taken $\mathcal{N}_s=25$ here and in all other results except otherwise stated. Additionally, in all cases we will consider a large momentum cutoff corresponding to $Hw=10^{-6}$, which easily satisfies the constraint $\mathcal{N}_s (Hw)<1$ (separation between scatterers $>$ width of scatterers; see (\ref{eq:Nsconst}))\footnote{\label{fn:Hw}We note a caveat to our choice of parameters. Note that $Hw=10^{-6}< \sqrt{\Delta^2_{\zeta,\rm Planck}}$ which violates the condition for ignoring the backreaction of the curvature perturbations on the evolution of $\chi$ (see discussion at the beginning of Section \ref{sec:spectfield}). Qualitatively, taking $Hw= 10^{-3}$ (for example) only moves the cutoff in the spectrum $k_f/(Hw)$ without affecting much else. We chose $Hw=10^{-6}$ because for larger $Hw$, the scaling behaviors in regions $\Re_3$-$\Re_4$ would be harder to see and understand separately.}, and which will allow us to clearly distinguish four regimes in $k$ depending on its magnitude compared to $k_0$, $k_f$ and $k_f/Hw$ (c.f.~Fig.~\ref{fig:Integrations}). We separate the discussion into three different regimes: the ``weak'' scattering regime, for which $\delta\Delta_{\zeta}^2\ll \Delta_{\zeta,0}^2$ for most members of the ensemble; ``moderate'' scattering, for which $\delta\Delta_{\zeta}^2\sim \Delta_{\zeta,0}^2$ is a generic outcome, and ``strong'' scattering, for a case in which $\delta\Delta_{\zeta}^2\gg \Delta_{\zeta,0}^2$ for the majority of the disorder realizations. In describing our results, we will use the following types of averages over $n$ realizations of the disorder,
\Beq
\langle \delta\Delta_{\zeta}^2\rangle &=\frac{1}{n}\sum_{a=1}^{n} \delta\Delta_{\zeta,a}^2\,,\qquad&\textrm{(arithmetic sample mean)}\\
\exp \langle \ln(\delta\Delta_{\zeta}^2) \rangle &=\left(\prod_{a=1}^n \delta\Delta_{\zeta,a}^2 \right)^{\!\!\frac{1}{n}}\,,\qquad&\textrm{(geometric sample mean)}
\Eeq
where the subindex $a$ denotes the $a$-th realization. If $n\rightarrow \infty$, we refer to the above quantities as ensemble means rather than sample ones.

 \par\bigskip

\subsubsection{Weak Stochastic Sourcing}

\begin{figure}[!t]
\centering
    \includegraphics[width= 0.98\textwidth]{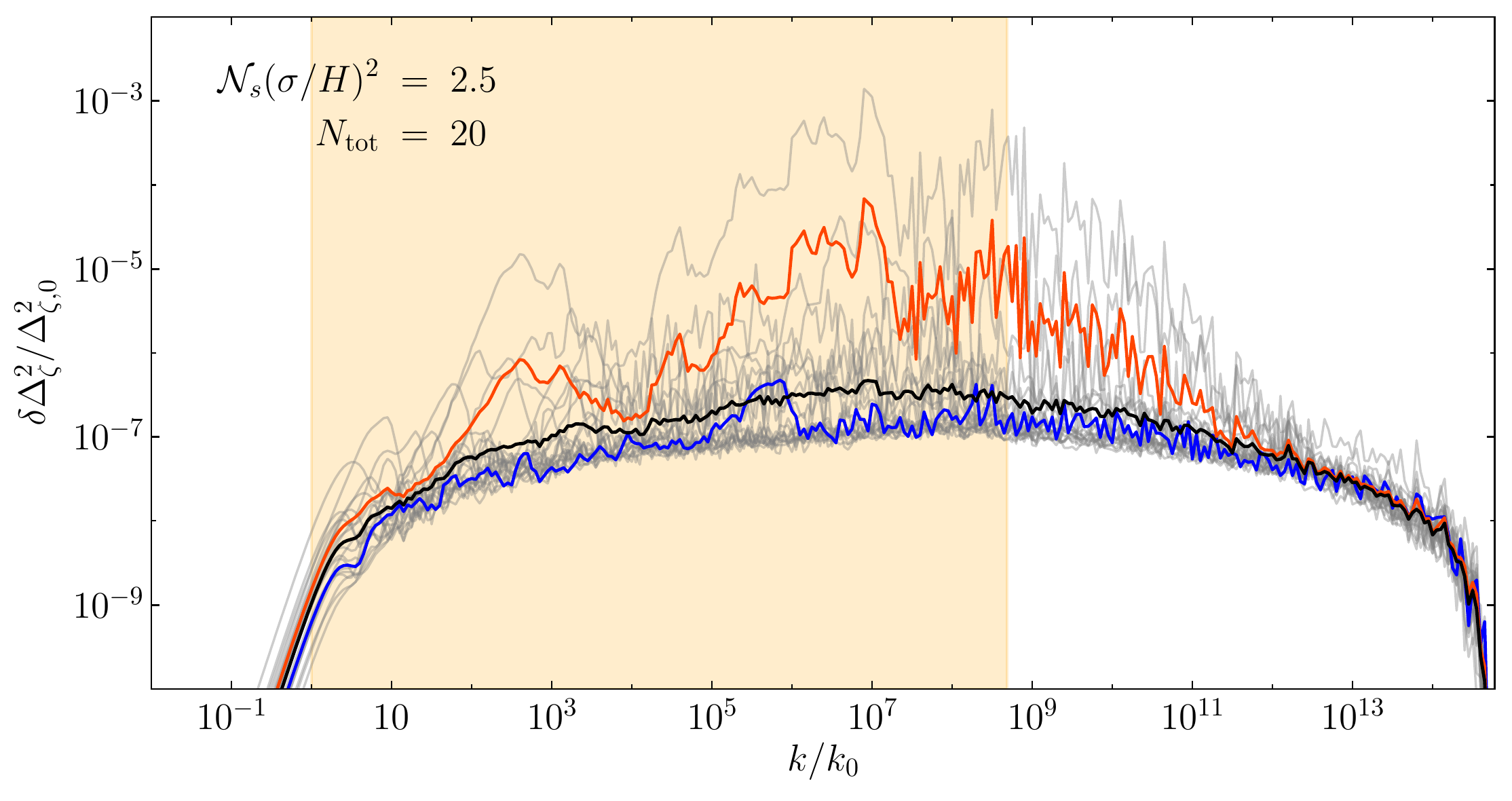}
    \caption{Relative correction to the curvature power spectrum for 20 unique realizations of the disorder (gray curves), in the `weak' scattering case, with the scattering parameter $\S=2.5$, number of $e$-folds of scattering $N_{\rm tot}=20$ ($k_f\simeq 5\times 10^8 k_0$), the effective temporal width of each scatterer $Hw=10^{-6}$, and $\Delta_{\zeta,0}^2=\Delta_{\zeta,\,{\rm Planck}}^2\simeq 2.1\times 10^{-9}$. The blue curve highlights one particular realization. The red curve corresponds to the arithmetic mean $\langle \delta\Delta_{\zeta}^2\rangle$ of the sample, while the black curve shows the geometric mean $\exp \langle \ln(\delta\Delta_{\zeta}^2)\rangle$ of the sample. The yellow region highlights those modes that leave the horizon during scatterings. Note that, the sourced spectrum is subdominant compared to the adiabatic one in this case, with the total curvature spectrum $\Delta_\zeta^2=\Delta_{\zeta,0}^2 + \delta\Delta_{\zeta}^2\approx \Delta_{\zeta,0}^2$. }
    \label{fig:2.5_20}
\end{figure}

Fig.~\ref{fig:2.5_20} shows the sourced correction to the power spectrum relative to its adiabatic value, $\delta\Delta_{\zeta}^2/\Delta_{\zeta,0}^2$, for 20 different realizations for $\S=2.5$ and $N_{\rm tot}=20$ ($k_f\simeq 5\times 10^8\,k_0$), shown in gray. 
For definiteness we have considered $\Delta_{\zeta,0}^2=\Delta_{\zeta,\,{\rm Planck}}^2$. 
\begin{enumerate}
\item[$\Re_1$:] For $k< k_0$, all realizations are almost perfectly parallel, and grow as $k^3$, as we expected from (\ref{eq:K1k0}).  Note that the arithmetic average of the sample of trajectories $\langle \delta\Delta_{\zeta}^2\rangle$ (shown in orange), and the geometric mean of that sample $\exp \langle \ln(\delta\Delta_{\zeta}^2)\rangle  $ (shown in black), also follow the same $k^3$ trend. The curve in blue shows the form of one particular realization of $m^2(t)$.

\item[$\Re_2$:] For $k_0\lesssim k \lesssim k_f$, the spectra are no longer parallel, and span four orders of magnitude in $\delta\Delta_{\zeta}^2/\Delta_{\zeta,0}^2$. Moreover, the relative accumulation of trajectories at low amplitudes indicate a heavily skewed probability distribution for $\delta\Delta_{\zeta}^2$ (for any fixed wavenumber, see the discussion that follows below). Nevertheless, most of the realizations have a similar qualitative behavior. The grand majority display a blue tilt in this domain, with a gentle slope. Note that the value of the arithmetic average $\langle \delta\Delta_{\zeta}^2\rangle$ of the sample is dominated by the largest outliers, while the geometric mean $\exp \langle \ln(\delta\Delta_{\zeta}^2)\rangle  $ appears to provide a better estimate of the behavior of the `typical' realization. If we focus on the geometric mean, we can infer that the estimate (\ref{eq:K1cont}) with $\gamma \sim -0.2$ roughly reproduces the magnitude and tilt of the stochastically sourced component of $\Delta_{\zeta}^2$. 
Besides this, perhaps the most interesting feature of the results is given by the fact that the general blue trend of the spectra is complemented by the presence of bumps and troughs of varied heights and widths partly reflecting the sensitivity of particle production in $\chi$ to different momentum modes. We will explore the related observational phenomenology of these in detail in Section~\ref{sec:cmb}. 

\item[$\Re_3$:] When $k_f\lesssim k$, we expect a red spectrum, as per (\ref{eq:K1kf}); this clearly seems to be the case. In this regime it is difficult to distinguish `real' features on the power spectrum from numerical artifacts, due to the approximations that the evaluation of (\ref{eq:deltaDeltafinal}) requires. 
\item[$\Re_4$:] At the cutoff scale $k\sim k_f/Hw\sim 10^{15}k_0$, the spectrum dies off roughly following the logarithmic scaling predicted in (\ref{eq:logdie}).
\end{enumerate}
None of the 20 realizations considered above have the sourced component of the curvature power spectrum $\delta \Delta_\zeta^2$ that is larger than the purely vacuum contribution.  This means that for such a weak stochastic sourcing, the correction to the vacuum power spectrum is likely unobservable.  However, this does not forbid the existence of rare ensemble members for which the correction can become observable. As we will show in Section~\ref{sec:averageps}, the extreme skewness of the distribution for $\delta\Delta_{\zeta}^2$ accommodates these rare but very large outliers. Moreover, it is also possible that the sourced aspects of the curvature perturbations make an appearance in higher point correlation functions.\par\bigskip

\subsubsection{Moderate Stochastic Sourcing}

\begin{figure}[!t]
\centering
    \includegraphics[width= 0.98\textwidth]{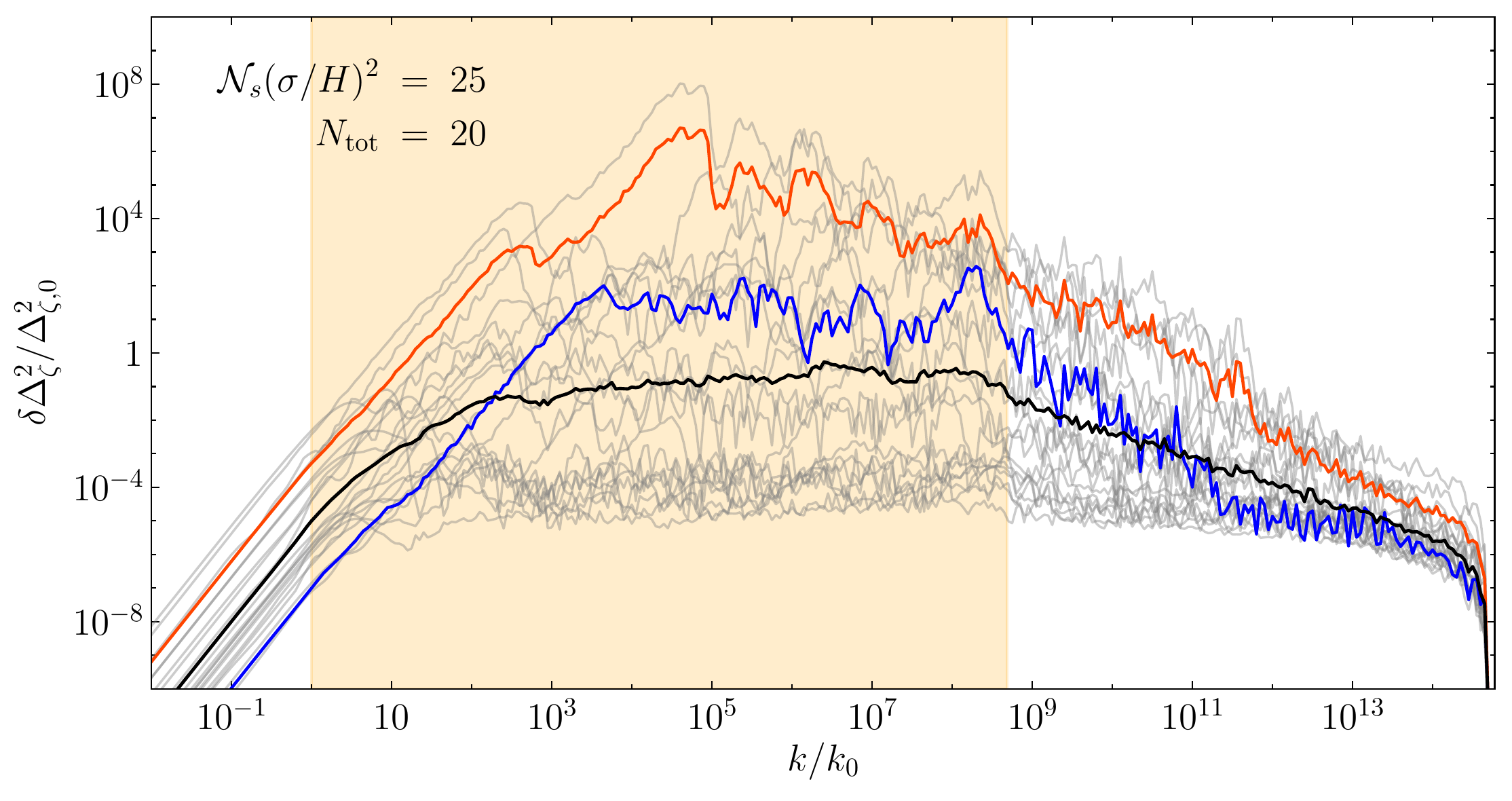}
    \caption{Relative enhancement of the curvature power spectrum for 20 unique realizations of the disorder (gray curves), in the `moderate' scattering case, with the scattering parameter $\S=25$, total $e$-folds of scattering $N_{\rm tot}=20$ ($k_f\simeq 5\times 10^8 k_0$), effective temporal width of scatterers $Hw=10^{-6}$, and $\Delta_{\zeta,0}^2=\Delta_{\zeta,\,{\rm Planck}}^2\simeq 2.1\times 10^{-9}$. The blue curve highlights one particular realization. The red curve corresponds to the arithmetic mean $\langle \delta\Delta_{\zeta}^2\rangle$ of the sample, while the black curve shows the geometric mean $\exp \langle \ln(\delta\Delta_{\zeta}^2)\rangle$ of the sample. The yellow region highlights those modes that leave the horizon during scatterings. Since the geometric mean lies near, but somewhat below $\Delta_{\zeta,\rm Planck}^2$, it is not unlikely for some of these samples to provide localized features in the observed curvature spectrum without violating the observational constraints. Note that even in cases where $\delta\Delta_\zeta^2\gg \Delta_{\zeta,0}^2$, perturbativity of $\zeta$ is not necessarily violated since the total curvature spectrum $\Delta_\zeta^2=\delta\Delta_{\zeta}^2+\Delta_{\zeta,0}^2\lesssim 1$. Furthermore, if we choose $\Delta_{\zeta,0}^2\ll \Delta_{\zeta,\rm Planck}^2$, then we would have $\Delta_\zeta^2\ll 1$.}
    \label{fig:25_20}
\end{figure}
The relative correction to the curvature spectrum for $\S=25$ is shown in Fig.~\ref{fig:25_20} for $N_{\rm tot}=20$ and $\Delta_{\zeta,0}^2=\Delta_{\zeta,\,{\rm Planck}}^2$. The similarities and differences relative to Fig.~\ref{fig:2.5_20} are evident. 
\begin{enumerate}
\item[$\Re_1$]: For $k\lesssim k_0$ the causality enforced $\delta\Delta_\zeta^2\propto k^3$ scaling can be seen near and to the left edge of the yellow band in Fig.~\ref{fig:25_20}.
\item[$\Re_2$]: For $k_0\lesssim k\lesssim k_f$  -- the regime where the curvature modes leave the horizon during scatterings (yellow band in  Fig.~\ref{fig:25_20}), the presence of an exponential enhancement  of the amplitude and features is clearly visible in Fig.~\ref{fig:25_20}. Note however that the coarse-grained spectral tilt does not necessarily have a monotonic behavior even for a given realization, as seen from the blue curve. In particular, the causality-related $k^3$ scaling appears to extend beyond the initial horizon scale. This suggests that for the largest members of the ensemble, the dominant contribution to the momentum integral  can be given by the early super-horizon modes even beyond the naively expected regime. Nevertheless, beyond a certain scale, we observe a relatively scale-invariant  spectrum, in the sense that neither positive nor negative coarse-grained tilt is preferred. Focusing on the black `typical' curve, in this case we would have $\gamma\lesssim 1$ in (\ref{eq:K1cont}), which appears to be provide a good approximation for the tilt for several of the trajectories shown. 
\item[$\Re_3$]: At $k\gtrsim k_f$, we observe the noisy decreasing spectrum. 
\end{enumerate}
Note that for the case we just discussed, in some realizations, the stochastic component ($\delta\Delta_{\zeta}^2$) can dominate over the vacuum one ($\Delta_{\zeta,0}^2\sim 10^{-9}$), although the most common realizations have $\delta\Delta_{\zeta}^2<\Delta_{\zeta,0}^2$. Even in the former case, the perturbativity assumption, $\Delta_\zeta^2\ll 1$ under which (\ref{eq:deltazeta}) was derived is not violated. Importantly, note that the geometric mean of the ensemble of curvature spectra lies near, but somewhat below $\Delta_{\zeta,\rm Planck}^2$ (see black curve in Fig.~\ref{fig:25_20}). This means that there is a reasonable probability for some of the realizations to provide localized features in the observed curvature spectrum without violating the observational constraints which are broadly consistent with scale invariance. This makes the parameter choices in this case phenomenologically interesting. \par\bigskip
\begin{figure}[!t]
\centering
    \includegraphics[width= 0.98\textwidth]{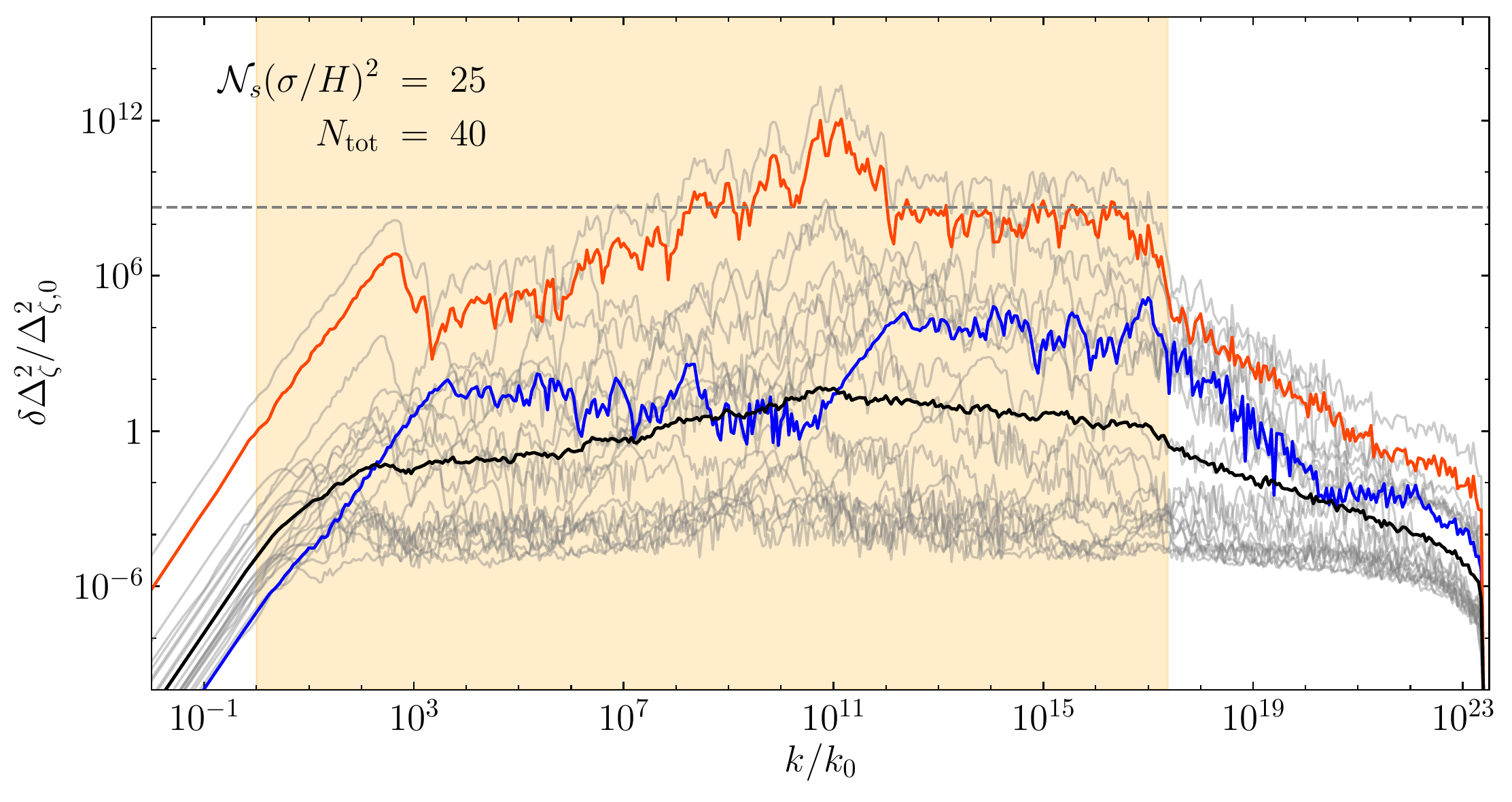}
    \caption{Relative enhancement of the curvature power spectrum for 20 unique realizations of the disorder (gray curves), in the moderate' scattering case, with the scattering strength parameter $\S=25$, number of $e$-folds of scattering $N_{\rm tot}=40$ ($k_f\simeq 2\times 10^{17}$), effective temporal width of scatterers $Hw=10^{-6}$ and $\Delta_{\zeta,0}^2=\Delta_{\zeta,\,{\rm Planck}}^2\simeq 2.1\times 10^{-9}$. The blue curve highlights one particular realization. The red curve corresponds to the sample mean $\langle \delta\Delta_{\zeta}^2\rangle$, while the black curve shows the sample geometric mean $\exp \langle \ln(\delta\Delta_{\zeta}^2)\rangle$. The yellow region highlights those modes that leave the horizon during scatterings. Since the geometric mean lies near $\Delta_{\zeta,\rm Planck}^2$, it is again possible for some of these samples to provide localized features in the observational window. However, it is more difficult to hide below the scale-invariant spectrum in this case.  Note that the horizontal dashed line corresponds to the perturbative bound $\delta\Delta_{\zeta}^2=1$, which is seemingly violated for the large valued samples. However, note that this bound would be much higher if we choose $\Delta_{\zeta,0}^2\ll\Delta_{\zeta,\,{\rm Planck}}^2$. As a result, $\Delta_\zeta^2=\delta\Delta_{\zeta}^2+\Delta_{\zeta,0}^2\ll 1$ can still be satisfied.}
    \label{fig:25_40}
\end{figure}
Fig.~\ref{fig:25_40} displays the corresponding stochastic component of $\Delta_{\zeta}^2$ for the previous case with $\S=25$, but the number of $e$-folds of non-adiabaticity are increased to $N_{\rm tot}=40$. As expected, for the larger the $N_{\rm tot}$ we get a larger $\delta \Delta_{\zeta}^2$. We also note that the cubic tilt for $k\lesssim k_0$ is still present, but now the tilt for larger $k$ appears even less monotonic than in the previous scenario. Note the spread over $\sim 20$ orders of magnitude in the signal. 

There is an important subtlety in interpreting the amplitude of  $\delta\Delta_\zeta^2/\Delta_{\zeta,0}^2$. We can see that a very large amount of power can be injected stochastically onto the curvature fluctuation even for not-so-large scattering strengths if the duration of the scattering epoch is sufficiently long. Some of the curves in Fig.~\ref{fig:25_40} go above the horizontal dashed line which corresponds to $\delta\Delta_{\zeta}^2=1$. Since the total curvature spectrum $\Delta_\zeta^2=\delta\Delta_{\zeta}^2+\Delta_{\zeta, 0}^2$, perturbativity in terms of curvature perturbations seems to be broken.  The apparent violation of $\delta\Delta_\zeta^2< 1$ is a result of our choice $\Delta_{\zeta,0}^2=\Delta_{\zeta,\,{\rm Planck}}^2$. If we had chosen $\Delta_{\zeta,0}^2\ll \Delta_{\zeta,\,{\rm Planck}}^2$, perturbativity is not violated and $\delta\Delta^2_\zeta\ll 1$.\par\bigskip

\subsubsection{Strong Stochastic Sourcing}

\begin{figure}[!t]
\centering
    \includegraphics[width= 0.98\textwidth]{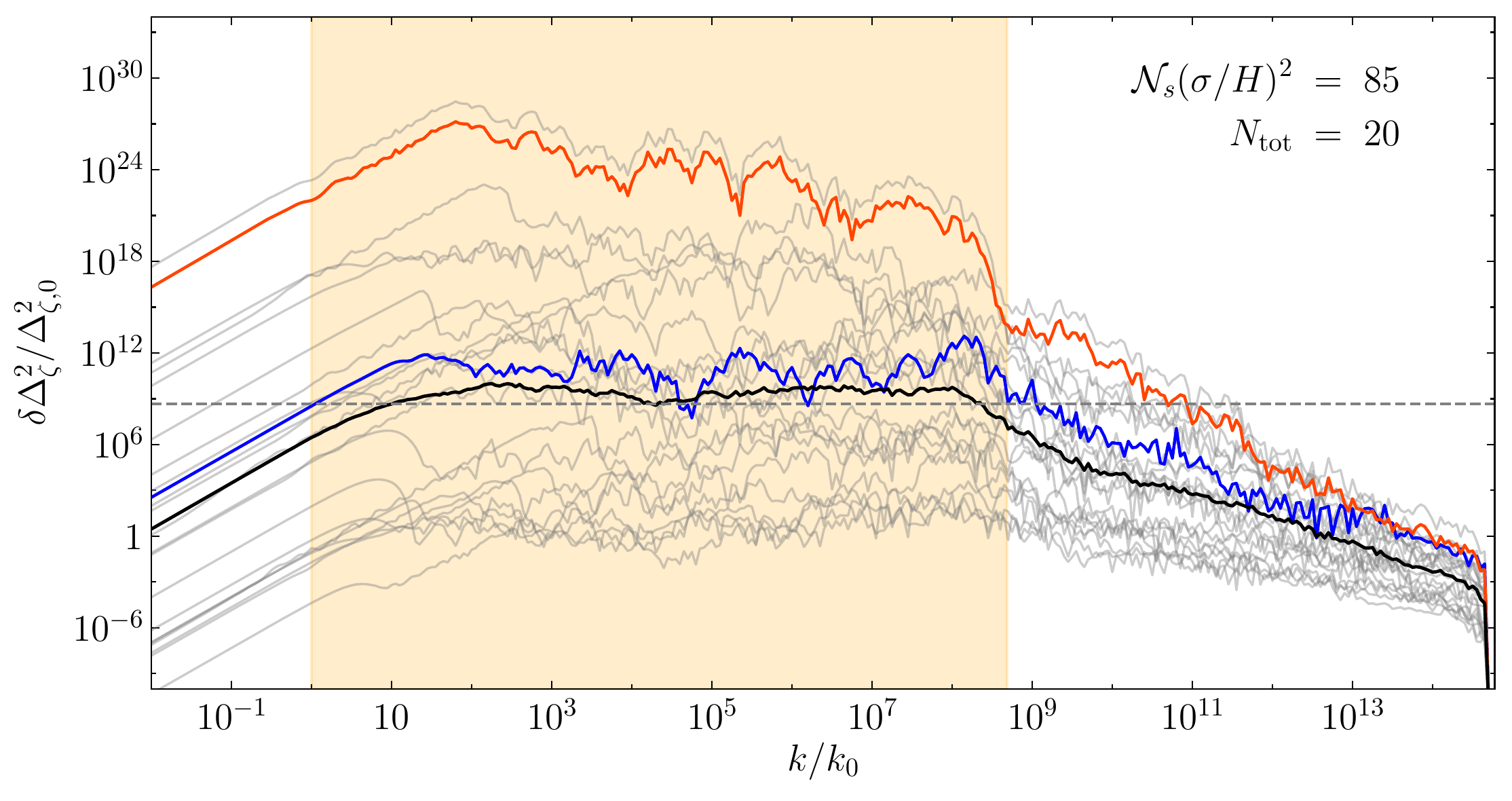}
    \caption{Relative enhancement of the curvature power spectrum for 20 unique realizations of the disorder (gray curves), in the `strong' scattering case, with $\S=85$, $N_{\rm tot}=20$ ($k_f\simeq 5\times 10^8k_0$), $Hw=10^{-6}$ and $\Delta_{\zeta,0}^2=\Delta_{\zeta,\,{\rm Planck}}^2\simeq 2.1\times 10^{-9}$. The blue curve highlights one particular realization. The red curve corresponds to the mean $\langle \delta\Delta_{\zeta}^2\rangle$, while the black curve shows the `typical' value $\exp \langle \ln(\delta\Delta_{\zeta}^2)\rangle$. The yellow region highlights those modes that leave the horizon during scatterings. The horizontal dashed line corresponds to the perturbative bound $\delta\Delta_{\zeta}^2=1$. This bound would be higher if instead $\Delta_{\zeta,0}^2\ll\Delta_{\zeta,\,{\rm Planck}}^2$. However, given that the geometric mean is way above the Planck constraint on the curvature amplitude, this set of parameters will not generically provide an observationally viable, almost scale invariant sourced spectrum in the CMB window.}
    \label{fig:85_20}
\end{figure}

The form of the stochastically sourced curvature powers spectrum for $\S=85$ and $N_{\rm tot}=20$ is shown in Fig.~\ref{fig:85_20} for 20 different realizations of the disorder. In this case, all trajectories show stochastic dominance on the spectral signal for at least a limited range of $k$. The sourcing is in fact so large that the geometric mean lies beyond the perturbativity bound (for $\Delta_{\zeta,0}^2=\Delta_{\zeta,\,{\rm Planck}}^2$), and the trajectories appear to show an exponentially enhanced variance, spreading over $\sim 30$ orders of magnitude. Evidently, unless $\Delta_{\zeta,0}^2\ll\Delta_{\zeta,\,{\rm Planck}}^2$, barring the weakest signals that lie below the dashed line, none of these curves is expected to accurately represent the shape of the power spectrum. Nevertheless, we note that the red tilt observed in many of them is consistent with our expectation (\ref{eq:K1cont}).

\subsection{Probability Distributions} \label{sec:dist}

As the numerical results discussed above attest, we have been successful in estimating the form and magnitude of $\delta\Delta_{\zeta}^2$ for a limited set of realizations of the disorder. From these results, we can immediately arrive to the conclusion that both the (geometric) mean and the spread of the corresponding power spectra realizations are functions of the disorder strength parametrized by $\S$. The arithmetic  mean of the power spectra is in turn overwhelmingly dominated by those realizations for which $\delta\Delta_{\zeta}^2$ is the largest. This indicates a non-trivial, highly skewed probability distribution function (pdf). In this section, we will discuss  the form of the pdf for the power spectrum enhancement as a function of $\S$ and $k$ in a mostly qualitative fashion. 

In order to construct the pdf, we first note that it is convenient to work in terms not of the sourced power spectrum itself, but in terms of its logarithm,
\beq\label{eq:xi}
\xi \;\equiv\; \ln\left(\frac{\delta\Delta_{\zeta}^2}{\Delta_{\zeta,0}^2} \right)\,,
\eeq
due to the span over several orders of magnitude for the sourced power spectrum within a given ensemble of $m^2(t)$ (for the same $\S$ and $N_{\rm tot}$). 
\begin{figure}[!t]
\centering
    \includegraphics[width= \textwidth]{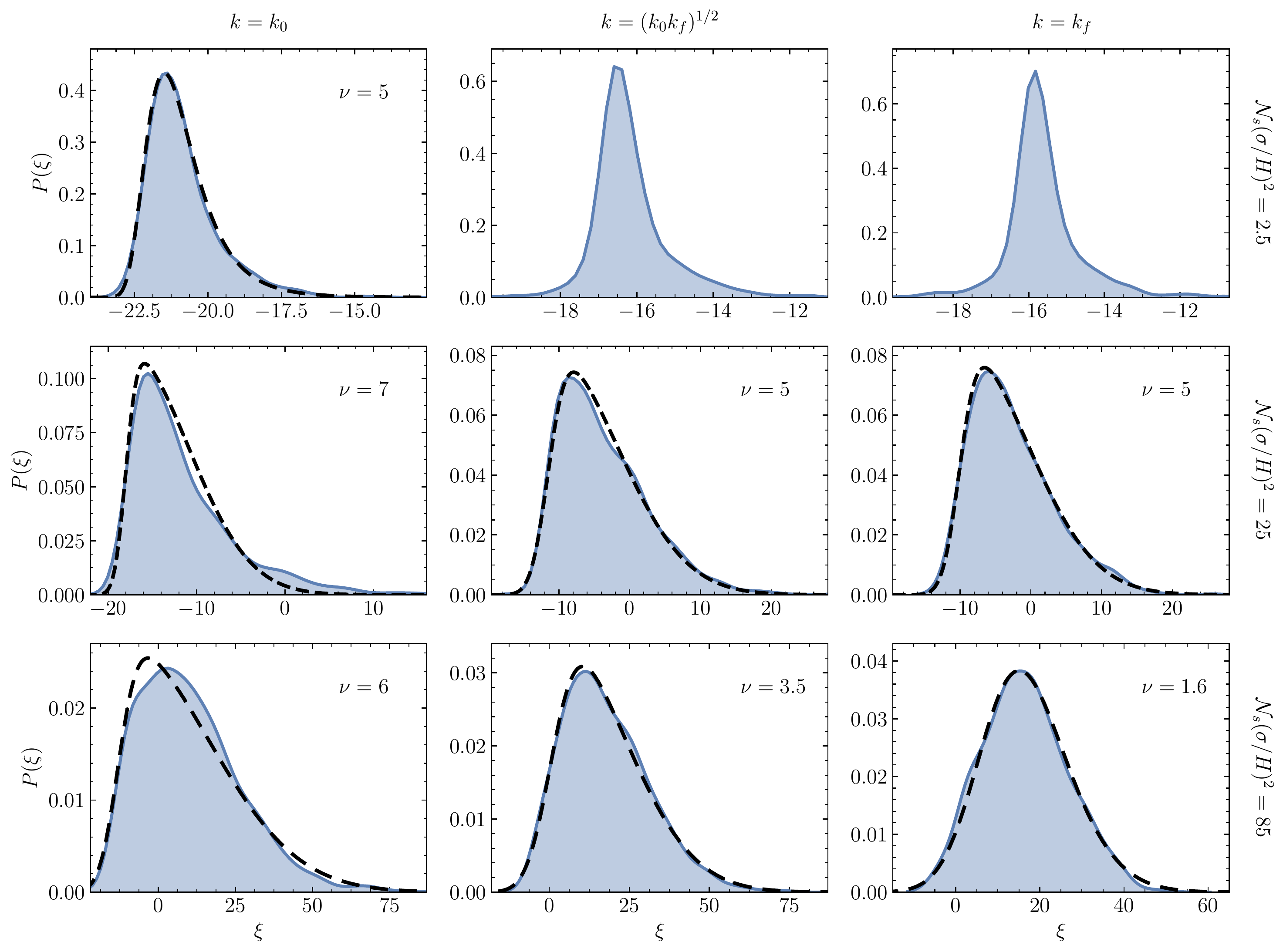}
    \caption{Probability density of $\xi = \ln (\delta \Delta_\zeta^2/\Delta^2_{\zeta,0})$ as a function of the co-moving wavenumber of the curvature perturbation and the scattering strength $\S$. Blue, solid: the empirical distribution for 1000 unique realizations of the $m^2(t)$. Black, dashed: skew-normal fit to the pdf with shape parameter $\nu$ (where adequate). }
    \label{fig:dist}
\end{figure}

In Fig.~\ref{fig:dist}, we show the form of the pdf for three different values of the comoving momentum $k$ and three different values of the scattering strength $\S$. In all cases we have considered $N_{\rm tot}=20$, and we have computed the stochastic power spectrum for a total of 1000 unique realizations. The pdfs are built using a Gaussian kernel density estimator of variable bin size~\cite{parzen1962}. 

The dashed black curves in Fig.~\ref{fig:dist} are a skew-normal fit to the distributions of $\xi$ (when appropriate). As a reminder, a random variable $x$ is skew-normal distributed if its pdf is of the form
\beq
P(x) \;=\; \frac{2}{\sqrt{2\pi\omega^2}} e^{-\frac{(x-x_0)^2}{2\omega^2}}\int_{-\infty}^{\nu\left(\frac{x-x_0}{\omega}\right)} e^{-\frac{t^2}{2}}\,dt\,,
\eeq
where $x_0$, $\omega$ and $\nu$ denote the location, scale and shape parameters, respectively~\cite{skew1,skew2}. The normal distribution is recovered in the limit $\nu=0$. Hence, $\delta\Delta_{\zeta}^2/\Delta_{\zeta,0}^2\sim e^{\xi}$ appears to have a {\em log-skew-normal} distribution. This result is not entirely unexpected, as it is known that log-skew-normal pdfs provide adequate approximations to the sum of lognormally distributed random variables~\cite{skew4,skew3}, which is $|X_k|^2$ in our case, cf.~Section~\ref{sec:spectfield}. The result is at the same time somewhat surprising, as these approximations rely on the assumption that the added random variables are independent, which is not the case here, since we add correlated random walks. 

The skewness of a lognormal distribution increases exponentially with the variance of the corresponding normal distribution. A log-skew-normal is, as its name suggests, even more skewed, with a very heavy tail~\cite{skew5}. Therefore, even if the geometric mean of the power spectrum, $\exp\langle \xi\rangle$ is $\ll 1$, the arithmetic average $\langle \delta\Delta_{\zeta}^2\rangle/\Delta_{\zeta,0}^2$ can be much larger, $\mathcal{O}(1)$ or more. We will refer back to this average when we compute it analytically in Section~\ref{sec:averageps}.
\par\bigskip
\noindent{\bf Weak Scattering}: The first row in Fig.~\ref{fig:dist} shows the resulting distributions for the weak scattering case with $\S=2.5$, which in turn correspond to the results shown in Fig.~\ref{fig:2.5_20}. In the first panel from the left, the pdf at $k=k_0$ is displayed. It can be immediately noted that the distribution is not symmetric around its mean, but it is skewed, sharply rising from the left of its maximum, and gently decreasing to its right. This skew form is consistent with the observation that no power spectrum curves are found below $\delta\Delta_{\zeta}^2/\Delta_{\zeta,0}^2\lesssim 10^{-10}$ in Fig.~\ref{fig:2.5_20}. 

The center and right panels of the top row in Fig.~\ref{fig:dist} correspond to the numerically obtained distributions for $k=(k_0k_f)^{1/2}$ and $k=k_f$, respectively, with $\S=2.5$. Again, note the skewness of the pdf. Note also the consistency with Fig.~\ref{fig:2.5_20}, in that there are not many realizations there that would fall on the right tail of the distribution. These pdfs are also consistent with our earlier conclusion that a realization that can overcome the suppression by the factor $(\Delta_{\zeta,0}^2)^2$ in (\ref{eq:deltaDeltafinal}) and be phenomenologically interesting is very unlikely for this value of the scattering parameter. We finally note in these two cases that the shape of the distribution is such that a skew-normal fit is inadequate, as we cannot both fit the shape of the peak and the shape of the tail of the numerical pdf. Nevertheless, we do not exclude the possibility that this is the result of our limited numerical precision and ensemble size.\par\bigskip
\noindent{\bf Moderate Scattering}: The middle row of Fig.~\ref{fig:dist} shows the pdf for the `moderate' scattering case with $\S=25$. Here we observe the same features that we discussed in the previous case: a distribution that spans many orders of magnitude in $\delta\Delta_{\zeta}^2$, which is of skew-normal form for $\xi$ and log-skew-form for the power spectrum correction, in this case for all three values of $k$ that we have considered. Note a cutoff in the distribution at around $\xi\sim -20$, or $\delta\Delta_{\zeta}^2/ \Delta_{\zeta,0}^2\sim 10^{-9}$ for $k=k_0$. This observation will be relevant for our discussion of what we call the $\mu_2$-suppressed mean power spectrum in Section~\ref{sec:averageps}. 

In turn, the skew-normal fit for the middle panel corresponds to  $\langle \xi\rangle \simeq -3.9$, or a geometric mean of the power spectrum $e^{\langle\xi\rangle}\simeq 10^{-2}$, roughly in agreement with Fig.~\ref{fig:25_20}. It also leads to $\langle \delta\Delta_{\zeta}^2\rangle/\Delta_{\zeta,0}^2 \gtrsim 10^{15}$, where the exact value is heavily dependent on the shape of the tail of the distribution. Note again that for this value of the scattering parameter a significant fraction of the realizations lie in a phenomenologically interesting range of $\xi$.\par\bigskip
\noindent{\bf Strong Scattering}: Finally, in the last row of Fig.~\ref{fig:dist} we observe the pdf for $\xi$ in the strong scattering case with $\S=85$. Besides the obvious difference with respect to the other two cases that is the noticeably larger values of $\xi$ shown in the horizontal axis, we also note that the pdfs in this scenario appear to be less skewed, and closer to normal distributions, in particular the last panel from the left. For all three values of the momentum considered a skew-normal pdf appears to be a good fit. If we focus on the middle panel, corresponding to the geometric mean of $k_0$ and $k_f$, we note that the numerical results then suggest that $e^{\langle\xi\rangle}\simeq 10^{7}$ and $\langle \delta\Delta_{\zeta}^2\rangle/\Delta_{\zeta,0}^2 \gtrsim 10^{105}$, the first value  well within the phenomenologically interesting regime, albeit in some tension with backreaction constraints (see Section~\ref{sec:backreact}), while the second value is well beyond the applicability of the perturbative scheme used here to compute the stochastically excited curvature power spectrum.

\section{The Ensemble-Averaged Power Spectrum}\label{sec:averageps}

In the previous section, we determined the form of the power spectrum correction in the $\mathcal{N}_s\gg 1$ limit, which we proceeded to evaluate numerically. Although the analytical approximation of these previously shown results is a tall task, we will now show that the Brownian property of the spectator fields at superhorizon scales, cf.~(\ref{eq:lnchirateconf}), allows for a relatively simple evaluation of the ensemble average of the stochastically sourced curvature power spectrum. This analytic estimation is possible because, as discussed in Section~\ref{sec:stochform}, the behavior of $X_k$ at superhorizon scales dominates the integral (\ref{eq:deltaDeltafinal}). Moreover, we will show how to bound the behavior of a typical member of the ensemble of solutions. 

Following the result (\ref{eq:deltaDeltafinal}), in the $\mathcal{N}_s\gg 1$ limit, the expectation value of the correction to $\Delta_{\zeta}^2$ can be written as follows, 
\begin{align} \notag
\langle \delta\Delta_{\zeta}^2(k) \rangle  \;&=\; 4\pi^2 (\Delta_{\zeta,0}^2)^2 \sum_{i,j} \bigg\langle \frac{m_i m_j}{H^2}(k\tau_i)^2(k\tau_j)^2 \mathcal{G}_k(\tau,\tau_i) \mathcal{G}_k(\tau,\tau_j) \\ \notag \displaybreak[0]
&\hspace{90pt}\times \int \frac{d^3\bp}{(2\pi)^{3}k}\, \left[  X_p(\tau_i)X_p^*(\tau_j) \right]_{\rm AS}  \big[  X_{|\bp-\bk|}(\tau_i)X_{|\bp-\bk|}^*(\tau_j) \big]_{\rm AS}\bigg\rangle_{\tau,m}\\ \notag
&=\; 4\pi^2 (\Delta_{\zeta,0}^2)^2 \sum_{i,j} \bigg\langle \int \frac{d^3\bp}{(2\pi)^{3}k}\,(k\tau_i)^2(k\tau_j)^2 \mathcal{G}_k(\tau,\tau_i) \mathcal{G}_k(\tau,\tau_j) \\ \label{eq:deltadeltaav0}
&\hspace{90pt}\times  \left\langle \frac{m_i m_j}{H^2}\left[  X_p(\tau_i)X_p^*(\tau_j) \right]_{\rm AS}  \big[  X_{|\bp-\bk|}(\tau_i)X_{|\bp-\bk|}^*(\tau_j) \big]_{\rm AS} \right\rangle_{m} \bigg\rangle_{\tau}\,,
\end{align} 
where the sub-indexes denote the variable with respect to which the expectation value is to be computed, $\tau$ for scattering locations and $m$ for scattering strengths. Recall that the we assume uniform distribution of scatterers in cosmic time, not conformal time.
To evaluate these expectation values, we note that the values of the $X$ mode functions must be continuous everywhere, in particular at each scatterer location (see Eq.~(\ref{eq:junct1})). This implies that the mode function at $\tau=\tau_i$ cannot depend on the value of the amplitude $m_i$ at the same scattering location, therefore $X_p(\tau_i)$ and $m_i$ can be treated as independent random variables. We can then factor the $m$-expectation value in (\ref{eq:deltadeltaav0}) and make use of $\langle m_i m_j\rangle=\sigma^2\delta_{ij}$ (see \eqref{eq:mstats}) to write\footnote{This result has been verified numerically. Up to the precision and number of realizations considered, off-diagonal terms in the sum in (\ref{eq:deltadeltaav0}) always are subdominant relative to the diagonal ($i=j$) terms.\label{foot:eqtime}}
\begin{align} \notag
\langle \delta\Delta_{\zeta}^2(k) \rangle  \;&=\; 4\pi^2 (\Delta_{\zeta,0}^2)^2 \left(\frac{\sigma}{H}\right)^2 \sum_{i} \bigg\langle (k\tau_i)^4 \mathcal{G}_k^2(\tau,\tau_i)  \int \frac{d^3\bp}{(2\pi)^{3}k}\, \left\langle |X_p(\tau_i)|^2_{\rm AS} |X_{|\bp-\bk|}(\tau_i)|^2_{\rm AS}  \right\rangle_{m} \bigg\rangle_{\tau}\\ \label{eq:deltadeltaav}
&\simeq \; 4\pi^2 (\Delta_{\zeta,0}^2)^2 \left(\frac{\sigma}{H}\right)^2 \sum_{i} (k\tau_i)^4 \mathcal{G}_k^2(\tau,\tau_i)  \int \frac{d^3\bp}{(2\pi)^{3}k}\, \left\langle |X_p(\tau_i)|^2_{\rm AS} |X_{|\bp-\bk|}(\tau_i)|^2_{\rm AS}  \right\rangle_{m,\tau}\,.
\end{align}
In the second line of the previous expression we have approximated the non-stochastic factors, $(k\tau_i)^4\mathcal{G}_k^2(\tau,\tau_i)$, by their values at the average scatterer location $\langle \tau_i\rangle$. For uniformly distributed $t_i=-\ln|H\tau_i|/H$, this is equivalent to the evaluation of the sum in (\ref{eq:deltadeltaav}) over a uniform grid in cosmic time.\footnote{In~\cite{StochasticSpectator} we verified that the average growth rates of the mode functions are independent of the stochasticity of $t_i$. } 
\par\bigskip

The analytical evaluation of (\ref{eq:deltadeltaav}) is still a lengthy calculation. Therefore, in order to avoid cluttering this section with mathematical computations, we have provided the full derivation of our results in Appendix~\ref{sec:intdom}, with Eq.~(\ref{eq:deltadeltaav}) as starting point. It is nevertheless worth mentioning that the most crucial aspects of the evaluation correspond to (i) the use of (\ref{eq:Xnpoint}) to compute the necessary Brownian ensemble averages, and (ii) the infrared cutoff provided by $k_0$, which corresponds to the mode that leaves the horizon at the beginning of scatterings. This infrared cutoff is needed to prevent divergent contributions from growth of superhorizon modes from the infinite past. We also note that in this computation, we have not considered the explicit introduction of an ultraviolet cutoff scale at $k\gg k_f$. This is because at large $k$ (even without the cutoff), the ensemble averaged power spectrum is subdominant compared to the adiabatic one. 

The result of the evaluation of (\ref{eq:deltadeltaav}) can be summarized as follows. For convenience, we note that the parameters $\alpha=\mu_1+(3/2)\mu_2$ and $\beta=\mu_1+(1/2)\mu_2$ appearing above are functions of $\S$, and are shown in Fig.~\ref{fig:f1f2}; recall that $\mu_1=\partial_{Ht}\langle \ln |X_k|^2\rangle$, and $\mu_2=\partial_{Ht}{\rm Var}\,[ \ln |X_k|^2]$ (see Sec.~\ref{sec:spectfield}).\footnote{The interpolation at the breaking points (e.g.~$\alpha+\beta=4$ for (\ref{eq:deltaDeltacase1})) require the exact expressions, which are provided in Eqs.~(\ref{eq:fullintk})-(\ref{eq:fullintk3}).}

\begin{enumerate}
\item[$\Re_1$:] For $k\ll k_0$,
\beq \label{eq:deltaDeltacase1}
\langle \delta\Delta_{\zeta}^2  \rangle  \;\simeq\; \frac{2}{9} (\Delta_{\zeta,0}^2)^2 \mathcal{N}_s\left(\frac{\sigma}{H}\right)^2 \left(\dfrac{k}{k_0}\right)^3 \times \begin{cases}
\dfrac{1}{3} \left(1+\dfrac{8}{\beta-4}-\dfrac{4}{\alpha+\beta-4}\right)\,, & \alpha+\beta<4\,,\\[10pt]
\left(\dfrac{\alpha+\beta}{\alpha+\beta-1}\right)\dfrac{e^{(\alpha+\beta-4)N_{\rm tot}}}{\alpha+\beta-4}\,, & \alpha+\beta>4\,.
\end{cases}
\eeq
In this case, the expected cubic scaling with momenta for any value of $\S$ is immediately evident, cf.~(\ref{eq:K1k0}). For $\alpha+\beta>4$ we note in addition the exponential dependence on the duration of the scattering phase. In the $\alpha+\beta<4$ regime, the lack of this exponential enhancement makes the stochastic component the subdominant piece of $\Delta_{\zeta}^2$, while for $\alpha+\beta>4$, it is clear that it can be easily dominant if $N_{\rm tot}$ is sufficiently large, as we will quantify below. A depiction of the dependence of the power spectrum on the scattering parameter $\S$ in this regime can be found in Fig.~\ref{fig:nsplots2} in Appendix~\ref{sec:meanps} (see also the accompanying discussion).

\item[$\Re_2$:]  For $k_0\ll k \ll k_f$,
\beq \label{eq:deltaDeltacase2}
\langle \delta\Delta_{\zeta}^2 \rangle  \;\simeq\; \frac{2}{9} (\Delta_{\zeta,0}^2)^2 \mathcal{N}_s\left(\frac{\sigma}{H}\right)^2  \times \begin{cases}
D_{\alpha\beta}\,, & \alpha+\beta<4\,,\\[10pt]
\dfrac{C_{\alpha\beta}(k/k_0)}{\alpha+\beta-4}\left(\dfrac{k}{k_0}\right)^{4-\alpha-\beta}e^{(\alpha+\beta-4)N_{\rm tot}} \,, & \alpha+\beta>4\,,
\end{cases}
\eeq
where the coefficient function $D_{\alpha\beta}$ is defined as follows,
\begin{align} \notag
D_{\alpha\beta} \;=\; &\frac{7}{48} +\frac{1}{\beta -2} +\frac{1}{12-3 \beta} -\frac{1}{3 (\alpha +\beta-1)}+\frac{\alpha }{8 (\alpha+\beta )} -\frac{1}{(1-\beta) (\alpha +\beta-4)} \Bigg[ \frac{2^{\alpha+\beta -2}}{\alpha +\beta -2} \\ \label{eq:dab}
&  -B_{\left(1/2,1\right)}(1-\alpha,2-\beta) -B_{(0,1)}(\alpha +\beta-2,2-\beta) \Bigg]\,,
\end{align}
and where
\begin{align}\notag
C_{\alpha\beta}(k/k_0)  \;&=\; \frac{1}{1-\beta}\Bigg[ \frac{ 2^{\alpha+\beta-2} }{\alpha+\beta-2}  - B_{(1/2,1)}(1-\alpha,2-\beta) - B_{(0,1)}(\alpha+\beta-2,2-\beta) \\ \label{eq:calphabeta}
&\hspace{60pt} - \beta \left(\frac{1-\beta}{2-\beta}\right)\left(\frac{k}{k_0}\right)^{\beta-2} \Bigg]\,.
\end{align}
Here $B$ denotes the generalized incomplete beta function (see Eq.~(\ref{eq:betafunct}) for its definition). In this case, the mode of interest leaves the horizon during scatterings and it is therefore of particular phenomenological interest. From  (\ref{eq:deltaDeltacase2}) one immediately notices that the stochastically sourced component of the power spectrum is on average scale invariant for weak scattering. Nevertheless, the mean excited spectrum is never the dominant component of the curvature power spectrum (see Fig.~\ref{fig:nsplots1}). For $\S\gtrsim 1.5$, scale invariance is lost, and the spectator field is in the regime of exponential excitation, in agreement with our expectation (\ref{eq:K1cont}) and our numerical results. One can verify that values beyond the assumption of perturbativity (for $\Delta_{\zeta,0}^2=\Delta_{\zeta,\,{\rm Planck}}^2$) are found for $\mathcal{O}(10)$ magnitudes of $\S$, as it can be explicitly observed in panel (b) of Fig.~\ref{fig:nsplots1}, found in Appendix~\ref{sec:meanps}. One must nevertheless reserve arriving to a conclusion yet, as the results presented here correspond to the arithmetic average over the ensemble of disorder realizations. As we found in Section~\ref{sec:numres}, this average is dominated by large outliers within the ensemble, and therefore does not represent a typical member of the ensemble. We will elaborate on this distinction in the following section.
\item[$\Re_3$:]  For $k_f\ll k$,
\beq \label{eq:deltaDeltacase3}
\langle \delta\Delta_{\zeta}^2 \rangle  \;\simeq\; \frac{1}{16} (\Delta_{\zeta,0}^2)^2 \left[\mathcal{N}_s\left(\frac{\sigma}{H}\right)^2\right]^2  \times \begin{cases}
\dfrac{4 }{4-\beta^2} \left(\dfrac{k}{k_0}\right)^{-\beta/2-1} e^ {(\beta/2+1) N_{\rm tot} }\,, & \beta<2\,,\\[10pt]
\dfrac{1}{\beta} \left(\dfrac{k}{k_0}\right)^{-2} e^{\beta N_{\rm tot}}\,, & \beta>2\,.
\end{cases}
\eeq
When the Goldstone mode ($\pi_k$) satisfies the $k_f\ll k$ condition, at least one of the $X$-mode functions in (\ref{eq:deltadeltaav}) is sub-horizon, which results in a suppression due to the AS. This results in the observed dependence on the square of $\S$ in (\ref{eq:deltaDeltacase3}). Note also that the stochastic component $\delta\Delta_{\zeta}^2$ is not scale invariant for any scattering strength. For weak scattering, $\langle \delta\Delta_{\zeta}^2\rangle \sim k^{-1}$, while for strong scattering, $\langle \delta\Delta_{\zeta}^2\rangle \sim k^{-2}$. The dependence of the ratio $\langle \delta\Delta_{\zeta}^2\rangle/\Delta_{\zeta,0}^2$ as a function of the scattering parameter for $k>k_f$ is shown in Fig.~\ref{fig:nsplots3} of Appendix~\ref{sec:meanps}.
\end{enumerate}
Fig.~\ref{fig:kplots1} shows the mean stochastic enhancement of the curvature power spectrum as a function of $k/k_0$, for a few selected values of $\mathcal{N}_s(\sigma/H)^2$ with $N_{\rm tot}=20$.  
\begin{figure}[!t]
\centering
    \includegraphics[width= \textwidth]{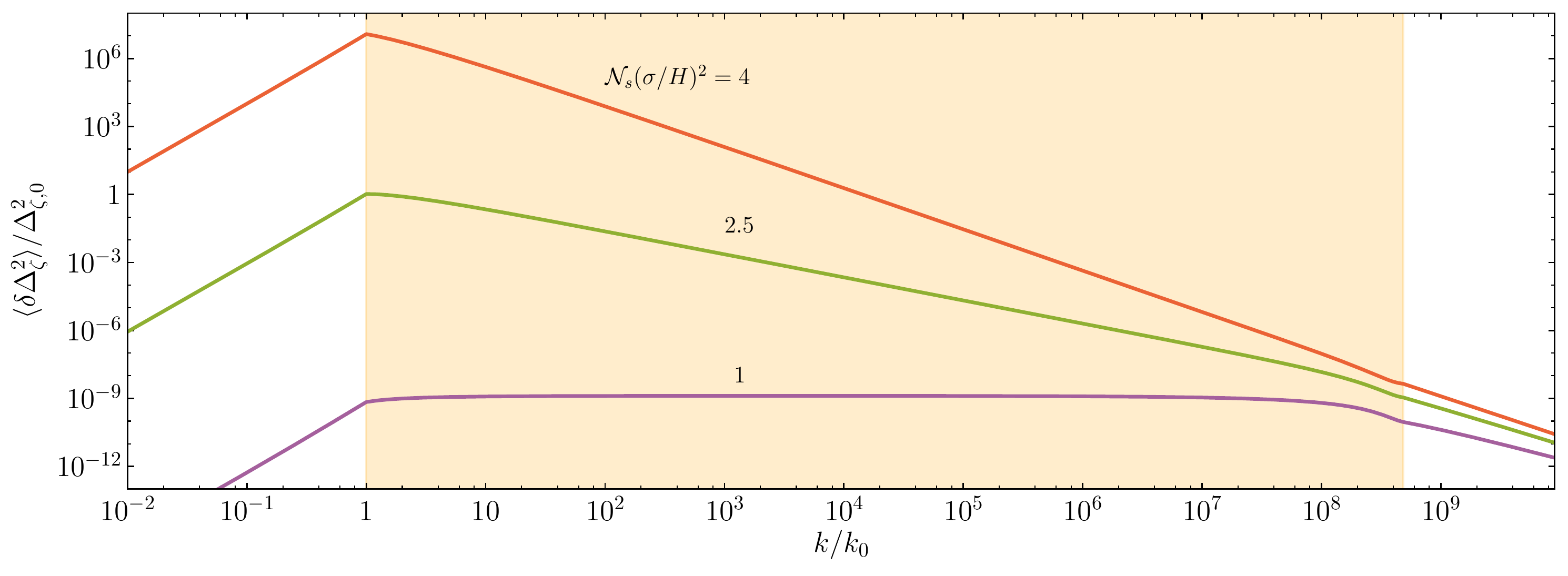}
    \caption{Ratio of the mean stochastic component of the curvature power spectrum to the adiabatic one, as a function of the wavenumber $k$, for $N_{\rm tot}=20$, $\Delta_{\zeta,0}^2=\Delta_{\zeta,\,{\rm Planck}}^2$, and $\S=1$, $2.5$ and $4$. The yellow region highlights those modes that leave the horizon during scatterings.}
    \label{fig:kplots1}
\end{figure}
The difference with respect to the numerical results discussed in Section~\ref{sec:numres} is evident. We must recall and emphasize first that our analytical approximation does not account for the physical cutoff at $k\gg k_f$, which is why we do not show the form of the mean power spectrum beyond a decade to the right of $k_f$. We note nevertheless the red tilt of the spectrum for $k>k_f$. The next feature we note is the difference in the tilt of the spectrum for $k<k_0$ and $k>k_0$. For the former, small momentum case, our analytical approximation displays the causality-enforced cubic dependence on $k$. For large momenta, the predicted scale-invariance for weak scattering, and the increasing red tilt as a function of $\S$ for strong scattering can be explicitly observed. It is however the overall magnitude of the spectrum that most notably clashes with our numerical results, for which the corresponding power spectrum realizations are shown in Fig.~\ref{fig:2.5_20}, with $\S=2.5$. Note that numerically we observe a near scale invariant result for most realizations, with a maximum enhancement that is $\mathcal{O}(10^{-3})$. Our analytically computed mean in turn displays a significant tilt, $\langle \delta\Delta_{\zeta}^2\rangle \sim k^{-1}$, and its maximum reaches the adiabatic value of the spectrum. 

As we previously discussed in Section~\ref{sec:dist}, the lognormality of the spectator field results in a heavily skewed distribution for the $\pi$ two-point function. In our present analytical construction, it manifests itself via the dependence of the expectation values of $|X|^2$ on the variance parameter $\mu_2$. For this field, the geometric mean $e^{\langle \ln|X_p|^2\rangle}$ can be identified with a typical member of the ensemble of realizations; $\ln|X_p|^2$ is normally distributed with mean proportional to $\mu_1$. However, the skewed variable $|X_p|^2$ possesses a arithmetic average that is dominated by large outliers: improbably large excursions of the spectator field for which $|X_p(t)|^2\sim e^{(\mu_1+\mu_2/2)Ht}$. Given that our derivation depends on these arithmetic averages, it is no surprise that our resulting mean power spectrum is determined by realizations living on the tail of the probability distribution. 

\subsubsection*{Variance-suppressed Power Spectrum}
\begin{figure}[!t]
\centering
    \includegraphics[width= \textwidth]{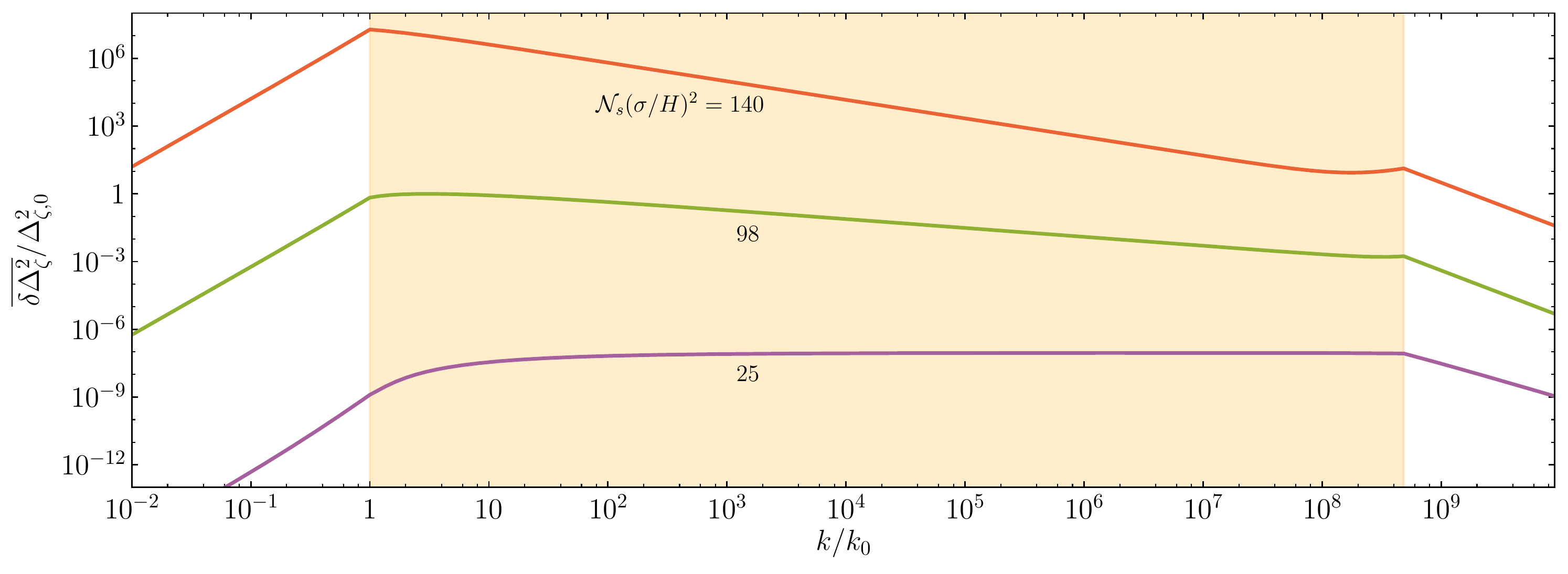}
    \caption{Ratio of the $\mu_2$-suppressed mean stochastically sourced component of the curvature power spectrum to the adiabatic one, as a function of the wavenumber $k$, for $N_{\rm tot}=20$ and $\S=25$, $98$ and $140$. The yellow region highlights those modes that leave the horizon during scatterings. }
    \label{fig:kplots2}
\end{figure}

In our study of the dynamics of the spectator $X$, we identified the limit $\mu_2\rightarrow 0$ of arithmetic means with the typical value of a member of the ensemble~\cite{StochasticSpectator}. Due to the lognormality of $X$, this identification is justified. In the case of the power spectrum, we have demonstrated that the probability distribution has a log-skew-normal form, which is not lognormal. Nevertheless, it is still of interest to explore the $\mu_2\rightarrow 0$ limit of our previous calculation. The analogous of Fig.~\ref{fig:kplots1} in this limit is shown in Fig.~\ref{fig:kplots2}, where we have defined
\beq
\label{eq:MeanNoVar}
\overline{\delta\Delta_{\zeta}^2} \;\equiv\; \langle \delta\Delta_{\zeta}^2 \rangle_{\mu_2\rightarrow 0}\,,
\eeq
in order to simplify the notation. We immediately note a major difference between the two figures. The $\mu_2$-suppressed result produces scale invariant spectra even for significantly larger values of the scattering parameter than the mean. It is only when $\S\sim\mathcal{O}(10^2)$ that the tilt turns noticeably red, and the enhancement can reach values comparable to the purely adiabatic result. If we compare this figure with the results in Sections~\ref{sec:numres} and \ref{sec:dist}, in particular Figs.~\ref{fig:25_20} and \ref{fig:dist}, we note that $\overline{\delta \Delta_{\zeta}^2}$ lies below the smallest realizations, and it has a similar shape. Therefore, although we cannot identify the limit $\mu_2\rightarrow 0$ with that of the typical member of the ensemble, our results suggest that, for a typical, not improbably large realization,
\beq\label{eq:range}
\overline{\delta \Delta_{\zeta}^2} \;\lesssim\; \delta \Delta_{\zeta}^2 \;\lesssim \; \langle \delta \Delta_{\zeta}^2\rangle\,.
\eeq

\section{Backreaction Constraints}\label{sec:backreact}

There are some natural restrictions to the applicability of our analysis. Regarding the spectator field $\chi$, its nature will be jeopardized if the growth of its magnitude due to non-adiabatic particle production is such that its energy density becomes comparable to that of the background. In addition, dissipation effects connected with the inverse sourcing $\pi\rightarrow\chi$ have been argued 
to be negligible (see discussion following equation \eqref{eq:eomchi}). Finally, the implicit perturbativity assumption behind the expansion (\ref{eq:spi}) will be violated if the stochastic component of the power spectrum exceeds $\delta \Delta^2_\zeta \gtrsim \mathcal{O}(1)$. In this section, we compile our results for the energy density of $\chi$ and the stochastic component of the power spectrum $\delta\Delta_{\zeta}^2$ to determine the parameter space within which our formalism is applicable, and in which it leads to potentially observable effects. We will make use of our analytical results, since they provide useful upper and lower bounds on the behavior of a typical disorder realization.\par\bigskip

\subsubsection*{Background Energy Density Constraint: $\rho_\chi\lesssim 3M_P^2H^2$}
We will first determine the region in parameter space where the energy density sourced by $\chi$, denoted by $\rho_\chi$, dominates the inflationary background. The  mean and typical energy densities of this spectator field have been computed in detail in~\cite{StochasticSpectator}. The backreaction constraint $\Omega_{\chi}<1$, with
\beq
\Omega_{\chi} \;\equiv\; \frac{\rho_{\chi}}{3H^2M_P^2}\,,
\eeq
can be considered as an equivalent constraint on the number of $e$-folds for active scatterings, as a function of the scattering parameter $\S$. For the arithmetic mean $\langle \rho_{\chi}\rangle$, the limit is saturated for a number of $e$-folds given by
\beq\label{eq:neff}
N_{e}(\tau) \;\simeq\; \frac{1}{\beta - 2} \ln\left[1 + 16 \pi^2\left(\frac{\beta-2}{\beta}\right) \frac{ M_P^2 }{H^2}\right]\,,
\eeq
which in our case should be at least equal to $N_{\rm tot}$. The parameter $\beta$ is shown as a function of $\S$ in Fig.~\ref{fig:f1f2}.

\begin{figure}[!t] 
   \centering
   \includegraphics[width=0.76\textwidth]{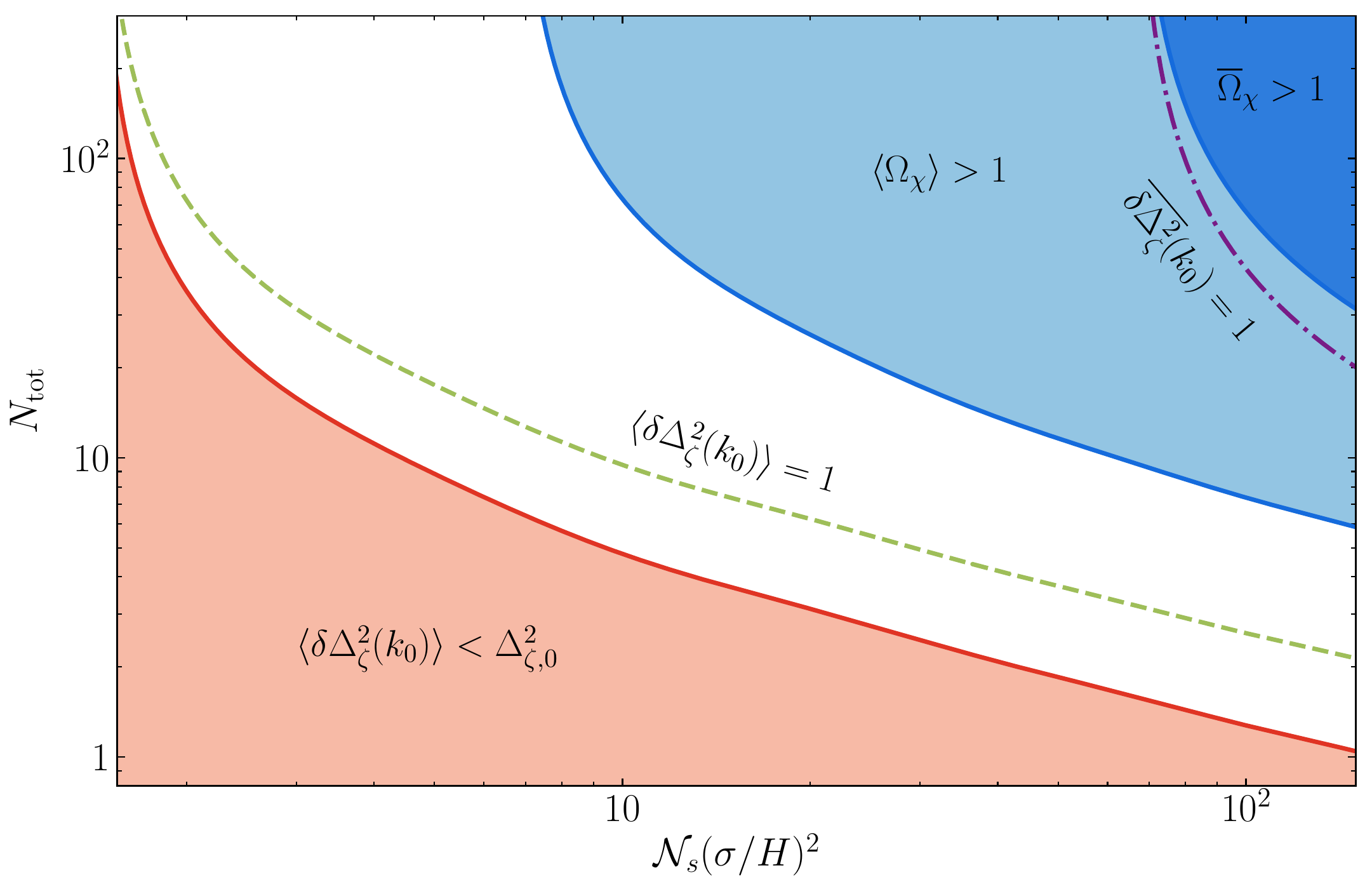} 
   \caption{The parameter space region of interest. The orange shaded region corresponds to parameters that yield corrections to the power spectrum that are smaller than their unperturbed value. The shaded light blue region is excluded by the backreaction constraint for the spectator field in the arithmetic average sense. The dark blue region is excluded by the analogous backreaction constraint in the geometric average sense. The dashed curve corresponds to the mean curvature power spectrum that reaches the perturbativity limit with $\Delta_{\zeta,0}^2=\Delta_{\zeta,\,{\rm Planck}}^2$. The dotted dashed curve is the analogous perturbativity limit for the $\mu_2$-suppressed mean spectrum.}
   \label{fig:constraints}
\end{figure}

In Fig.~\ref{fig:constraints}, we show in light blue the region of parameter space that is excluded by the $\langle \Omega_{\chi}\rangle<1$ constraint, assuming $H=10^{13}\, {\rm GeV}$. Interestingly, we find that for $\S \lesssim 6.8$, the duration of the non-adiabatic particle production can be arbitrarily long, and can in principle last throughout the entire inflationary epoch.
However, as the scattering strength increases, the allowed number of scattering $e$-folds sharply decreases, varying from $N_{\rm tot}\lesssim 10^2$ for $\S\sim10$ to $N_{\rm tot}\lesssim 7$ for $\S\gtrsim 10^2$. We caution the reader that, since the mean of the ensemble is always much larger than its typical values, this bound is not as stringent as it naively suggests. Higher values of $N_{\rm tot}$ are therefore still possible for a typical realization of the ensemble, and is the reason behind our consideration of $\S=85$ in Section~\ref{sec:numres}.

On the other hand, the dark blue region in Fig.~\ref{fig:constraints} illustrates the domain excluded by the backreaction constraint on the typical closure fraction $\overline{\Omega}_{\chi}$, obtained in the $\mu_2\rightarrow 0$ limit of the arithmetic mean result (see Eq.~\eqref{eq:MeanNoVar}). Note that, as expected, the constraint in this case is much milder, inexistent for $\S\lesssim 70$ and only important for very strong scattering -- a regime in which the perturbativity assumption on the power spectrum is already badly violated. \par\bigskip

\subsubsection*{Curvature Perturbation Constraint: $\Delta_\zeta^2\lesssim 1$}
The orange region in Fig.~\ref{fig:constraints} represents the domain in the parameter space where the arithmetic mean of the sourced power spectrum, $\langle \delta \Delta_\zeta^2 \rangle$, is subdominant relative to its unsourced counterpart, which we take to be $\Delta_{\zeta,0}^2=\Delta_{\zeta,\,{\rm Planck}}^2$. For definiteness, we compare this with the maximum value of $\langle \delta \Delta_\zeta^2 \rangle$, which we saw in Fig.~\ref{fig:kplots1} is located at $k=k_0$. As expected, for extremely weak scatterings $\S\lesssim 1$, no amount of expansion is sufficient to lead to an observable effect. Note again that for a typical sample of the ensemble, the correction to the power spectrum is likely to be even smaller. \par\bigskip

The remaining uncolored strip in Fig.~\ref{fig:constraints} therefore represents the region in which the stochastic excitation of $\zeta$ can be the dominant contribution to the power spectrum without running into backreaction issues for most realizations. Note that the green dashed curve represents the contour for which the perturbativity bound is saturated in the arithmetic average sense, i.e. $\langle \delta \Delta^2_\zeta \rangle = 1$, and it would in principle correspond to the upper bound in the allowed range in $\S$ and $N_{\rm tot}$. Nevertheless, the arithmetic mean is expected to significantly overestimate the amplitude of a typical member of the ensemble. Moreover, in the observable wavenumber window, the vacuum power spectrum may be re-normalized, so that $\Delta_{\zeta}^2\simeq \delta\Delta_{\zeta}^2 \gg\Delta_{\zeta,0}^2$. The dash-dotted purple curve shows the saturation of the perturbativity bound for the $\mu_2$-suppressed mean of $\delta\Delta_{\zeta}^2$ for $\Delta_{\zeta,0}^2=\Delta_{\zeta,\,{\rm Planck}}^2$, and can be interpreted as the limit in which our assumptions are violated for most, if not all members of the ensemble of realizations. \par\bigskip

We therefore conclude that, despite the sharp dependence of the stochastic excitation of the power spectrum and the $\chi$ energy density on the number of $e$-folds and the scattering parameter, a wide phenomenologically interesting region in the parameter space can be identified. We now proceed to study the potential observational consequences of the existence of disorder in this allowed window.

\section{Observational Implications}\label{sec:cmb}

\begin{figure}[!t] 
   \centering
   \includegraphics[width=\textwidth]{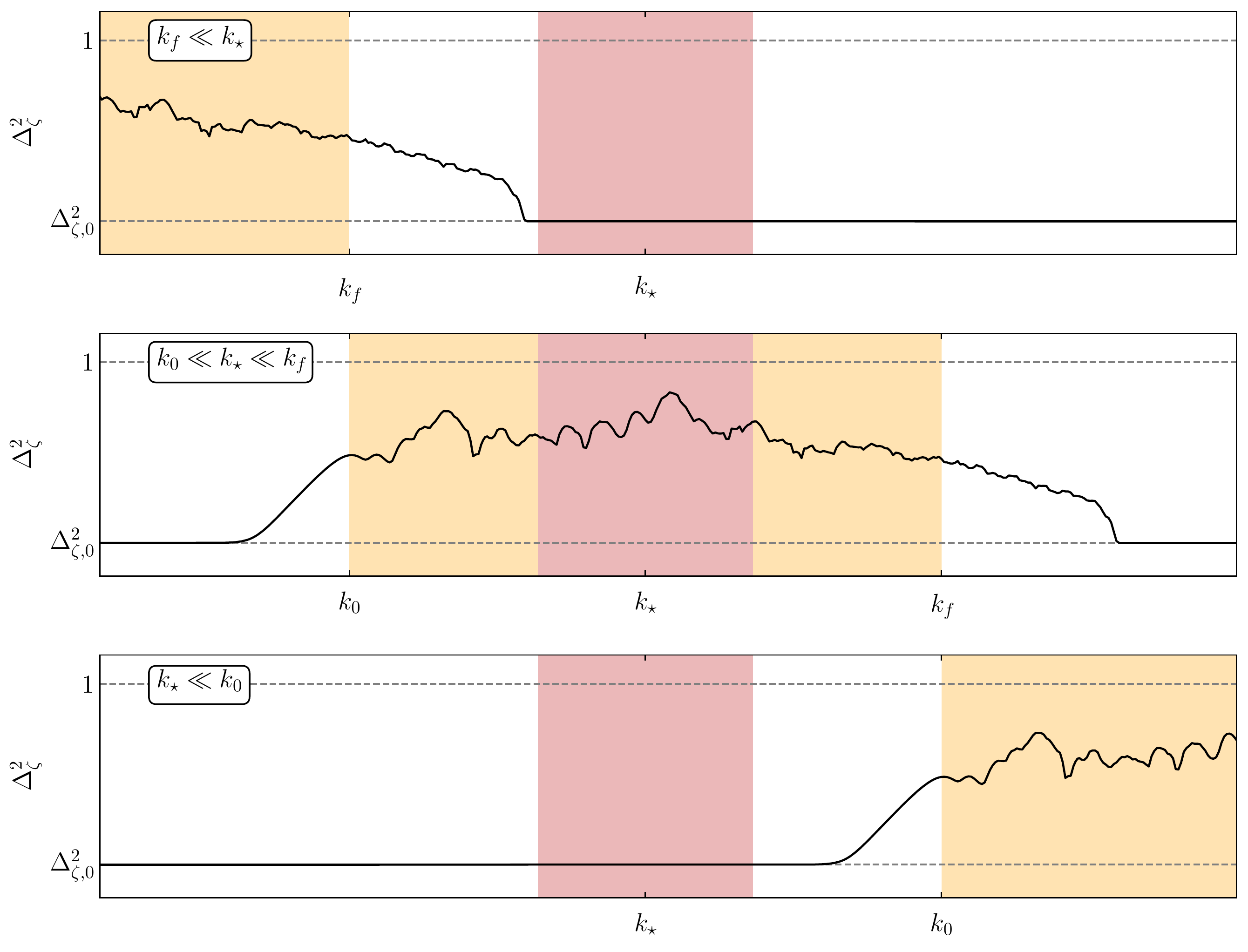} 
   \caption{Relative position of the observable co-moving momentum band (pink) relative to the domain spanned by the stochastically sourced power spectrum (yellow), in the case $\delta\Delta_{\zeta}^2\gg \Delta_{\zeta,0}^2$. Here $k_{\star}$ denotes the (Planck) pivot scale, $k_0=|\tau_0|^{-1}$ and $k_f=|\tau_f|^{-1}$. The tilt of the vacuum contribution $\Delta_{\zeta,0}^2$ has been omitted for clarity.}
   \label{fig:pheno1}
\end{figure}

\begin{figure}[!t]  
   \centering
   \includegraphics[width=\textwidth]{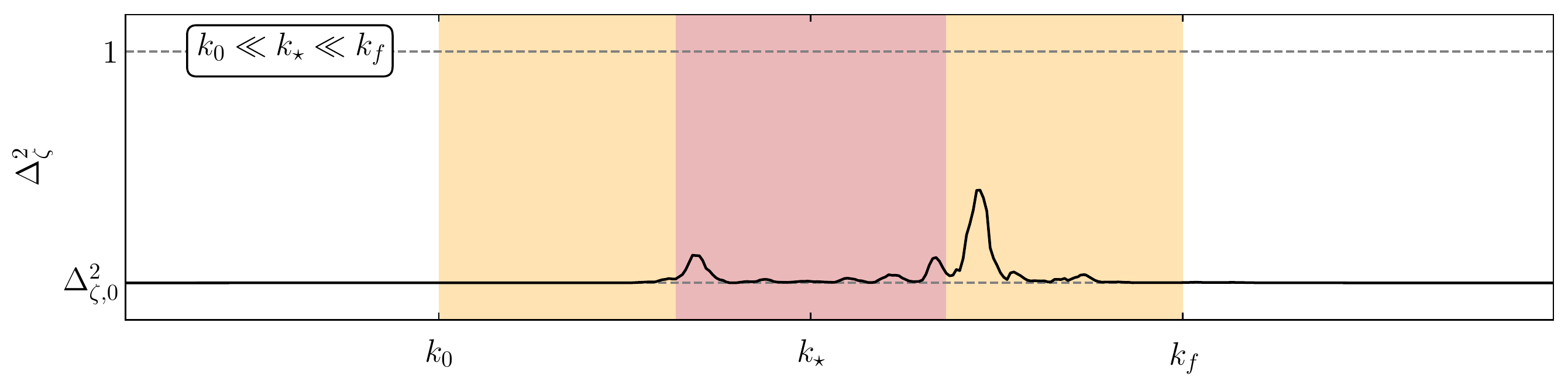} 
   \caption{Position of the observable co-moving momentum band (pink) relative to the domain spanned by the stochastically sourced power spectrum (yellow), in the case $\delta\Delta_{\zeta}^2\simeq \Delta_{\zeta,0}^2$. Here $k_{\star}$ denotes the (Planck) pivot scale, $k_0=|\tau_0|^{-1}$ and $k_f=|\tau_f|^{-1}$. The tilt of the vacuum contribution $\Delta_{\zeta,0}^2$ has been omitted for clarity.}
   \label{fig:pheno2}
\end{figure}

In this section, we discuss the observational consequences of an epoch of non-adiabatic particle production during inflation. We will focus on the power spectrum of the primordial curvature fluctuation, as observed through the temperature and polarization fluctuations of the CMB and the distribution of matter in the low redshift universe (and tracers thereof). The scenarios that potentially lead to observable effects on the curvature power spectrum are summarized in Figs.~\ref{fig:pheno1} and \ref{fig:pheno2} for convenience. 

While inflation typically lasts at least 50-60 $e$-folds, observations directly probe $<10$ $e$-folds.  Over this range of scales, observations of the CMB and galaxy surveys require the primordial power spectrum to be very nearly scale-invariant.  Localized deviations of scale invariance are constrained at the one-percent level from the Planck~\cite{Akrami:2018odb} and BOSS~\cite{Beutler:2019ojk} data.  Indirect constraints from spectral distortions~\cite{Nakama:2017ohe} or primordial black holes~\cite{Sato-Polito:2019hws} constrain the amplitude of fluctuations on smaller scales but still allow for large deviations from scale invariance on such scales.  

We have shown that stochastic particle production introduces, in general, a scale-dependent correction to the primordial power spectrum, and is most prominent for the curvature modes that cross the horizon during the era of stochastic particle production.  The amplitude of this correction depends on $\S$, and also on the particular realization of the disorder for a given $\S$, varying significantly from realization to realization of the stochastic masses and distribution of scatterers in time. 

If the amplitude of the stochastic piece of the power spectrum on a given range of scales is much smaller than the vacuum production of adiabatic modes, $\delta\Delta_{\zeta}^2\ll \Delta_{\zeta,0}^2$, then the effect of stochastic particle production will of course be difficult to observe. Analytically, the results presented in Section~\ref{sec:averageps} suggest that this will be the case if the disorder strength $\S\lesssim \mathcal{O}(1)$. This is a statistical result based on the mean of a very skewed distribution. While even for fixed $\S$ we cannot predict with certainty if the stochastic sourcing will be observable or not, based on the probability distributions discussed in Section \ref{sec:dist}, a large correction to the adiabatic spectrum is unlikely for $\S\lesssim \mathcal{O}(1)$.\par\bigskip

Fig.~\ref{fig:pheno1} shows three possibilities that can arise if the stochastic power spectrum dominates over the adiabatic one, $\delta\Delta_{\zeta}^2\gg \Delta_{\zeta,0}^2$ which we expect for $\S\gtrsim \mathcal{O}(10)$. In our study, the scales corresponding to the beginning ($k_0$) and the end ($k_f$) of the particle production epoch are unconstrained relative to the total duration of inflation. As a result, we can consider at least three distinct possibilities for their relation with the observable range $\Delta k_{\rm obs}$ roughly centered around the pivot scale $k_{\star}$ of CMB measurements. These three possibilities include: (i) The particle-production era may have occurred very early during inflation, spanning wavenumbers much smaller than $k_{\star}$ (top panel). (ii) The non-adiabaticity could have occurred in a temporal window overlapping with the observationally reachable range, $k_0<k_{\star}<k_f$ (middle panel). (iii) The stochastic enhancement could have been active very late during inflation, at wavenumbers much higher than $k_\star$. (bottom panel).

\begin{enumerate}
\item[(i)]The top scenario, with $k_f\ll k_{\star}$ corresponds to an enhancement of power at superhorizon scales that are not reachable by direct observations. Note that here we would need $\Delta_{\zeta,{\rm Planck}}^2\simeq\Delta_{\zeta,0}^2$ for the total $\Delta_\zeta^2$ to be consistent with observations within $\Delta k_{\rm obs}$. Nevertheless, we could have observational implications in the higher point correlation functions. The statistics in our Hubble volume could also be biased by the long wavelength background~\cite{Bramante:2013moa}. Due to the complexity of the calculations leading to the two-point correlation function of $\zeta$ presented here, we leave the discussion of such effects on higher point correlation functions for future work. We believe that the effect of the stochastic sourcing on higher point functions may be even more dramatic given the marked skewness of the field amplitude distribution functions. Although we did not present/assume an explicit inflationary model in which copious particle production with $k_f\ll k_{\star}$ would be realized, models with a quasi-de Sitter complex ``pre-inflationary'' stage have been discussed in the literature~\cite{Ellis:2017jcp}.

\item[(ii)]The middle scenario of Fig.~\ref{fig:pheno1} corresponds to the case when we would have the observed {\it amplitude} of the curvature spectrum be dominated by the stochastically sourced contribution: $\Delta_{\zeta,{\rm Planck}}^2\simeq\delta\Delta_{\zeta}^2\gg \Delta_{\zeta,0}^2$. The stochastically sourced contribution is typically highly scale dependent with many features. Although this is an exciting prospect, given the strength of constraints related to deviations from scale invariance of $\Delta_\zeta^2$ on $\Delta k_{\rm obs}$ (7-10 $e$-folds)~\cite{Akrami:2018odb,Aghanim:2018eyx,Beutler:2019ojk,Chluba:2019kpb}, it is highly unlikely that this scenario is realized for our universe. This $k$ range however is still interesting, and can lead to observationally interesting results, albeit when $\Delta_{\zeta,{\rm Planck}}^2\simeq\Delta_{\zeta}^2\simeq \Delta_{\zeta,0}^2$, in the observational window. Such cases are shown in Fig.~\ref{fig:pheno2}, and discussed further below.  

\item[(iii)]The bottom panel of Fig.~\ref{fig:pheno1} depicts a situation where large deviations from scale invariance occur outside the currently observable window. The most obvious reason for this is if the observable modes cross the horizon before the era of stochastic particle production, with $k_0\gg k_{\star}$. The late-time stochastic excitation of the spectator field may be tied to the end of inflation and the eventual beginning of the (p)reheating epoch~\cite{Kofman:1997yn,Amin:2014eta}. Depending on how close the stochastic sourcing occurs to the end of inflation, the deviation from scale invariance may still be detectable indirectly, through CMB spectral distortions \cite{Chluba:2019kpb}, primordial black holes, etc.~\cite{Sato-Polito:2019hws}.

\par\bigskip
\end{enumerate}

Fig.~\ref{fig:pheno2} depicts what could be the most surprising and interesting possibility. In this scenario, particle production is happening during the interval when the observed modes crossed the horizon, however for $\Delta k_{\rm obs}$ we have here $\delta \Delta_{\zeta}^2 \simeq \Delta_{\zeta,0}^2 \simeq \Delta_{\zeta,{\rm Planck}}^2$. There might be some mild features in the power spectrum on observational scales, and potentially larger ones on scales outside the observational window. What is particularly intriguing in this case is that such highly scale dependent effects can arise, even though the underlying physics is time translation invariant.  In particular, the probability distribution for the stochastic masses is independent of time and thus a large change to the power spectrum on small scales occurs only by chance.  In this sense, indirect observations of the power spectrum may still yield dramatic surprises, even if inflation itself made no effort to hide these effects. 
As an explicit example, as discussed in the Introduction in the context of Figs.~\ref{fig:3D} and \ref{fig:planck}, as well as in Section~\ref{sec:dist}, the realizations for which $\delta\Delta_{\zeta}^2\simeq \Delta_{\zeta,0}^2$ are likelier than that for which $\delta\Delta_{\zeta}^2\gg \Delta_{\zeta,0}^2$, for the set of chosen parameters, namely $\S=25$, $N_{\rm tot}=20$.\footnote{The complete distribution also includes realizations that are ruled out by present data.} Although we do not elaborate on the details of their production, the case depicted in Fig.~\ref{fig:pheno2} is also of relevance for CMB spectral distortions \cite{Chluba:2019kpb}, primordial black holes etc. \cite{Sato-Polito:2019hws}.

\section{Conclusions}\label{sec:conc}

In this paper we have studied the imprints on the primordial curvature power spectrum ($\Delta_\zeta^2$) due to the repeated, non-adiabatic and stochastic excitation of a spectator field ($\chi$) in a de Sitter (inflating) background. Our general approach was to model the complexity of the underlying background field content and dynamics as an effective mass of spectator fields, $m^2(t)$, which changes repeatedly and non-adiabatically. This leads to particle production in the spectator fields, which in turn sources the curvature perturbation. We investigated how the curvature spectrum is sourced by an individual realization of the repeatedly changing effective mass, as well as the statistical properties of the curvature power spectrum over an ensemble of such realizations.  We summarize our formalism, results and implications below.
\subsubsection*{Formalism}We first set up the formalism necessary to calculate the curvature power spectrum  sourced by the repeated excitation of a spectator field. Within our formalism, the leading order coupling between the curvature perturbation and the spectator field was naturally determined in terms of the time derivative of the effective mass of the spectator fields: $\sim dm^2(t)/dt \,\zeta\, \chi^2$ . To simplify the analysis we restricted our attention to a conformally massive spectator field, and modeled the non-adiabatic changes in the effective mass by a series of Dirac-delta functions $m^2(t)=\sum_i m_i \delta(t-t_i)$. We argued that in the limit of a large number of scatterers per $e$-fold ($\mathcal{N}_s\gg 1$), the internal structure of the non-adiabatic changes in the effective mass (which we approximated as Delta-functions -- but more realistically have a temporal with $\sim w$) does not modify our results. Within this limit, an adiabatic subtraction scheme \cite{Parker:1974qw,Birrell:1977,Anderson:1987yt,Fulling:1989nb,parker2009quantum,Kohri:2017iyl} regulates the usual (unsourced) UV divergences in momentum integrals. 

Our formalism allows for controlled calculations even when the sourced curvature spectrum  dominates over the unsourced one ($\delta \Delta_\zeta^2\gg \Delta_{\zeta,0}^2$), as long as the total curvature perturbations remain small compared to unity. That is, $\Delta_\zeta^2=\delta \Delta_\zeta^2+\Delta_{\zeta,0}^2\ll 1$. Another natural restriction of the domain of applicability of our formalism is that the energy density of the spectator fields remains sub-dominant compared to the background energy density ($\rho_\chi\ll 3M_P^2H^2$).

\subsubsection*{Numerical Investigations}
 We first carried out detailed numerical calculations using the transfer matrix approach \cite{StochasticSpectator}, to follow the evolution of the sourcing spectator field for a wide range of parameters. A particularly useful parameter which determines the behavior of the fields is $\S$, where $\sigma^2$ characterizes the strengths of the variations in the effective mass: $\langle m_i m_j\rangle=\sigma^2\delta_{ij}$. Our investigations included strong, $\S\gg 1$, and weak scattering $\S\lesssim 1$. The evolution of spectator fields was already investigated in earlier work \cite{StochasticSpectator}. The evaluation of the curvature spectrum sourced by these excited spectator fields, however, was new to this present work. 

The evaluation of the curvature spectrum required repeated numerical evaluation of momentum integrals over the spectator field modes at unequal times and double-sums over the series of non-adiabatic changes in the effective mass.  This numerically intensive integration was carried out in full for a limited number of cases, but after a series of checks, a number of simplifying assumptions were made to capture the leading order contributions to the curvature spectrum. In particular, we made use of a numerical integration strategy inspired by the fact that the bulk of the curvature perturbation is sourced by super-horizon modes of the spectator fields, and that the equal time contributions to the integrals captured the important qualitative behavior (see Fig.~\ref{fig:25_ij_pq}).

We find that the part of the curvature power spectrum sourced by the spectator fields is enhanced on an interval of wavenumbers $\Delta k_{\rm stoc.}\equiv (k_0,k_f/Hw)$, where $k_0$ is determined by the inverse size of the co-moving horizon at the beginning of the time interval when the effective mass is changing repeatedly, and the upper limit $k_f$ by the inverse size of the co-moving horizon at the end of this duration. $w$ is the typical width in time of a single non-adiabatic event (and sets a momentum cut-off for modes that can be excited non-adiabatically). The shape of the spectrum below $k_0$ is always $\delta \Delta_\zeta^2\propto k^3$, and is essentially determined by causality considerations. The rest of the spectrum, amplitude and shape, is determined by $\S$ and the duration of scatterings in $e$-folds, $N_{\rm tot}$. For the purpose of the summary below we take $N_{\rm tot}\sim 20-40$.

The main results regarding the amplitude of the power spectrum can be summarized as follows: 
\begin{itemize}
\item In the weak scattering regime, with $\S\lesssim \mathcal{O}(1)$, the stochastic sourcing of the power spectrum is insufficient to significantly enhance the adiabatic (vacuum) component of the spectrum $\delta \Delta^2_\zeta\ll \Delta^2_{\zeta,0}$. This case is correlated with the absence of significant particle production in the $\chi$ field, as expected. See Fig.~\ref{fig:2.5_20}.
\item For larger values of the scattering parameter, $\mathcal{O}(1)\lesssim \S \lesssim \mathcal{O}(10^2)$, the enhancement of the power spectrum can be significant, with the possibility of $\delta\Delta_\zeta^2\sim \Delta_{\zeta,0}^2$, or even $\delta \Delta_\zeta^2\gg \Delta_{\zeta,0}^2$. See Figs.~\ref{fig:25_20}--\ref{fig:85_20}. For larger $\S$ backreaction constraints tend to get severe (see Fig.~\ref{fig:constraints}).

\item The shape of the power spectrum shows a lot bumps, which can change $\delta\Delta_\zeta^2$ by orders of magnitude. Moreover, there is a large variation (again by orders of magnitude) between different realizations of the effective mass (even if $\S$ and $N_{\rm tot}$ are fixed). See Fig.~\ref{fig:planck}.

\item For a given $\S$ and $N_{\rm tot}$, different realizations of the power spectra form an ensemble. The amplitudes of the power spectra in this ensemble (at some fixed wavenumber) show a highly skewed distribution which is in most cases well described by a skew-log-normal distribution (see Fig.~\ref{fig:dist}). For a finite sample size from this ensemble, we find that the sample mean overestimates the typical power spectrum. A better estimate for a typical member is given by the geometric mean
. See Fig.~\ref{fig:3D}.
\end{itemize}
\subsubsection*{Analytic Results} 
We exploited the geometric random walking nature of the spectator field magnitude to construct closed-form solutions for the ensemble averaged curvature power spectrum.  A key simplification arises in the calculation of the curvature spectrum: under the ensemble average, only terms evaluated at equal times contribute. We calculated the usual ensemble average, as well as a somewhat modified ensemble average where we ignored the variance of the spectator field perturbations. We showed that for $\S>1$, the ensemble mean is dominated by the tail of the distribution of power spectra amplitudes, and is thus significantly larger than the typical curvature spectrum from the ensemble. On the other hand the modified ensemble average is lower in amplitude than a typical curvature spectrum. 

The shape of the ensemble averaged power spectrum is determined $\S$ and $N_{\rm tot}$. For $k<k_0$, we find a $k^3$ behavior as expected from causality. For strong scattering, we find the slope for $k_0<k<k_f$ is red -- this is primarily related to the time spent by the curvature mode outside the horizon during the duration of scattering. For weak scattering, this slope is blue and logarithmic. See Fig.~\ref{fig:kplots1}.

The analytic calculations were quite useful for us to understand some qualitative aspects of the results for the curvature spectra. However, the high level of variability from realization to realization, and the very skewed distribution of power spectrum amplitudes (especially when $\S\gg1$) make the analytic calculations only marginally useful in estimating the behavior of any particular realization of the curvature power spectrum.

The power spectra (in an ensemble averaged sense) show a break from scale invariance even though we impose scale invariance on the statistical properties of the effective mass for the duration of scattering. Part of this can be explained by the assumption of a finite duration of scattering, and a ultraviolet cut-off in the wavenumbers that can be non-adiabatically excited. Even on scales away from these limits, the lack of scale invariance in the power spectrum comes from the fact that curvature perturbations continue to grow after horizon crossing for the duration that scattering continues (see Fig.~\ref{fig:3D}). The time spent outside the horizon during the scattering duration translates to a scale dependence.

\subsubsection*{Observational Implications} 
To understand the observational implications there are two intervals in $k$ that are relevant. First, is the observational window $\Delta k_{\rm obs}$ spanned by modes that can be probed by the CMB anisotropies and LSS (a few decades near the pivot scale $k_\star$). The second is the interval $\Delta k_{\rm stoc.} = (k_0,k_f/Hw)$ where the sourced part of the curvature power spectrum ($\delta\Delta_\zeta^2$) deviates significantly from scale invariance. For brevity, below we only discuss cases  where $\Delta k_{\rm obs}\cap \Delta k_{\rm stoc.}=\varnothing$ or $\Delta k_{\rm obs}\cap \Delta k_{\rm stoc.}=\Delta k_{\rm obs}$.

\begin{itemize}
\item For $\Delta k_{\rm obs}\cap \Delta k_{\rm stoc.}=\varnothing$, we must have the vacuum part of the curvature power spectrum  be dominant in the observational window: $\Delta_\zeta^2\simeq\Delta_{\zeta,0}^2$. While the power spectrum in the observational window would not directly contain features of the stochastic particle production, the higher-point correlation functions might still contain an imprint. See Fig.~\ref{fig:pheno1}.

If $k_f/Hw$ is to the left of the observational window (top panel in Fig.~\ref{fig:pheno1}), or if $k_0$ is to the right of the observational window (bottom panel Fig.~\ref{fig:pheno1}), it is possible for $\delta\Delta_\zeta^2\gg \Delta_{\zeta,0}^2$. The latter would bias our background, whereas the former could potentially lead to CMB spectral distortions and formation primordial black holes. Such large stochastic components are possible with $\S>1$ and/or $N_{\rm tot}$ being sufficiently large.

\item For $\Delta k_{\rm obs}\cap \Delta k_{\rm stoc.}=\Delta k_{\rm obs}$, given the constraints from CMB anisotropies and LSS, we would have to be in a regime where $\delta\Delta_\zeta^2\lesssim \Delta_{\zeta,0}^2$, at least within $\Delta k_{\rm obs}$. See middle panel of Fig.~\ref{fig:pheno1} and Fig.~\ref{fig:pheno2}. This restricts $\S\lesssim \mathcal{O}(10)$ if we want almost all realizations to have $\delta\Delta_\zeta^2\ll \Delta_{\zeta,0}^2$. 

\item There is high variability between different members of the same ensemble (especially for $\S>1$). This makes the inference from the measured power spectrum to definite underlying (statistical) parameters non-trivial. See Fig.~\ref{fig:3D} and Fig.~\ref{fig:planck}.

\end{itemize}
\subsubsection*{Future Directions} 
We have provided a detailed analysis of the two-point correlation function of the curvature spectrum sourced by spectator fields -- it would be natural to carry out similar calculations for higher-point correlation functions \cite{Flauger:2016idt}. These higher-point correlators may reveal evidence for stochastic particle production and the complexity of the inflationary dynamics, even when the additionally sourced curvature power spectrum is subdominant compared to the vacuum contribution. Another possible avenue to pursue is to work in a regime where the backreaction of the curvature perturbations on the particle production, and the backreaction of particle production on the background evolution need to be taken into account. We hope these analyses will be performed in future works. Similarly, we postpone the study of stochastic gravitational wave production \cite{Caprini:2018mtu}, as well as the application of the stochastic framework to the early stages of non-perturbative reheating \cite{Amin:2014eta}.

\section*{Acknowledgements}
The authors especially thank Daniel Baumann, Scott Carlsten and Horng Sheng Chia  for early interest, insights, continued discussion and work on this project. We would also particularly like to thank Horng Sheng Chia for a thorough reading of the manuscript and a number of helpful conversations and suggestions that improved the manuscript. The authors would also like to thank Dick Bond, Jonathan Braden, Mehrdad Mirbabayi and Eva Silverstein for helpful discussions.  Numerical results were obtained from a custom Fortran code utilizing the thread-safe arbitrary precision package MPFUN-For written by David H. Bailey.  MA is supported by the US Dept.~of Energy grant DE-SC0018216. The work of MG was supported by the US Dept.~of Energy grant DE-SC0018216, and by the Spanish Agencia Estatal de Investigaci\'on through the grants FPA2015-65929-P (MINECO/FEDER, UE), PGC2018095161-B-I00, IFT Centro de Excelencia Severo Ochoa SEV-2016-0597, and Red Consolider MultiDark FPA2017-90566-REDC. DG is supported by the US Dept.~of Energy grant DE-SC0019035.  MA acknowledges the hospitality at the KITP workshop ``From Inflation to the Hot Big Bang" supported in part by the National Science Foundation under Grant No. NSF PHY-1748958, where this work for completed.
MG would like to thank the Laboratoire de Physique Th\'eorique at Universit\'e Paris-Sud for their hospitality and financial support while completing this work. MG also acknowledges support by Institut Pascal at Universit\'e Paris-Saclay with the support of the P2I and SPU research departments and the P2IO Laboratory of Excellence (program ``Investissements d'avenir'' ANR-11-IDEX-0003-01 Paris-Saclay and ANR-10-LABX-0038), as well as the IPhT.


\appendix

\section{Derivation of the Sourced Power Spectrum for Dirac-delta Scatterers}\label{app:diracdelta}


In this appendix we derive in detail Eq.~(\ref{eq:deltazetaK}) for the correction to the curvature power spectrum due to the stochastic excitation of a spectator field by Dirac-delta scatterers. Our calculation will require the evaluation of the $X$ mode functions and their derivatives at each scatterer location. Recall from (\ref{eq:eomXk}) that the equation of motion for the mode functions of the canonically normalized spectator field has the form
\beq
X_k''(\tau) + \left[ k^2 - \frac{a''}{a} + a^2M^2 + \sum_i m_i a(\tau_i) \delta (\tau-\tau_i) \right] X_k(\tau)\;=\;0\,.
\eeq
It follows that, at each scatterer location, the junction conditions must be satisfied,
\begin{align}\label{eq:junct1}
X_k(\tau_i^+) \;&=\; X_k(\tau_i^-)\,,\\ \label{eq:junct2}
X^{\,\prime}_k(\tau_i^+) \;&=\;  X^{\,\prime}_k(\tau_i^-) - m_i a(\tau_i) X_k(\tau_i)\,.
\end{align}
Note the discontinuity of the derivative. We use the junction conditions above and Eq.~(\ref{eq:deltadeltazeta}) as our starting points. Distributing the conformal time derivatives, we split (\ref{eq:deltadeltazeta}) into three pieces, corresponding to how many derivatives act on the $X$ mode functions, $\delta \Delta_{\zeta}^2 = \delta \Delta_{\zeta}^2\big|_0 + \delta \Delta_{\zeta}^2\big|_1 + \delta \Delta_{\zeta}^2\big|_2$.

\subsection*{No $X$-derivative}

The first term in the expansion (\ref{eq:deltazetaK}) arises from the term in (\ref{eq:deltadeltazeta}) that contains no derivatives acting on $X$. In this case, the continuity of the mode functions allows for a straightforward evaluation,
\begin{align}\notag
\delta\Delta_{\zeta}^2\big|_{0} \;&=\; 4\pi^2 (\Delta_{\zeta,0}^2)^2 \frac{k^3}{H^6} \sum_{i,j} m_i m_j \int d\tau' \,d\tau''\, \frac{\delta(\tau'-\tau_i)}{a(\tau')} \frac{\delta(\tau''-\tau_j)}{a(\tau'')}  \frac{d}{d\tau'} \frac{d}{d\tau''} \bigg\{ \frac{G_k(\tau,\tau')}{a(\tau')}\frac{G_k(\tau,\tau'')}{a(\tau'')} \bigg\} \\ \notag \displaybreak[0]
&\hspace{140pt} \times \int \frac{d^3\bp}{(2\pi)^{3}}\, \left[  X_p(\tau')X_p^*(\tau'') \right]_{\rm AS}  \big[  X_{|\bp-\bk|}(\tau')X_{|\bp-\bk|}^*(\tau'') \big]_{\rm AS}  \\ \notag
&=\; 4\pi^2 (\Delta_{\zeta,0}^2)^2 \frac{k^3}{H^2} \sum_{i,j} m_i m_j \int d\tau' \,d\tau''\, \delta(\tau'-\tau_i) \delta(\tau''-\tau_j) (\tau')^2(\tau'')^2   \mathcal{G}_k(\tau,\tau') \mathcal{G}_k(\tau,\tau'')  \\ \notag \displaybreak[0]
&\hspace{140pt} \times \int \frac{d^3\bp}{(2\pi)^{3}}\, \left[  X_p(\tau')X_p^*(\tau'') \right]_{\rm AS}  \big[  X_{|\bp-\bk|}(\tau')X_{|\bp-\bk|}^*(\tau'') \big]_{\rm AS}\\ \notag
&=\; 4\pi^2 (\Delta_{\zeta,0}^2)^2  \sum_{i,j} \frac{m_i m_j}{H^2}  (k\tau_i)^2(k\tau_j)^2   \mathcal{G}_k(\tau,\tau_i) \mathcal{G}_k(\tau,\tau_j)  \\ \notag \displaybreak[0]
&\hspace{140pt} \times \int \frac{d^3\bp}{(2\pi)^{3}k}\, \left[  X_p(\tau_i)X_p^*(\tau_j) \right]_{\rm AS}  \big[  X_{|\bp-\bk|}(\tau_i)X_{|\bp-\bk|}^*(\tau_j) \big]_{\rm AS}\\
&\equiv\; 4\pi^2 (\Delta_{\zeta,0}^2)^2  \sum_{i,j} \frac{m_i m_j}{H^2}  (k\tau_i)^2(k\tau_j)^2   \mathcal{K}_{ij}^{\rm I}\,.
\end{align}

\subsection*{One $X$-derivative}

Let us now collect all terms proportional to $\partial_{\tau}X$ in (\ref{eq:deltadeltazeta}). They can be written in the following form,
\begin{align}\notag
\delta\Delta_{\zeta}^2\big|_{1} \;&=\; 4\pi^2 (\Delta_{\zeta,0}^2)^2 \frac{k^3}{H^6} \sum_{i,j} m_i m_j \int d\tau' \,d\tau''\, \frac{\delta(\tau'-\tau_i)}{a(\tau')} \frac{\delta(\tau''-\tau_j)}{a(\tau'')}  \frac{d}{d\tau'} \bigg\{ \frac{G_k(\tau,\tau')}{a(\tau')} \bigg\} \frac{G_k(\tau,\tau'')}{a(\tau'')} \\ \notag \displaybreak[0]
&\hspace{80pt} \times  \frac{d}{d\tau''} \int \frac{d^3\bp}{(2\pi)^{3}}\, \left[  X_p(\tau')X_p^*(\tau'') \right]_{\rm AS}  \big[  X_{|\bp-\bk|}(\tau')X_{|\bp-\bk|}^*(\tau'') \big]_{\rm AS} \;+\; \tau'\leftrightarrow\tau''   \\ \notag
&=\; 4\pi^2 (\Delta_{\zeta,0}^2)^2  \sum_{i,j} \frac{m_i m_j}{H^2} (k\tau_i)^2 \mathcal{G}_k(\tau,\tau_i) \int d\tau''\, \delta(\tau''-\tau_j) (k\tau'')^2    G_k(\tau,\tau'')  \\ \label{eq:K2int}
&\hspace{80pt} \times \frac{d}{d\tau''}  \int \frac{d^3\bp}{(2\pi)^{3}k}\, \left[  X_p(\tau_i)X_p^*(\tau'') \right]_{\rm AS}  \big[  X_{|\bp-\bk|}(\tau_i)X_{|\bp-\bk|}^*(\tau'') \big]_{\rm AS} \;+\; \tau'\leftrightarrow\tau'' \,.
\end{align}
Note that now the result depends on the derivative of the mode functions of $X$ at the discontinuity (see (\ref{eq:junct2})). Nevertheless, an unambiguous value for the conformal time integral in (\ref{eq:K2int}) can be computed~\cite{diracdelta}. Approximating the delta function by a sequence of functions of area one, e.g.
\beq
\delta(\tau''-\tau_j) \;=\; \lim_{\epsilon\rightarrow 0}\, \frac{1}{2\epsilon}\times \begin{cases}
1\,, & |\tau''-\tau_j|<\epsilon\\
0\,, & \text{otherwise}
\end{cases}\,,
\eeq
and applying the continuity conditions on the mode function and its derivative across the interface, it is straightforward to verify that the following replacement rule applies inside the integral sign,
\begin{align} \notag
\delta(\tau''-\tau_j)  \frac{d}{d\tau''} X_p^*(\tau'') \;&=\; \frac{1}{2} \delta(\tau''-\tau_j) \left[ X_p^{*\prime}(\tau_j^+) + X_p^{*\prime}(\tau_j^-)  \right]\\ \label{eq:rrule}
&=\;  \delta(\tau''-\tau_j) \left[ X_p^{*\prime}(\tau_j^-) - \frac{1}{2}m_j a(\tau_j)X_p^*(\tau_j) \right]\,.
\end{align}
Substitution of this outcome into (\ref{eq:K2int}) gives the following final expression for the single-derivative contribution,
\begin{align}\notag
\delta\Delta_{\zeta}^2\big|_{1} \;&=\; 4\pi^2 (\Delta_{\zeta,0}^2)^2  \sum_{i,j} \frac{m_i m_j}{H^2} (k\tau_i)^2 (k\tau_j)^2 \mathcal{G}_k(\tau,\tau_i) G_k(\tau,\tau_j)     \\ \notag
&\hspace{20pt} \times   \int \frac{d^3\bp}{(2\pi)^{3}k}\, \bigg\{ \Big( \big[  X_p(\tau_i)X_p^{*\prime}(\tau_j^{-}) \big]_{\rm AS}  \big[  X_{|\bp-\bk|}(\tau_i)X_{|\bp-\bk|}^{*}(\tau_j) \big]_{\rm AS} \\ \notag
&\hspace{100pt} + \big[  X_p(\tau_i)X_p^{*}(\tau_j) \big]_{\rm AS}  \big[  X_{|\bp-\bk|}(\tau_i)X_{|\bp-\bk|}^{*\prime}(\tau_j^{-}) \big]_{\rm AS} + {\rm h.c.}\Big)\\ \notag
&\hspace{100pt} + \left(\frac{m_i}{2H\tau_i} + \frac{m_j}{2H\tau_j} \right) \Big( X_p(\tau_i)X_p^{*}(\tau_j)  \big[  X_{|\bp-\bk|}(\tau_i)X_{|\bp-\bk|}^{*}(\tau_j) \big]_{\rm AS}\\ \label{eq:dDz21}
&\hspace{148pt} + \big[  X_p(\tau_i)X_p^{*}(\tau_j) \big]_{\rm AS} X_{|\bp-\bk|}(\tau_i)X_{|\bp-\bk|}^{*}(\tau_j)  \Big) \bigg\}\,.
\end{align}

\subsection*{Two $X$-derivatives}

We finish by calculating the contribution from $\delta \Delta_{\zeta}^2\big|_2$. Using the replacement rule (\ref{eq:rrule}), this expansion is straightforward, albeit somewhat lengthy,
\begin{align}\notag
\delta\Delta_{\zeta}^2\big|_{2} \;&=\; 4\pi^2 (\Delta_{\zeta,0}^2)^2 \frac{k^3}{H^6} \sum_{i,j} m_i m_j \int d\tau' \,d\tau''\, \frac{\delta(\tau'-\tau_i)}{a(\tau')} \frac{\delta(\tau''-\tau_j)}{a(\tau'')}  \frac{G_k(\tau,\tau')}{a(\tau')}  \frac{G_k(\tau,\tau'')}{a(\tau'')} \\ \notag \displaybreak[0]
&\hspace{20pt} \times  \frac{d}{d\tau'}\frac{d}{d\tau''} \int \frac{d^3\bp}{(2\pi)^{3}}\, \left[  X_p(\tau')X_p^*(\tau'') \right]_{\rm AS}  \big[  X_{|\bp-\bk|}(\tau')X_{|\bp-\bk|}^*(\tau'') \big]_{\rm AS} \\[10pt] \displaybreak[0] \notag
&=\; 4\pi^2 (\Delta_{\zeta,0}^2)^2  \sum_{i,j} \frac{m_i m_j}{H^2} (k\tau_i)^2 (k\tau_j)^2 G_k(\tau,\tau_i) G_k(\tau,\tau_j)  \\ \displaybreak[0] \notag
&\hspace{20pt} \times  \int \frac{d^3\bp}{(2\pi)^{3}k}\, \bigg\{ \big[  X_p^{\prime}(\tau_i^{-})X_p^{*\prime}(\tau_j^{-}) \big]_{\rm AS}  \big[  X_{|\bp-\bk|}(\tau_i)X_{|\bp-\bk|}^*(\tau_j) \big]_{\rm AS} \\ \displaybreak[0] \notag
&\hspace{90pt} + \big[  X_p^{\prime}(\tau_i^{-})X_p^{*}(\tau_j) \big]_{\rm AS}  \big[  X_{|\bp-\bk|}(\tau_i)X_{|\bp-\bk|}^{*\prime}(\tau_j^{-}) \big]_{\rm AS}\\ \displaybreak[0] \notag
&\hspace{90pt} + \big[  X_p(\tau_i)X_p^{*\prime}(\tau_j^{-}) \big]_{\rm AS}  \big[  X_{|\bp-\bk|}^{\prime}(\tau_i^{-})X_{|\bp-\bk|}^*(\tau_j) \big]_{\rm AS}\\ \displaybreak[0] \notag
&\hspace{90pt} + \big[  X_p(\tau_i)X_p^{*}(\tau_j) \big]_{\rm AS}  \big[  X_{|\bp-\bk|}^{\prime}(\tau_i^{-})X_{|\bp-\bk|}^{*\prime}(\tau_j^{-}) \big]_{\rm AS}\\ \displaybreak[0] \notag
&\hspace{90pt} + \frac{m_j}{2H\tau_j} \Big( X_p^{\prime}(\tau_i^{-})X_p^{*}(\tau_j) \big[  X_{|\bp-\bk|}(\tau_i)X_{|\bp-\bk|}^*(\tau_j) \big]_{\rm AS}\\ \displaybreak[0] \notag
&\hspace{140pt}  + X_p(\tau_i)X_p^{*}(\tau_j)   \big[  X_{|\bp-\bk|}^{\prime}(\tau_i^{-})X_{|\bp-\bk|}^*(\tau_j) \big]_{\rm AS}\\ \displaybreak[0] \notag
&\hspace{140pt} + \big[  X_p^{\prime}(\tau_i^{-})X_p^{*}(\tau_j) \big]_{\rm AS}  X_{|\bp-\bk|}(\tau_i)X_{|\bp-\bk|}^*(\tau_j) \\ \displaybreak[0] \notag
&\hspace{140pt} + \big[  X_p(\tau_i)X_p^{*}(\tau_j) \big]_{\rm AS}  X_{|\bp-\bk|}^{\prime}(\tau_i^{-})X_{|\bp-\bk|}^*(\tau_j) + {\rm h.c.}\Big)\\ \displaybreak[0] \notag
&\hspace{90pt} + \frac{m_i m_j}{4H^2\tau_i\tau_j} \Big( X_p(\tau_i)X_p^{*}(\tau_j) \big[  X_{|\bp-\bk|}(\tau_i)X_{|\bp-\bk|}^*(\tau_j) \big]_{\rm AS}\\ \displaybreak[0] \notag
&\hspace{150pt} + \big[  X_p(\tau_i)X_p^{*}(\tau_j) \big]_{\rm AS}  X_{|\bp-\bk|}(\tau_i)X_{|\bp-\bk|}^*(\tau_j) \\ \displaybreak[0]
&\hspace{150pt} + 2 X_p(\tau_i)X_p^{*}(\tau_j)  X_{|\bp-\bk|}(\tau_i)X_{|\bp-\bk|}^*(\tau_j)  \Big) \bigg\} \,.
\end{align}
Our derivation is completed by noting that
\beq
\delta\Delta_{\zeta}^2\big|_{1} + \delta\Delta_{\zeta}^2\big|_{2} \;=\; 4\pi^2 (\Delta_{\zeta,0}^2)^2 \sum_{i,j} \frac{m_i m_j}{H^2}(k\tau_i)^2(k\tau_j)^2 \left( \mathcal{K}^{\rm II}_{ij} + \mathcal{K}^{\rm III}_{ij} \right)\,,
\eeq
with $\mathcal{K}^{\rm II}_{ij}$ and $\mathcal{K}^{\rm III}_{ij}$ given by (\ref{eq:K2}) and (\ref{eq:K3}).


\section{Cutoff-Dependence of the Power Spectrum}
\label{sec:cutoff}

Although the Dirac-delta approximation for the stochastic mass of the spectator field is a convenient mathematical tool for the computation of cosmological observables, the unphysical vanishing widths for the scattering events is a source of divergences that need to be addressed to obtain a physically sensible result. In this appendix, we study the dependence of the stochastic power spectrum on the UV cutoff that is required to regularize the momentum integral
in the Dirac-delta approximation for $m^2(\tau)$. First, we discuss the nature of the cutoff, and then we proceed to compute its contribution to $\Delta_{\zeta}^2$ in the large $\mathcal{N}_s$ limit.

\subsection{UV Sensitivity in the Dirac-delta Approximation}\label{app:cutoff1}

Consider for simplicity the ``diagonal'', equal-time terms in the sum in Eq.~(\ref{eq:deltazetaK}). As it was described in Section~\ref{sec:spectfield}, $X_p$ modes that are deep inside the horizon are near their vacuum state. Therefore, at a given conformal time $\tau_i$, any mode for which $|p\tau_i|\gg1$ can be approximated by the vacuum mode function plus a correction,
\begin{flalign}
& \text{($|p\tau_i|\gg 1$)} & \Cen{3}{X_p(\tau_i) \;\simeq\; X_p^0(\tau_i)+\delta X_p(\tau_i)\,.}      &&  
\end{flalign}
For a conformally massive spectator, $X_p^0(\tau_i)=e^{-ip\tau_i}/\sqrt{2p}$. The form of $\delta X_p(\tau_i)$ can be estimated by a ``last scatterer'' approximation; inside the horizon the mode will be most strongly sourced by the latest non-adiabatic event. Using the transfer matrix formalism (see~\cite{StochasticSpectator}),
\begin{flalign}\label{eq:deltaXp}
& \text{($|p\tau_i|\gg 1$)} & \Cen{3}{\delta X_p(\tau_i) \;=\; \frac{i m_i a(\tau_i)}{2p}e^{-2ip\tau_i}X_p^{0\,*}(\tau_i)\,.}      &&  
\end{flalign}
We introduce a momentum cutoff ($\Lambda$) beyond the horizon scale. Then, at large momenta,
\begin{align} \displaybreak[0] \label{eq:K1sc}
\mathcal{K}_{ij}^{\rm I} \;&\sim\;  \int \frac{d^3\bp}{(2\pi)^{3}}\, \left(X_p^0(\tau_i) \delta X_p(\tau_i) \right)^2 \;\propto\; \Lambda^{-1}\,,\\ \displaybreak[0]
\mathcal{K}_{ij}^{\rm II} \;&\sim\;  \int \frac{d^3\bp}{(2\pi)^{3}}\, \left(X_p^0(\tau_i) \right)^3 \delta X_p(\tau_i)  \;\propto\; \ln(\Lambda)\,,\\ \label{eq:K3sc}
\mathcal{K}_{ij}^{\rm III} \;&\sim\;  \int \frac{d^3\bp}{(2\pi)^{3}}\, \left(X_p^0(\tau_i) \right)^4 \;\propto\; \Lambda\,.
\end{align}
In the absence of a cutoff, the correction (\ref{eq:deltazetaK}) appears to exhibit a leading linear divergence, and a sub-leading logarithmic divergence, with only $\mathcal{K}^{\rm I}$ converging to a finite result. As advertised earlier, the source of this apparent failure of the AS regularization scheme is the singular nature of the effective mass (\ref{eq:mdelta}). For any finite-duration events, the non-adiabatic excitation of a mode $X_p(\tau_i)$ is subject to the condition that
\beq\label{eq:ddcut}
p_{\rm phys}w \;=\; |p\tau_i|Hw <1\,,
\eeq
where $p_{\rm phys}$ is the physical momentum at $\tau_i$, and $w$ denotes the temporal ``width'' of the scatterer. As an example, in the case of a ``sech'' effective mass,
\beq\label{eq:msech}
m^2(t) \;=\; \sum_j \frac{m_j}{2w} \,{\rm sech}^2\left(\frac{t-t_j}{w}\right)\,,
\eeq
which in the limit $w\rightarrow 0$ reduces to (\ref{eq:mdelta}), the excitation of the sub-horizon mode in the last-scatterer approximation takes the form~\cite{Amin:2015ftc}
\beq
|\delta X_p(\tau_i)| \;\simeq\; \left|\frac{\cos \left(\frac{\pi}{2}\sqrt{2 m_j w+1}\right)}{\sinh (\pi  |p\tau_j|  Hw)} X_p^{0*}(\tau_j)\right|
\eeq
in the narrow width limit. For $|p\tau_i|Hw \ll 1$, this expression reduces to the Dirac-delta result (\ref{eq:deltaXp}), while for $|p\tau_i|Hw \gg 1$ the excitation is exponentially suppressed, $\delta X_p(\tau_i)\propto e^{-\pi |p\tau_i|Hw}$.

Extrapolating\footnote{We have verified these claims numerically for a limited number of scatterers using the sech approximation (\ref{eq:msech}). In particular, for any $w\neq0$ the resulting power spectrum correction is always finite, and is consistent with the scalings (\ref{eq:K1sc})-(\ref{eq:K3sc}).} these results, we identify the comoving momentum cutoff with the inverse scattering width; more precisely
\beq
\Lambda_i \;\equiv\; (Hw\tau_i)^{-1}\,,
\eeq
i.e.~Eq.~(\ref{eq:cutoff}). Note that this cutoff is time-dependent, ensuring that for sufficiently narrow scatterers no super-horizon modes are ever suppressed from the calculation. Although the previous discussion appears to be valid only for the equal-time terms of the sum (\ref{eq:deltazetaK}), we have verified that our conclusions also apply in the unequal time case.

\subsection{Cutoff-Independence in the Large $\mathcal{N}_s$ Limit}\label{app:cutoff2}

We now show that in the large $\mathcal{N}_s$ limit, the cutoff dependence of $\Delta_{\zeta}^2$ is subdominant with respect to the stochastic sourcing from super-horizon modes. As we learned in the previous Section, $\mathcal{K}^{\rm III}$ contains a linear divergence in the UV due to the Dirac-delta approximation, that must be tamed with a scatterer width cutoff. Given that this divergence dominates over the logarithmic dependence on $\Lambda$ of $\mathcal{K}^{\rm II}$, we therefore find that the simplest way to explore the cutoff dependence of $\delta\Delta_{\zeta}^2$ is to consider the equal-time contribution from the last term of (\ref{eq:K3}), given that it lacks any AS terms that would regularize the integral in the ultraviolet.\par\bigskip

Let us then compute the equal-time contribution to the power spectrum coming from the last term of (\ref{eq:K3}). Recalling (\ref{eq:notcurlyg}), (\ref{eq:cutoff}) and (\ref{eq:K3sc}), we can schematically approximate the partial sum in (\ref{eq:deltazetaK}) as follows,
\begin{align}\notag
\delta\Delta_{\zeta}^2(k)\big|_{{\rm III},m^4} \;&\simeq \; 2\pi^2 (\Delta_{\zeta,0}^2)^2 \sum_{i} \left(\frac{m_i}{H}\right)^4(k\tau_i)^4 \left(\frac{G_k(\tau,\tau_i)}{\tau_i}\right)^2 \int \frac{d^3\bp}{(2\pi)^{3}k}\, |X_p(\tau_i)|^2  |X_{|\bp-\bk|}(\tau_i)|^2 \\ \label{eq:K3cont0}
&\sim\; (\Delta_{\zeta,0}^2)^2 (Hw)^{-1} \left\{ \sum_{|k\tau_i|\geq 1} \left(\frac{m_i}{H}\right)^4 |k\tau_i|^{-1} + \frac{1}{9} \sum_{|k\tau_i|<1} \left(\frac{m_i}{H}\right)^4 |k\tau_i|^{3} \right\}\,.
\end{align}
Note that we assume here that deep sub-horizon modes dominate the momentum integral, as it certainly is for a sufficiently small scatterer width $w$. Also note that we have split the sum over locations in two terms. The first adds over the time during which the $\pi$ $k$-mode is inside the horizon. If this mode remains always inside the horizon while scatterings are active, one can then disregard the second term inside the brackets. Note in passing that this result suggests a $k^{-1}$ scaling for the power spectrum correction for modes that are always sub-horizon. The second term inside the brackets adds over the times after the $k$-mode has left the horizon. If the mode is always outside the horizon while scatterings are active, a $k^3$ dependence arises for $\delta\Delta_{\zeta}^2(k)\big|_{{\rm III},m^4}$. 

We now estimate the instantaneous value of $m_i^4$ by the square of its variance, $m_i^4\sim \sigma^4$, as per (\ref{eq:mstats}). In this approximation, the evaluation of the sums in (\ref{eq:K3cont0}) is straightforward. Our present goal is just to show that each sum schematically combines with two powers of $m_i$ into the scattering parameter $\mathcal{N}_s(\sigma/H)^2$. Nevertheless, we evaluate these sums explicitly here as this will lead to a significant simplification of algebraic steps in the future (see Appendix~\ref{sec:intdom}). We approximate the locations of the scattering events by a uniform grid in cosmic time, with separation between sites $\delta t=1/H\mathcal{N}_s \ll 1/H$; this estimate is justified in Section~\ref{sec:averageps}. We denote by $i_*$ the (approximate) scattering site at which $\pi_k$ crosses the horizon, and for definiteness we assume that the mode leaves the horizon while scatterings are active, that is $|k\tau_0|\gg 1$ and $|k\tau_f|\ll 1$, where $\tau_0$ and $\tau_f$ denote the conformal times at the beginning and end of scattering, respectively (see (\ref{eq:definitions}) for a complete list of definitions). With all this in mind, the sums can then be approximated as follows,
\begin{align} \notag \displaybreak[0]
\sum_{|k\tau_i|\geq 1} \left(\frac{m_i}{H}\right)^4 |k\tau_i|^{-1} + &\frac{1}{9} \sum_{|k\tau_i|<1} \left(\frac{m_i}{H}\right)^4 |k\tau_i|^{3}\\ \notag \displaybreak[0]
&\sim\; \left(\frac{\sigma}{H}\right)^4 \left\{ |k\tau_0|^{-1}\sum_{i=0}^{i_*} \left(\frac{\tau_0}{\tau_i}\right) + \frac{|k\tau_0|^3}{9} \sum_{i=i_*}^{N_s} \left(\frac{\tau_i}{\tau_0}\right)^{3} \right\}\\ \notag \displaybreak[0]
&\simeq\; \left(\frac{\sigma}{H}\right)^4 \left\{ |k\tau_0|^{-1}\sum_{i=0}^{i_*} \left(e^{H\delta t}\right)^i + \frac{|k\tau_0|^3}{9} \sum_{i=i_*}^{N_s} \left(e^{-3H\delta t}\right)^{i} \right\}\\ \notag \displaybreak[0]
&\simeq\; \left(\frac{\sigma}{H}\right)^4 \left\{ |k\tau_0|^{-1} \left(\frac{\tau_0/\tau_*-1}{H\delta t}\right) + \frac{|k\tau_0|^3}{9} \left(\frac{(\tau_f/\tau_0)^3-(\tau_*/\tau_0)^3}{H\delta t}\right) \right\}\\ \notag \displaybreak[0]
&\simeq\; \mathcal{N}_s\left(\frac{\sigma}{H}\right)^4 \left\{  \left(1-|k\tau_0|^{-1}\right) + \frac{1}{27} \left(1-|k\tau_f|^3\right) \right\}\\ \label{eq:howtosum}
&\sim\; \mathcal{N}_s\left(\frac{\sigma}{H}\right)^4\,.
\end{align} 
Note that this result implies that the correction (\ref{eq:K3cont0}) to the power spectrum is approximately scale invariant for momenta $k_0\ll k \ll k_f$, where $k_0=|\tau_0|^{-1}$ and $k_f=|\tau_f|^{-1}$. All in all, we can write
\beq\label{eq:K3cont}
\delta \Delta_{\zeta}^2\big|_{{\rm III},m^4} \;\sim\;  \left(\mathcal{N}_s\left(\frac{\sigma}{H}\right)^2\right)^2\frac{(\Delta_{\zeta,0}^2)^2}{\mathcal{N}_sHw} \,.
\eeq
From the previous expression one can immediately read the $\mathcal{N}_s^{-1}$ suppression. Note in passing the quadratic dependence of $\delta\Delta_{\zeta}^2$ on $\S$, a hallmark of sub-horizon mode contributions. Moreover, as mentioned earlier, this contribution is scale-invariant. 

Comparing this result to that corresponding to the $\mathcal{K}^{\rm I}$ contribution in the $k_0<k< k_f$ regime, shown in Eq.~(\ref{eq:K1cont}), we find that
\beq\label{eq:K3K1rat}
\frac{\delta \Delta_{\zeta}^2\big|_{{\rm III},m^4}}{\delta \Delta_{\zeta}^2\big|_{{\rm I}}} \;\sim\; \mathcal{N}_s\left(\frac{\sigma}{H}\right)^2 \frac{e^{-|\gamma| N_{\rm tot}}}{\mathcal{N}_sHw} \left(\frac{k}{k_0}\right)^{\gamma}\,.
\eeq
From this expression one is tempted to read the $\mathcal{N}_s^{-1}$ suppression as a justification to disregard this cutoff-dependent correction in the limit of a large scatterer density. However, the width of the scatterers and their density in time are not unrelated. Indeed, the constraint
\beq\label{eq:Nsconst}
\mathcal{N}_s (Hw)<1\,,
\eeq
must be satisfied (separation between scatterers $>$ width of scatterers). It is in fact the saturation regime $\mathcal{N}_s (Hw)\sim 1$ with $\mathcal{N}_s\gg 1$ to what we refer as the large $\mathcal{N}_s$ limit. Note nevertheless that, at $k\sim k_0$, the $\mathcal{K}^{\rm I}$ contribution will not be observable unless $\Delta_{\zeta,0}^2\S e^{|\gamma| N_{\rm tot}} \gtrsim 1$. If this is the case,
\beq
\frac{\delta \Delta_{\zeta}^2\big|_{{\rm III},m^4}}{\delta \Delta_{\zeta}^2\big|_{{\rm I}}} \;\lesssim\; \left( \mathcal{N}_s\left(\frac{\sigma}{H}\right)^2 \right)^2 \Delta_{\zeta,0}^2\,,
\eeq
which is $\ll 1$ for any value of the scattering parameter that avoids the backreaction constraint for $N_{\rm tot}\gtrsim 1$ (see Section~\ref{sec:backreact}). We therefore conclude that the cutoff-dependent $m^4$ contribution to the power spectrum correction can be disregarded in the large $\mathcal{N}_s$ limit. \par\bigskip

The previous result applies only to the last term in (\ref{eq:K3}). An analogous argument based on momentum power-counting can be constructed for the cutoff-dependent (sub-horizon) component of the entire $\mathcal{K}^{\rm II}$ and $\mathcal{K}^{\rm III}$ terms in (\ref{eq:deltazetaK}). However, one must also take into account the cutoff-independent, super-horizon contribution to these two terms, which can in principle be as important as that given by $\mathcal{K}^{\rm I}$. Instead of a lengthly, semi-quantitative argument, we present a brief numerical exploration of our findings.

\begin{figure}[!t]
\centering
    \includegraphics[width= \textwidth]{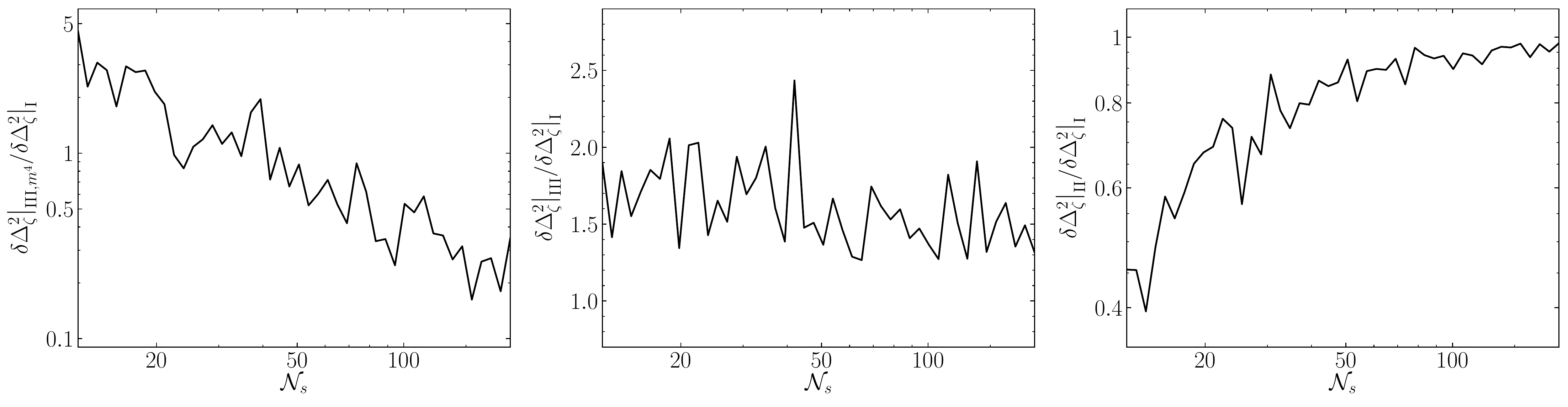}
    \caption{Ratios of various contributions to the total power spectrum correction (\ref{eq:deltazetaK}) relative to the $\mathcal{K}^{\rm I}$-dependent fraction (\ref{eq:K1}) as functions of $\mathcal{N}_s$ for $Hw=10^{-3}$, $k=k_0$, $N_{\rm tot}=20$ and $\S=25$. Left: the relative correction for the $m^4$-dependent term of $\mathcal{K}^{\rm III}$ (\ref{eq:K1}). Center: the relative correction for the full $\mathcal{K}^{\rm III}$ contribution. Right: the relative correction for the $\mathcal{K}^{\rm II}$ contribution (\ref{eq:K2}). Each curve corresponds to the geometric mean of 50 unique realizations of the disorder. For the numerical method used see Appendix~\ref{app:numerics}.}
    \label{fig:conv_Ns}
\end{figure} 
Fig.~\ref{fig:conv_Ns} shows the geometric mean of the ratio (\ref{eq:K3K1rat})  and its corresponding counterparts for the full $\mathcal{K}^{\rm II}$ and $\mathcal{K}^{\rm III}$ contributions, computed as a function of $\mathcal{N}_s$ for 50 realizations of the effective mass $m^2(t)$, with $k=k_0$, $N_{\rm tot}=20$, $\S=25$ and $Hw=10^{-3}$. The left panel clearly depicts the inverse dependence of (\ref{eq:K3K1rat}) with $\mathcal{N}_s$. The center and right panels of demonstrate that this behavior is not shared by the full $\mathcal{K}^{\rm II}$ and $\mathcal{K}^{\rm III}$ terms, suggesting that the cutoff dependence is irrelevant at large $\mathcal{N}_s$. Moreover, both contributions can be as large or larger than that given by $\mathcal{K}^{\rm I}$, although the results seemingly confirm our claim made in Section~\ref{sec:ddscatterers} that they can be regarded as $\mathcal{O}(1)$ corrections to the $\mathcal{K}^{\rm I}$ result.\par\bigskip

\begin{figure}[!t]
\centering
    \includegraphics[width= \textwidth]{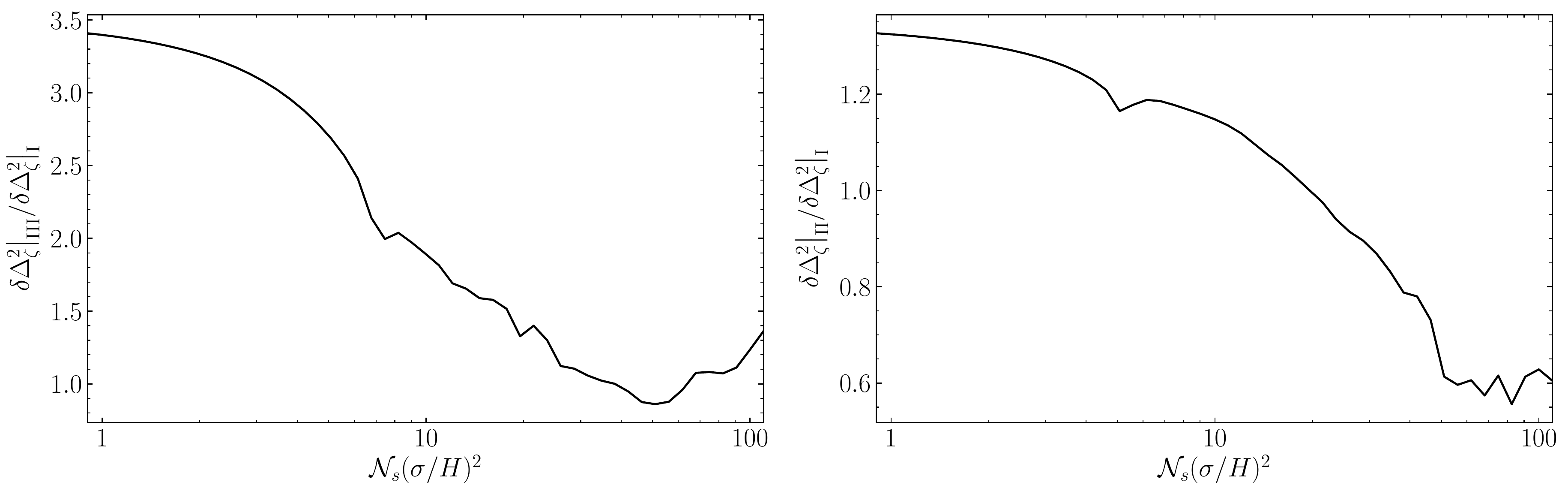}
    \caption{Ratios of various contributions to the total power spectrum correction (\ref{eq:deltazetaK}) relative to the $\mathcal{K}^{\rm I}$-dependent fraction (\ref{eq:K1}) as functions of $\mathcal{N}_s(\sigma/H)^2$ for $Hw=10^{-3}$, $k=k_0$, $N_{\rm tot}=20$ and $\mathcal{N}_s=50$. Left: the relative correction for the full $\mathcal{K}^{\rm III}$ contribution  (\ref{eq:K1}). Right: the relative correction for the $\mathcal{K}^{\rm II}$ contribution (\ref{eq:K2}). Each curve corresponds to the geometric mean of 50 unique realizations of the disorder. For the numerical method used see Appendix~\ref{app:numerics}.}
    \label{fig:conv_Nss2}
\end{figure}

For both $\mathcal{K}^{\rm II}$ and $\mathcal{K}^{\rm III}$ the integrand contains conformal time derivatives of the $X$ Fourier modes. We therefore expect that, in the regime of exponential excitation of the spectator field, the contribution from small momentum, super-horizon modes to the integral will be suppressed relative to that in $\mathcal{K}^{\rm I}$. The result of our numerical exploration in this regime is presented in Fig.~\ref{fig:conv_Nss2}, which shows the ratios of the power spectrum components as a function of scattering strength, with $Hw=10^{-3}$, $k=k_0$, $N_{\rm tot}=20$ and $\mathcal{N}_s=50$. The left panel corresponds to the ratio of the $\mathcal{K}^{\rm III}$ term of (\ref{eq:deltazetaK}) to the $\mathcal{K}^{\rm I}$ term. Note that at  the lower end of the considered range of scattering strengths, the ``III'' contribution dominates over the ``I'' one, but is nevertheless only enhanced by an $\mathcal{O}(1)$ factor. As $\S$ is increased, the ratio decreases, until it becomes $\lesssim 1$ for $\S\simeq 40$. For $\S\gtrsim 60$, the decreasing trend is reverted, and the ratio increases, albeit not significantly for scattering strengths $\lesssim 100$. Larger values of $\S$ typically lead to the breakdown of the perturbative expansion (\ref{eq:spi}) (see Section~\ref{sec:backreact}). 

The right panel of Fig.~\ref{fig:conv_Nss2} shows the ratio of the $\mathcal{K}^{\rm II}$ term to $\mathcal{K}^{\rm I}$ as a function of $\S$. Note that in this case the decreasing trend is also present, and there is no growth at large $\S$. For all scattering strengths considered, the result implies that the ``II'' component of $\delta\Delta_{\zeta}^2$ is at most 1.3 times larger than the ``I'' component, and it is smaller than it for $\S\gtrsim 20$. We therefore conclude that (\ref{eq:deltaDeltafinal}) is indeed an adequate approximation to $\delta\Delta_{\zeta}^2$ at large $\mathcal{N}_s$.

%


\section{Numerical Method}\label{app:numerics}

In this appendix we discuss the approximations that we have made in order to evaluate numerically the stochastic component of the power spectrum (\ref{eq:deltaDeltafinal}) in Section~\ref{sec:numres}. We first note that the validity of our results relies on the assumption that the number of scatterers per Hubble time is large, $\mathcal{N}_s\gg 1$. For even a modest amount of expansion, $N_{\rm tot}=20$, this requires thousands of numerical operations, first to compute the excited spectator momentum modes, second to numerically evaluate the momentum integral, and third to compute the sum over scatterers. This complexity, coupled with the exponentially increasing or decreasing factors in  (\ref{eq:deltaDeltafinal}) requires a numerical code capable of handling the extremely high precision that is required. To achieve this, we have built our (Fortran) code making use of the thread-safe arbitrary precision package MPFUN-For~\cite{mpfun}. We have confirmed that our precision of choice (500 digits) is sufficient to ensure that the Wronskian condition $X_k(\tau)X_k^{*\prime}(\tau) - X_k'(\tau)X_k^{*}(\tau)=i$ is satisfied at all times.

The need for high precision translates into long code run times and significant memory usage. For $\mathcal{N}_s\sim 25$, the evaluation of (\ref{eq:deltaDeltafinal}) for a single value of $k$ requires measuring run times in units of core-days, even when we use a simple trapezoidal estimator for the momentum integral. Therefore, in order to explore the dependence of the power spectrum correction on the wavenumber of the Goldstone mode, the duration of scatterings, and $\S$ we are forced to rely on a few simplifications. For the first we notice that only the diagonal $i=j$ terms in the sum are positive definite. Off-diagonal contributions alternate signs stochastically, and for $\mathcal{N}_s\gg1$ we expect them to approximately cancel each other due to the non-correlation of the scatterer amplitudes $m_i$ (see (\ref{eq:mstats})). We therefore assume that the equal-time approximation (\ref{eq:K1cont0}) encodes the bulk of the enhancement of $\Delta_{\zeta}^2$. 

The second approximation that we use is more subtle. For small wavenumbers $k\leq (k_0 k_f)^{1/2}$, the momentum integral in (\ref{eq:K1cont0}) is dominated by modes that leave the horizon before or during the time that scatterings are active, for which $p,q\lesssim |\tau_i|^{-1}$ with $|\tau_i|^{-1}\gg k$. We can therefore approximate the integral by its value with $p=q$. For $k\geq (k_0 k_f)^{1/2}$ the integral will instead be dominated by super-horizon $X$-modes with $p,q\ll k$. In this case, we note that the integral may be approximated by its value with $q=p+k$ and $q=p-k$, which are at the edges of the integration domain (see Fig.~\ref{fig:Integrations} and Figs.~\ref{fig:regions_irx1} and \ref{fig:regions_irx2}). In order to work with a uniform grid, we consider the rotated variables $x$ and $y$,
\beq
\begin{aligned}
p \;&=\; k\left(\frac{x-y}{\sqrt{2}} + \frac{1}{2}\right)\,,\\
q \;&=\; k\left(\frac{x+y}{\sqrt{2}} + \frac{1}{2}\right)\,,
\end{aligned}
\eeq
and define the rescaled mode function $Y_{p}(\tau)\equiv \sqrt{2p}X_p(\tau)$. In terms of these, our numerical approximation to the power spectrum correction can be written as follows,
\begin{align} \notag \displaybreak[0]
 \delta\Delta_{\zeta}^2(k)  \;&=\; (\Delta_{\zeta,0}^2)^2 \sum_{i,j}^{N_s}  \frac{m_i m_j}{H^2}(k\tau_i)^2(k\tau_j)^2 \mathcal{G}_k(\tau,\tau_i) \mathcal{G}_k(\tau,\tau_j)k^{-2}  \\ \notag \displaybreak[0]
&\hspace{70pt}\times \int_{0}^{\infty}dp\, p\int_{|p-k|}^{p+k}dq\,q\, \left[  X_p(\tau_i)X_p^*(\tau_j) \right]_{\rm AS}  \big[  X_{q}(\tau_i)X_{q}^*(\tau_j) \big]_{\rm AS}\\ \notag \displaybreak[0]
&=\; \frac{1}{4}(\Delta_{\zeta,0}^2)^2 \sum_{i,j}^{N_s}  \frac{m_i m_j}{H^2}(k\tau_i)^2(k\tau_j)^2 \mathcal{G}_k(\tau,\tau_i) \mathcal{G}_k(\tau,\tau_j)  \\ \notag \displaybreak[0]
&\hspace{70pt}\times \int_{0}^{\infty}dx \int_{-\frac{1}{\sqrt{2}}}^{ \frac{1}{\sqrt{2}} }dy\,   \left[  Y_p(\tau_i)Y_p^*(\tau_j) \right]_{\rm AS}  \big[  Y_{q}(\tau_i)Y_{q}^*(\tau_j) \big]_{\rm AS}\\ \notag \displaybreak[0]
&=\; \frac{1}{4}(\Delta_{\zeta,0}^2)^2 \sum_{i,j}^{N_s}  \frac{m_i m_j}{H^2}(k\tau_i)^2(k\tau_j)^2 \mathcal{G}_k(\tau,\tau_i) \mathcal{G}_k(\tau,\tau_j)  \\ \notag \displaybreak[0]
&\hspace{70pt}\times \int_{x_{\rm super}}^{x_{\rm sub}} dx \int_{-\frac{1}{\sqrt{2}}}^{ \frac{1}{\sqrt{2}} }dy\,   \left[  Y_p(\tau_i)Y_p^*(\tau_j) \right]_{\rm AS}  \big[  Y_{q}(\tau_i)Y_{q}^*(\tau_j) \big]_{\rm AS}\\ \notag \displaybreak[0]
&=\; \frac{1}{4}(\Delta_{\zeta,0}^2)^2 \sum_{i,j}^{N_s}  \frac{m_i m_j}{H^2}(k\tau_i)^2(k\tau_j)^2 \mathcal{G}_k(\tau,\tau_i) \mathcal{G}_k(\tau,\tau_j)  \\  \displaybreak[0] \label{eq:discretefull}
&\hspace{70pt}\times \sum_{l=0}^{N_x} \sum_{m=0}^{N_y}   w_{l,m} x_l   \left[  Y_{p_{l,m}}(\tau_i)Y_{p_{l,m}}^*(\tau_j) \right]_{\rm AS}  \big[  Y_{q_{l,m}}(\tau_i)Y_{q_{l,m}}^*(\tau_j) \big]_{\rm AS} \Delta_{\ln(x)} \Delta_y\\  \displaybreak[0] \label{eq:discretepart}
&\approx\; \frac{1}{4}(\Delta_{\zeta,0}^2)^2 \sum_{i}^{N_s}  \frac{m_i^2}{H^2}(k\tau_i)^4 \mathcal{G}_k^2(\tau,\tau_i)  \sum_{j=0}^{N_x} \sum_{l=0}^{N_y}   w_{j,l} x_l |Y_{p_{j,l}}(\tau_i)|^2_{\rm AS}  |Y_{q_{j,l}}(\tau_i)|^2_{\rm AS} \Delta_{\ln(x)} \Delta_y\,.
\end{align} 
In the third line we have limited the integration over $x$ to the domain $(x_{\rm super},x_{\rm sub})$, where $x_{\rm super}$ ($x_{\rm sub}$) denotes a momentum scale sufficiently deep inside (outside) the horizon. In the third line we have discretized the integral over a uniform grid with $(N_x+1)\times (N_y+1)$ points in $\ln(x)$ and $y$. Therein $w_{j,l}$ denote the weights of the integration routine, in this case the composite trapezoidal rule. We note then that integration over the lines $p=q$, $q=p+k$ and $q=p-k$ is equivalent to taking the number of nodes along the $y$-direction $N_y=3$. Note that in any case, the equation of motion for the spectator field must be solved for a total of $N_s$ times for $(N_x+1)\times (N_y+1)$ different momenta.\par\bigskip
\begin{figure}[!t]
\centering
    \includegraphics[width= 0.85\textwidth]{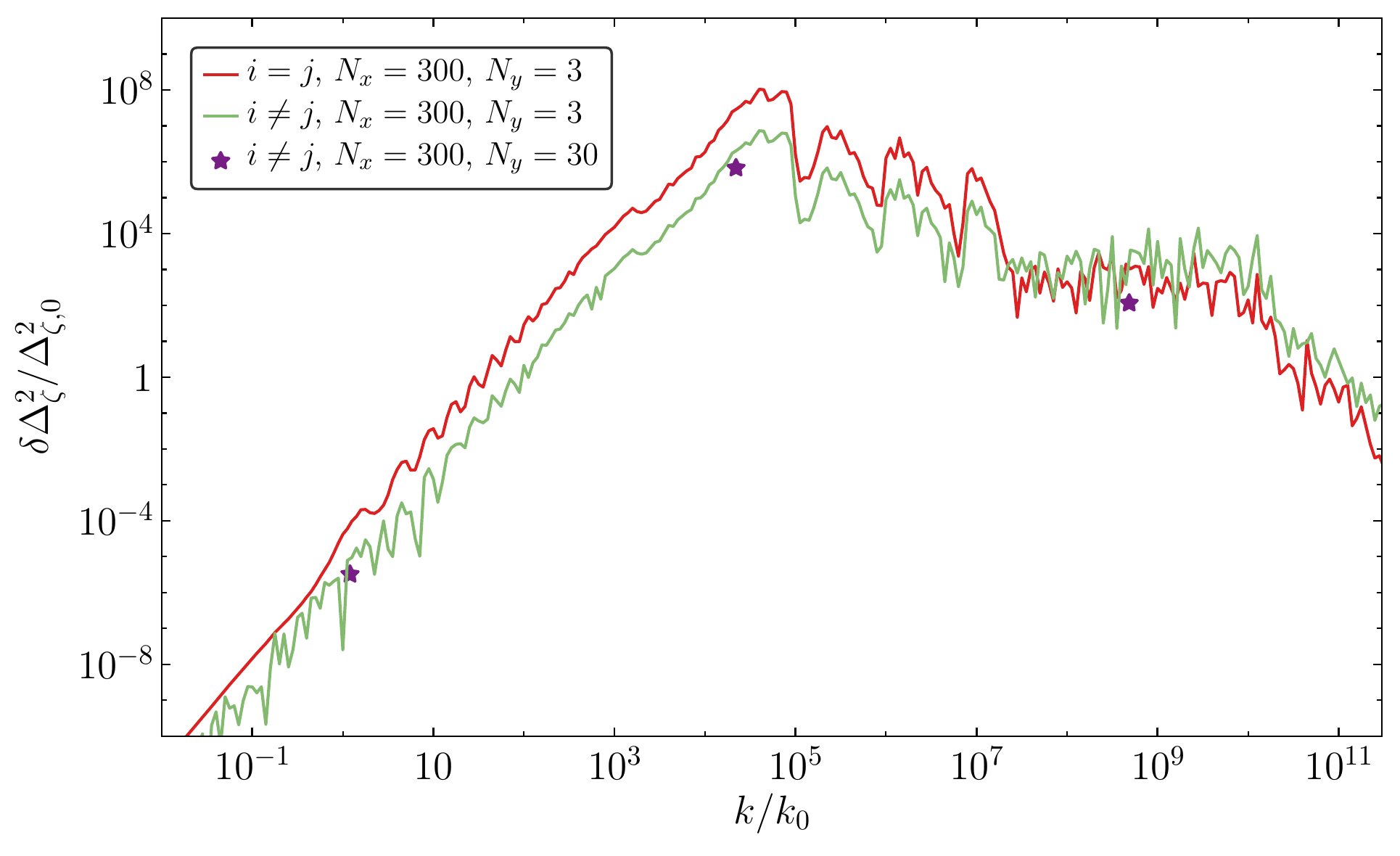}
    \caption{Relative correction to the curvature power spectrum for a unique realization of the disorder, with $\S=25$, $N_{\rm tot}=20$ and $N_s=980$. The purple stars correspond to the full approximation (\ref{eq:discretefull}) with $N_x=300$ and $N_y=30$. The green curve corresponds also to the full approximation (\ref{eq:discretefull}) but with $N_y=3$. The red curve shows the result of the equal time approximation (\ref{eq:discretepart}) with $N_x=300$ and $N_y=3$.}
    \label{fig:25_ij_pq}
\end{figure}
    
Figure~\ref{fig:25_ij_pq} compares three different approximations to the power spectrum correction for a unique realization of the disorder, with $\S=25$. This realization is the same as the largest outlier in Fig.~\ref{fig:25_20}. The purple stars correspond to the ``full'' approximation given by (\ref{eq:discretefull}) for $k=k_0,(k_0k_f)^{1/2}$ and $k_f$. Here the number of scatterers $N_s\approx 1000$ and $N_x=300$, $N_y=30$ in momenta. In this case it is difficult to increase the density of points in the plot due to the large amount of computational resources needed. 

The continuous green curve in Fig.~\ref{fig:25_ij_pq} corresponds also to (\ref{eq:discretefull}), but in this case we have made use of the simplifying approximation $N_y=3$. As it can be appreciated in the figure, no significant difference in values with respect to the previous case can be appreciated, which justifies our assumptions. Although for this scenario we have been able to present a full power spectrum shape, the computation is still very time consuming; each point in the curve requires $\mathcal{O}(1)$  days to be evaluated. Given that our hope is to characterize the value of $\delta\Delta_{\zeta}^2$ for different realizations, number of scatterers and scattering strengths, we are forced to further simplify the calculation.

The final approximation that we consider, which we have used extensively in Section~\ref{sec:numres} and Appendix~\ref{app:cutoff2}, is given by (\ref{eq:discretepart}) and is shown as the red curve in Fig.~\ref{fig:25_ij_pq}. In this case, we make use of the $N_y=3$ approximation and the equal time approximation, $i=j$. When this is the case, a few hours are sufficient to obtain the whole power spectrum correction. The shape of this red curve is very similar to the green curve discussed above, which we interpret again as a justification for our estimates. However, note that the separation between these curves can be as large as an order of magnitude for the largest values of the correction. Nevertheless, given that the spread between realizations is significantly more pronounced that this difference (see Fig.~\ref{fig:25_20}), we consider that these approximations are adequate to statistically address the parametric dependence of $\delta\Delta_{\zeta}^2$ on $N_{\rm tot}$ and $\S$.

\section{Computation of the Mean Power Spectrum}\label{sec:intdom}

In this Appendix we provide the full derivation of the mean power spectrum $\langle \delta\Delta_{\zeta}^2\rangle$, starting from Eq.~(\ref{eq:deltadeltaav}). To allow for a simpler reading of the procedure we have divided the computation in three, starting with the evaluation of the ensemble expectation values, to continue with the integration over momentum and finishing with the summation over scatterers.

\subsection{Brownian Ensemble Averages}\label{sec:brownian}

In order to evaluate the momentum integral in (\ref{eq:deltadeltaav}), it is convenient first to calculate the expectation values of the required powers of $|X_p(\tau)|^2$. This task can be readily completed by invoking the geometric (Brownian) random walk nature of the spectator field. For two momenta $p$ and $q$, the $n$-point function (\ref{eq:Xnpoint}) reduces to
\Beq
\langle |X_p(\tau)|^2|X_{q}(\tau)|^2\rangle \;=\; e^{\left\langle\ln|X_p|^2\right\rangle}e^{\langle\ln|X_q|^2\rangle}e^{\frac{1}{2}\left(\langle Z_p^2\rangle+\langle Z_q^2\rangle+2\langle Z_pZ_q\rangle\right)}\,.
\Eeq
Note that here we have suppressed the $m,\tau$ subindexes in the expectation value for simplicity. Without loss of generality, let us assume that $p>q$. Recall that we also assume that scatterings start when $\tau=\tau_0$. If we further assume that $X_p$ leaves the horizon before the beginning of scatterings, then $X_q$ also does. The mean and the variance of both modes grow immediately when $\tau=\tau_0$ and we can therefore write
\begin{align}\notag
\langle |X_p(\tau)|^2|X_{q}(\tau)|^2\rangle  \; &\simeq\; \left|X_p^0\left(\tau_0\right)\right|^2 \left|X_q^0\left(\tau_0\right)\right|^2 e^{2\mu_1 H(t-t_0)} e^{2\mu_2H(t-t_0)}\\ 
&=\; \frac{1}{4pq}\left(\frac{\tau_0}{\tau}\right)^{2\mu_1+2\mu_2}\,,
\end{align}
and
\beq
\langle |X_p(\tau)|^2_{\rm AS}|X_{q}(\tau)|^2_{\rm AS}\rangle  \; \simeq\; \frac{1}{4pq}\left[|k_0\tau|^{-2\mu_1-2\mu_2} -2 |k_0\tau|^{-\mu_1-\frac{1}{2}\mu_2} +1\right]\,.
\eeq

If, instead, $X_p$ leaves the horizon while scatterings are active, $|p\tau_0|>1$, but $|q\tau_0|<1$, then, at a given time we can have either $|p\tau|<1$ or $|p\tau|>1$. Evaluating the former case first, we obtain 
\begin{align} \notag
\langle |X_p(\tau)|^2|X_{q}(\tau)|^2\rangle  \; &\simeq\; \left|X_p^0\left(\tau_p\right)\right|^2 \left|X_q^0\left(\tau_0\right)\right|^2e^{\mu_1 H(t-t_p)} e^{\mu_1 H(t-t_0)} e^{\frac{3}{2}\mu_2H(t-t_p)} e^{\frac{1}{2}\mu_2H(t-t_0)}\\
&=\; \frac{1}{4pq}\left(\frac{\tau_0}{\tau}\right)^{\mu_1+\frac{1}{2} \mu_2} \left(\frac{\tau_p}{\tau}\right)^{\mu_1+\frac{3}{2} \mu_2}\,,
\end{align}
and 
\beq \label{eq:2pc2}
\langle |X_p(\tau)|_{\rm AS}^2|X_{q}(\tau)|^2_{\rm AS}\rangle  \; \simeq\; \frac{1}{4pq} \left[  |k_0\tau|^{-\mu_1-\frac{1}{2} \mu_2} |p\tau|^{-\mu_1-\frac{3}{2} \mu_2} - |k_0\tau|^{-\mu_1-\frac{1}{2} \mu_2} - |p\tau|^{-\mu_1-\frac{3}{2} \mu_2} + 1 \right]\,.
\eeq
If instead $|p\tau|>1$, the result will be suppressed by AS, given that $X_p$ will be near its vacuum state. Nevertheless, it is possible to estimate the result, recalling that correlations between sub- and super-horizon modes are negligible. Using (\ref{eq:lnchisubh}), this allows us to factor the two-point function as follows,
\begin{align} \notag
\langle |X_p(\tau)|^2_{\rm AS}|X_{q}(\tau)|^2_{\rm AS}\rangle  \; &\simeq\; \left(\langle |X_p(\tau)|^2\rangle - |X_p^0(\tau)|^2 \right)\left(\langle |X_q(\tau)|^2\rangle - |X_q^0(\tau)|^2 \right)\\ \notag
&\simeq\; \frac{1}{4pq}\left( e^{\frac{1}{8}\mathcal{N}_s(\sigma/H)^2 |p\tau|^{-2}}-1 \right)\left( e^{\mu_1 H(t-t_0)} e^{\frac{1}{2}\mu_2 H(t-t_0)}  -1 \right)\\ \label{eq:cD}
&\simeq\; \frac{1}{32pq} \mathcal{N}_s(\sigma/H)^2 |p\tau|^{-2} \left( |k_0\tau|^{-\mu_1-\frac{1}{2} \mu_2}  -1 \right)\,.
\end{align}

 Finally, if $X_p$ and $X_q$ leave the horizon during scatterings, for only those times for which $|p\tau|,|q\tau|<1$ the AS will not result in a large suppression. It is clear that we can recover the corresponding expression for the two-point function via the replacement $k_0\rightarrow q$ in (\ref{eq:2pc2}). A similar argument applies for the case where $|p\tau|>1$ and $|q\tau|<1$ with (\ref{eq:cD}). From Eq.~(\ref{eq:cD}) one can also immediately deduce the result for the doubly AS-suppressed case $|p\tau|,|q\tau|>1$. We therefore summarize our results for the Brownian ensemble averages for $p>q$ in the following way,
 \begin{align}\notag
 \langle |X_p(\tau)|_{\rm AS}^2& |X_{q}(\tau)|^2_{\rm AS}\rangle \\ \label{eq:XpXq}
 &\simeq\; \frac{1}{4pq}\times
 \begin{cases}
|k_0\tau|^{-\alpha-\beta} - 2|k_0\tau|^{-\beta} +1 \,, & 1>|p\tau_0| \hspace{102.5pt}\textrm{(A)}\\[5pt]
|p\tau|^{-\alpha}|k_0\tau|^{-\beta}- |p\tau|^{-\beta} - |k_0\tau|^{-\beta} + 1 \,, & |p\tau_0|>1>|p\tau|,|q\tau_0|\hspace{44.5pt}  \textrm{(B)}\\[5pt]
|p\tau|^{-\alpha}|q\tau|^{-\beta}- |p\tau|^{-\beta} - |q\tau|^{-\beta} + 1\,, & |p\tau_0|,|q\tau_0|>1>|p\tau|,|q\tau|\qquad  \textrm{(C)}\\[7pt]
\dfrac{1}{8}\mathcal{N}_s\left(\dfrac{\sigma}{H}\right)^2|p\tau|^{-2}\left( |k_0\tau|^{-\beta}  -1 \right)\,, & |p\tau|>1>|q\tau_0|\hspace{70.5pt}  \textrm{(D)}\\[7pt]
\dfrac{1}{8}\mathcal{N}_s\left(\dfrac{\sigma}{H}\right)^2|p\tau|^{-2}\left( |q \tau|^{-\beta}  -1 \right)\,, & |p\tau|,|q\tau_0|>1>|q\tau|  \hspace{48pt} \textrm{(E)}\\[7pt]
\dfrac{1}{64}\mathcal{N}_s^2\left(\dfrac{\sigma}{H}\right)^4 |p\tau|^{-2} |q\tau|^{-2} \,, & |p\tau|,|q\tau|>1 \hspace{84pt}\textrm{(F)}\,.
 \end{cases}
 \end{align}
We have labeled each case with a letter for future convenience. The exponents $\alpha$ and $\beta$ were defined in (\ref{eq:alphabeta}). Note that if $q>p$, the corresponding two-point function can be obtained by taking $p\leftrightarrow q$ in the previous expression.

\subsection{The Momentum Integral}\label{sec:momint}

We now make use of (\ref{eq:XpXq}) to evaluate the momentum integral in (\ref{eq:deltadeltaav}). We first note that the integral can be rewritten in the following way,
\begin{align}\notag
 \int \frac{d^3\bp}{(2\pi)^{3}k}\, \left\langle |X_p(\tau_i)|^2_{\rm AS} |X_{|\bp-\bk|}(\tau_i)|^2_{\rm AS}  \right\rangle \;&=\;  \frac{1}{(2\pi)^2k^2}\int_0^{\infty} dp\, p \int_{|p-k|}^{p+k} dq\, q\, \left\langle |X_p(\tau_i)|^2_{\rm AS} |X_{q}(\tau_i)|^2_{\rm AS}  \right\rangle\\ \notag
 &=\;\frac{1}{(2\pi)^2k^2} \Big[\int_{k/2}^{\infty} dp\, p \int_{|p-k|}^p dq\, q \, \left\langle |X_p(\tau_i)|^2_{\rm AS} |X_{q}(\tau_i)|^2_{\rm AS}  \right\rangle \\ \notag
&\hspace{60pt}  +\int_{k/2}^{\infty} dq\, q \int_{|q-k|}^q dp\, p \,\left\langle |X_p(\tau_i)|^2_{\rm AS} |X_{q}(\tau_i)|^2_{\rm AS}  \right\rangle \Big]\\ \label{eq:pqint}
&=\; \frac{1}{2\pi^2k^2}\int_{k/2}^{\infty} dp \, p \int_{|p-k|}^p dq \, q\,\left\langle |X_p(\tau_i)|^2_{\rm AS} |X_{q}(\tau_i)|^2_{\rm AS}  \right\rangle\,.
\end{align}
In arriving to the last line we have made use of the fact that the two-point function has the same functional form for $q>p$ as it has for $p>q$ with the change $p\leftrightarrow q$. Therefore, it suffices to consider the $p>q$ case shown in (\ref{eq:XpXq}).\par\bigskip

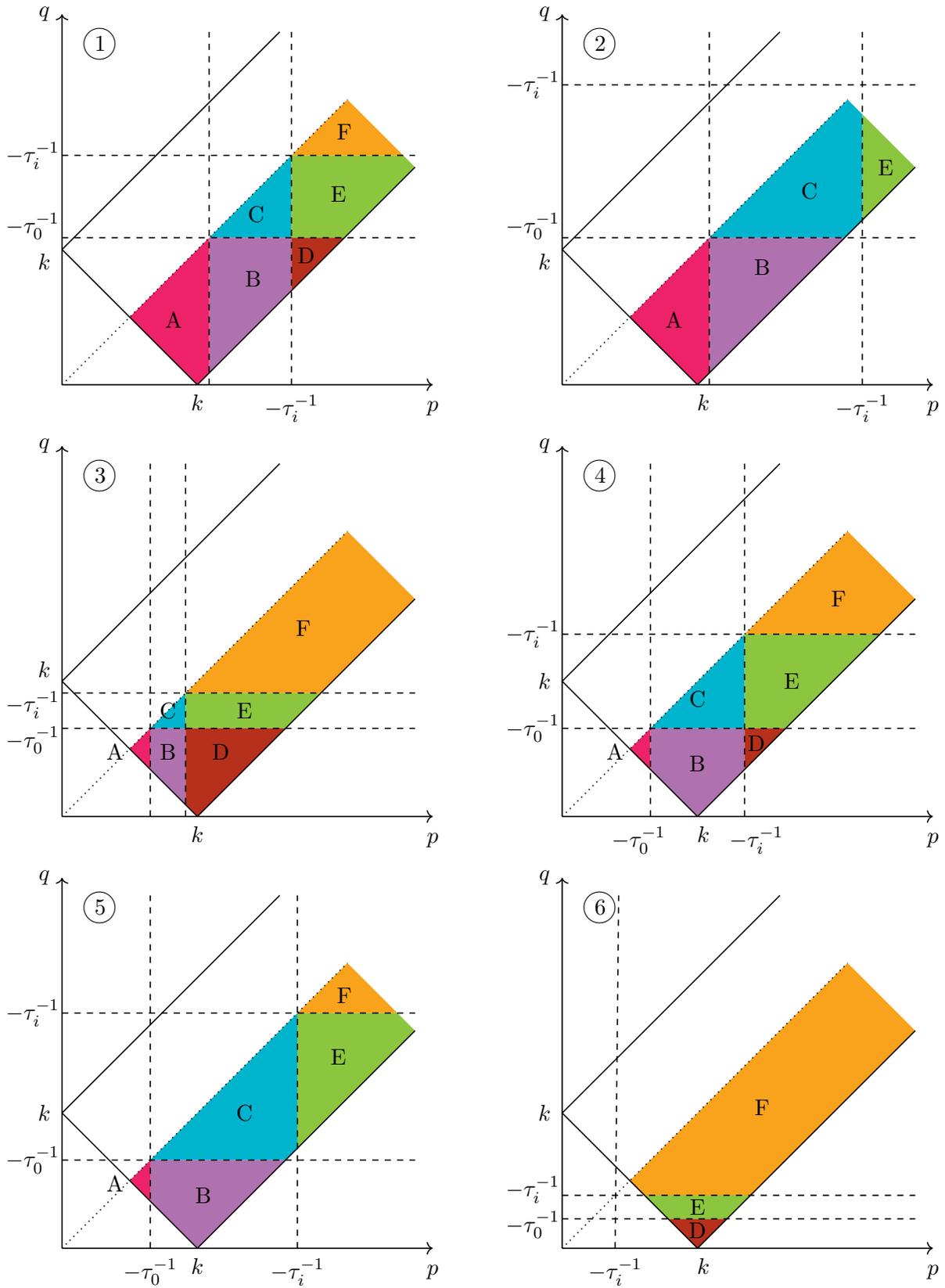
\begin{figure}[!t]
\begin{tikzpicture}%
    		\fill[Turquoise] (2.5,2.5) - - (3.9,2.5) - - (3.9,3.9) - - cycle;
    		\fill[Orchid] (2.5,2.5) - - (3.9,2.5) - - (3.9,1.6) - - (2.5,0.2)- - cycle;
    		\fill[LimeGreen](3.9,3.9) - - (5.8,3.9) - - (6,3.7) - - (4.8,2.5) - - (3.9,2.5) - - cycle;
    		\fill[YellowOrange](3.9,3.9) - - (5.8,3.9) - - (4.85,4.85) - - cycle;
		\fill[BrickRed] (3.9,2.5) - - (3.9,1.6) - - (4.8,2.5) - - cycle;
		\fill[WildStrawberry] (1.15,1.15) - - (2.5,2.5) - - (2.5,0.2) - - (2.3,0) - - cycle;
    		\draw[->, line width=0.6pt] (0,0) -- (6.3,0);%
    		\draw[->, line width=0.6pt] (0,0) -- (0,6.3);%
    		\draw[line width=0.6pt] (0,2.3) -- (2.3,0);%
    		\draw[line width=0.6pt] (0,2.3) -- (3.7,6);%
    		\draw[line width=0.6pt] (2.3,0) -- (6,3.7);%
    		\draw[dotted,line width=0.6pt] (0,0) -- (4.85,4.85);%
    		\draw[dashed,line width=0.6pt] (3.9,0) -- (3.9,6);%
    		\draw[dashed,line width=0.6pt] (0,3.9) -- (6,3.9);%
    		\draw[dashed,line width=0.6pt] (0,2.5) -- (6,2.5);%
		\draw[dashed,line width=0.6pt] (2.5,0) -- (2.5,6);%
		\node at (-0.5,2.7) {$-\tau_0^{-1}$};
    		\node at (4.15,2.2) {D};
    		\node at (2.3,-0.3) {$k$};
    		\node at (-0.3,2.1) {$k$};
    		\node at (6.3,-0.4) {$p$};
    		\node at (-0.3,6.3) {$q$};
    		\node at (3.9,-0.4) {$-\tau_i^{-1}$};
    		\node at (-0.5,3.9) {$-\tau_i^{-1}$};
    		\node at (3.3,2.9) {C};
    		\node at (4.7,3.25) {E};
    		\node at (4.8,4.3) {F};
		\node at (3.25,1.8) {B};
		\node at (1.9,1.1) {A};
		\node at (0.9,5.8) {\circle{15}};
		\node at (0.65,5.8) {1};
    		%
    		%
    		%
    		%
    		%
    		\fill[WildStrawberry] (9.65,1.15) - - (10.8,0) - - (11,0.2) - - (11,2.5) - - cycle;
    		\fill[LimeGreen](13.6,2.8) - - (13.6,4.6) - - (14.5,3.7) - - cycle;
    		\fill[Orchid] (11,2.5) - - (13.3,2.5) - - (11,0.2) - - cycle;
    		\fill[Turquoise](11,2.5) - - (13.3,2.5) - - (13.6,2.8) - - (13.6,4.6) - - (13.35,4.85) - - cycle;
    		\draw[->, line width=0.6pt] (8.5,0) -- (14.8,0);
    		\draw[->, line width=0.6pt] (8.5,0) -- (8.5,6.3);%
    		\draw[line width=0.6pt] (8.5,2.3) -- (10.8,0);%
    		\draw[line width=0.6pt] (8.5,2.3) -- (12.2,6);%
    		\draw[line width=0.6pt] (10.8,0) -- (14.5,3.7);%
    		\draw[dotted,line width=0.6pt] (8.5,0) -- (13.35,4.85);%
    		\draw[dashed,line width=0.6pt] (13.6,0) -- (13.6,6);%
    		\draw[dashed,line width=0.6pt] (8.5,5.1) -- (14.5,5.1);%
    		\draw[dashed,line width=0.6pt] (8.5,2.5) -- (14.5,2.5);%
		\draw[dashed,line width=0.6pt] (11,0) -- (11,6);%
    		\node at (8,2.7) {$-\tau_0^{-1}$};
    		%
    		\node at (10.9,-0.3) {$k$};
    		\node at (8.2,2.1) {$k$};
    		\node at (14.8,-0.4) {$p$};
    		\node at (8.2,6.3) {$q$};
    		\node at (13.6,-0.4) {$-\tau_i^{-1}$};
    		\node at (8,5.1) {$-\tau_i^{-1}$};
    		\node at (10.4,1.1) {A};
    		\node at (11.9,2.0) {B};
    		\node at (12.7,3.3) {C};
		
    		\node at (14.0,3.7) {E};
		\node at (9.4,5.8) {\circle{15}};
		\node at (9.15,5.8) {2};
    		%
    		%
  \end{tikzpicture}
  \begin{tikzpicture}%
    		\fill[Turquoise] (1.5,1.5) - - (2.1,2.1) - - (2.1,1.5) - - cycle;
    		\fill[Orchid] (1.5,1.5) - - (2.1,1.5) - - (2.1,0.2) - - (1.5,0.8)- - cycle;
    		\fill[LimeGreen](2.1,2.1) - - (4.4,2.1) - - (3.8,1.5) - - (2.1,1.5) - - cycle;
    		\fill[YellowOrange](2.1,2.1) - - (4.85,4.85)  - - (6,3.7) - - (4.4,2.1) - - cycle;
		\fill[BrickRed] (2.1,1.5) - - (3.8,1.5) - - (2.3,0) - - (2.1,0.2) - - cycle;
		\fill[WildStrawberry] (1.15,1.15) - - (1.5,1.5) - - (1.5,0.8) - - cycle;
    		\draw[->, line width=0.6pt] (0,0) -- (6.3,0);%
    		\draw[->, line width=0.6pt] (0,0) -- (0,6.3);%
    		\draw[line width=0.6pt] (0,2.3) -- (2.3,0);%
    		\draw[line width=0.6pt] (0,2.3) -- (3.7,6);%
    		\draw[line width=0.6pt] (2.3,0) -- (6,3.7);%
    		\draw[dotted,line width=0.6pt] (0,0) -- (4.85,4.85);%
    		\draw[dashed,line width=0.6pt] (2.1,0) -- (2.1,6);%
    		\draw[dashed,line width=0.6pt] (0,2.1) -- (6,2.1);%
    		\draw[dashed,line width=0.6pt] (0,1.5) -- (6,1.5);%
		\draw[dashed,line width=0.6pt] (1.5,0) -- (1.5,6);%
    		\node at (-0.5,1.3) {$-\tau_0^{-1}$};
    		\node at (2.7,1.1) {D};
    		\node at (2.3,-0.3) {$k$};
    		\node at (-0.3,2.5) {$k$};
    		\node at (6.3,-0.4) {$p$};
    		\node at (-0.3,6.3) {$q$};
    		%
    		\node at (-0.5,1.9) {$-\tau_i^{-1}$};
    		\node at (1.8,1.8) {C};
    		\node at (3.1,1.8) {E};
    		\node at (4.1,3.2) {F};
		\node at (0.9,1.1) {A};
		\node at (1.8,1.1) {B};
		\node at (0.9,5.8) {\circle{15}};
		\node at (0.65,5.8) {3};
    		%
    		%
    		%
    		%
    		%
    		\fill[WildStrawberry] (9.65,1.15) - - (10,1.5) - - (10,0.8) - - cycle;
    		\fill[LimeGreen](11.6,1.5) - - (11.6,3.1) - - (13.9,3.1) - - (12.3,1.5) - - cycle;
    		\fill[Orchid] (10,1.5) - - (11.6,1.5) - - (11.6,0.8) - - (10.8,0) - - (10,0.8) - - cycle;
    		\fill[Turquoise](10,1.5) - - (11.6,3.1) - - (11.6,1.5) - - cycle;
		\fill[BrickRed] (11.6,1.5) - - (12.3,1.5) - - (11.6,0.8) - - cycle;
    		\fill[YellowOrange] (11.6,3.1) - - (13.35,4.85)  - - (14.5,3.7) - - (13.9,3.1) - - cycle;
    		\draw[->, line width=0.6pt] (8.5,0) -- (14.8,0);
    		\draw[->, line width=0.6pt] (8.5,0) -- (8.5,6.3);%
    		\draw[line width=0.6pt] (8.5,2.3) -- (10.8,0);%
    		\draw[line width=0.6pt] (8.5,2.3) -- (12.2,6);%
    		\draw[line width=0.6pt] (10.8,0) -- (14.5,3.7);%
    		\draw[dotted,line width=0.6pt] (8.5,0) -- (13.35,4.85);%
    		\draw[dashed,line width=0.6pt] (11.6,0) -- (11.6,6);%
    		\draw[dashed,line width=0.6pt] (8.5,3.1) -- (14.5,3.1);%
    		\draw[dashed,line width=0.6pt] (8.5,1.5) -- (14.5,1.5);%
		\draw[dashed,line width=0.6pt] (10,0) -- (10,6);%
    		\node at (8,1.5) {$-\tau_0^{-1}$};
    		\node at (9.8,-0.4) {$-\tau_0^{-1}$};
    		%
    		\node at (10.9,-0.3) {$k$};
    		\node at (8.2,2.3) {$k$};
    		\node at (14.8,-0.4) {$p$};
    		\node at (8.2,6.3) {$q$};
    		\node at (11.8,-0.4) {$-\tau_i^{-1}$};
    		\node at (8,3.1) {$-\tau_i^{-1}$};
    		\node at (9.4,1.1) {A};
    		\node at (10.8,0.9) {B};
    		\node at (10.8,2.0) {C};
    		\node at (12.4,2.3) {E};
    		\node at (11.8,1.25) {D};
    		\node at (13.2,3.7) {F};
		\node at (9.4,5.8) {\circle{15}};
		\node at (9.15,5.8) {4};
    		%
    		%
  \end{tikzpicture}
  \begin{tikzpicture}%
    		\fill[Turquoise] (1.5,1.5) - - (4,4) - - (4,1.7) - - (3.8,1.5)  - - cycle;
    		\fill[Orchid] (1.5,1.5) - - (3.8,1.5) - - (2.3,0) - - (1.5,0.8) - - cycle;
    		\fill[LimeGreen](4,4) - - (5.7,4) - - (6,3.7) - - (4,1.7) - - cycle;
    		\fill[YellowOrange] (4,4) - - (5.7,4) - - (4.85,4.85) - - cycle;
		\fill[WildStrawberry] (1.15,1.15) - - (1.5,1.5) - - (1.5,0.8) - - cycle;
    		\draw[->, line width=0.6pt] (0,0) -- (6.3,0);%
    		\draw[->, line width=0.6pt] (0,0) -- (0,6.3);%
    		\draw[line width=0.6pt] (0,2.3) -- (2.3,0);%
    		\draw[line width=0.6pt] (0,2.3) -- (3.7,6);%
    		\draw[line width=0.6pt] (2.3,0) -- (6,3.7);%
    		\draw[dotted,line width=0.6pt] (0,0) -- (4.85,4.85);%
    		\draw[dashed,line width=0.6pt] (4.0,0) -- (4.0,6);%
    		\draw[dashed,line width=0.6pt] (0,4.0) -- (6,4.0);%
    		\draw[dashed,line width=0.6pt] (0,1.5) -- (6,1.5);%
		\draw[dashed,line width=0.6pt] (1.5,0) -- (1.5,6);%
    		\node at (-0.5,1.5) {$-\tau_0^{-1}$};
		\node at (1.5,-0.4) {$-\tau_0^{-1}$};
    		%
    		\node at (2.3,-0.3) {$k$};
    		\node at (-0.3,2.3) {$k$};
    		\node at (6.3,-0.4) {$p$};
    		\node at (-0.3,6.3) {$q$};
    		\node at (4.0,-0.4) {$-\tau_i^{-1}$};
    		\node at (-0.5,4.0) {$-\tau_i^{-1}$};
    		\node at (3.1,2.3) {C};
    		\node at (4.7,3.25) {E};
    		\node at (4.8,4.3) {F};
		\node at (2.4,0.9) {B};
		\node at (0.9,1.1) {A};
		\node at (0.9,5.8) {\circle{15}};
		\node at (0.65,5.8) {5};
    		%
    		%
    		%
    		%
    		%
    		\fill[BrickRed] (10.3,0.5) - - (11.3,0.5) - - (10.8,0) - - cycle;
    		\fill[LimeGreen] (10.3,0.5) - - (9.9,0.9) - - (11.7,0.9)  - - (11.3,0.5) - - cycle;
    		\fill[YellowOrange] (9.9,0.9) - - (9.65,1.15) - - (13.35,4.85) - - (14.5,3.7) - - (11.7,0.9) - - cycle;
    		\draw[->, line width=0.6pt] (8.5,0) -- (14.8,0);
    		\draw[->, line width=0.6pt] (8.5,0) -- (8.5,6.3);%
    		\draw[line width=0.6pt] (8.5,2.3) -- (10.8,0);%
    		\draw[line width=0.6pt] (8.5,2.3) -- (12.2,6);%
    		\draw[line width=0.6pt] (10.8,0) -- (14.5,3.7);%
    		\draw[dotted,line width=0.6pt] (8.5,0) -- (13.35,4.85);%
    		\draw[dashed,line width=0.6pt] (9.4,0) -- (9.46,6);%
    		\draw[dashed,line width=0.6pt] (8.5,0.9) -- (14.5,0.9);%
    		\draw[dashed,line width=0.6pt] (8.5,0.5) -- (14.5,0.5);%
    		\node at (8,0.4) {$-\tau_0^{-1}$};
    		%
    		\node at (10.9,-0.3) {$k$};
    		\node at (8.2,2.3) {$k$};
    		\node at (14.8,-0.4) {$p$};
    		\node at (8.2,6.3) {$q$};
    		\node at (9.4,-0.4) {$-\tau_i^{-1}$};
    		\node at (8,1.0) {$-\tau_i^{-1}$};
		\node at (10.8,0.3) {D};
    		\node at (10.8,0.7) {E};
    		\node at (11.9,2.4) {F};
    		%
		\node at (9.4,5.8) {\circle{15}};
		\node at (9.15,5.8) {6};
    		%
    		%
  \end{tikzpicture}
  \caption{Domains for the momentum integral (\ref{eq:pqint}). Labels coincide with those in (\ref{eq:XpXq}) and (\ref{eq:intcases}).} \label{fig:regions_irx1}
  \vspace{-20pt}
\end{figure}

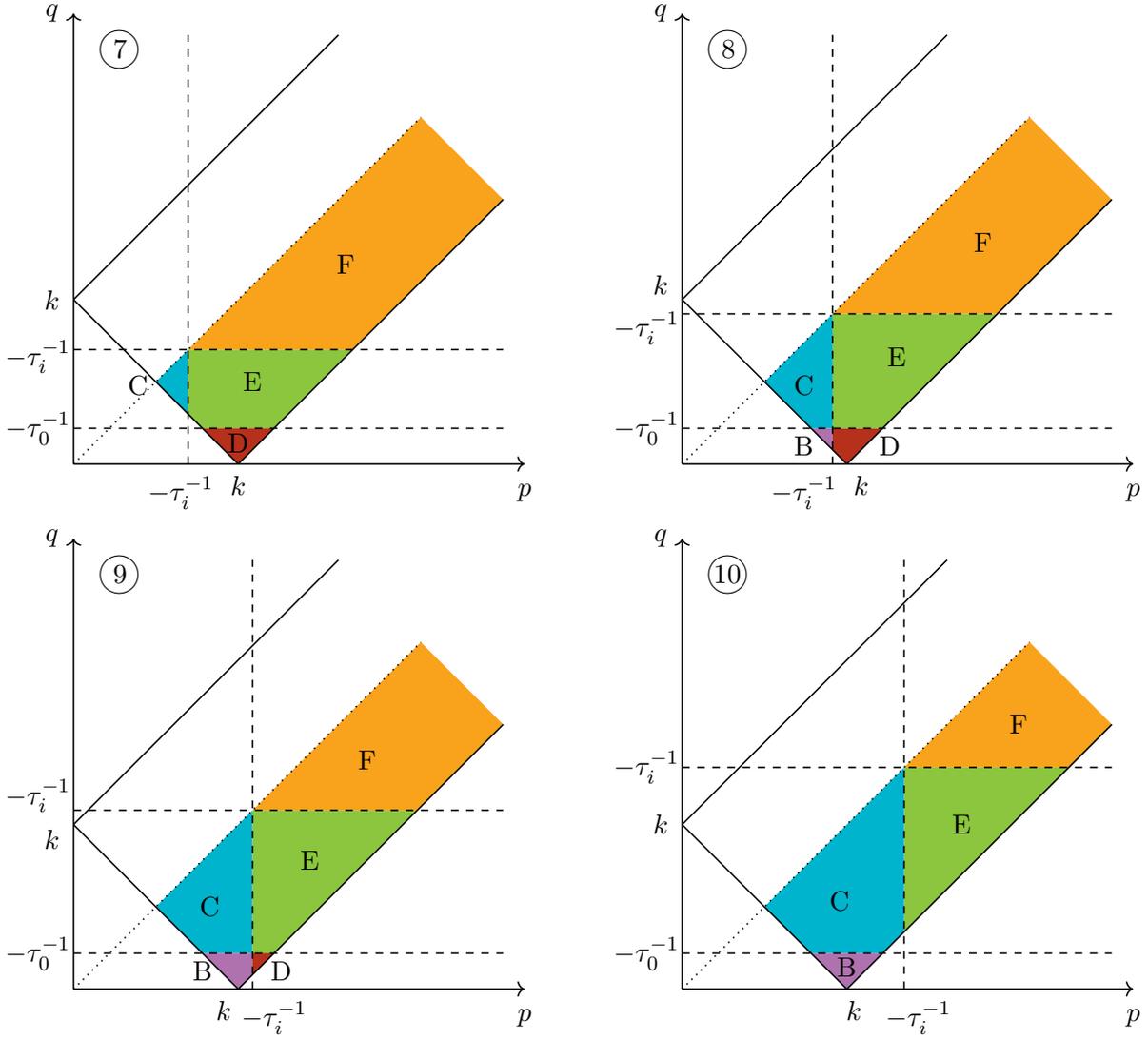
\begin{figure}[!h]
\begin{tikzpicture}%
    		\fill[Turquoise] (1.15,1.15) - - (1.6,1.6) - - (1.6,0.7) - - cycle;
    		\fill[LimeGreen](1.8,0.5) - - (1.6,0.7) - - (1.6,1.6) - - (3.9,1.6) - - (2.8,0.5) - - cycle;
    		\fill[YellowOrange](1.6,1.6) - - (3.9,1.6) - - (6,3.7) - - (4.85,4.85) - - cycle;
		\fill[BrickRed] (1.8,0.5) - - (2.8,0.5) - - (2.3,0) - - cycle;
    		\draw[->, line width=0.6pt] (0,0) -- (6.3,0);%
    		\draw[->, line width=0.6pt] (0,0) -- (0,6.3);%
    		\draw[line width=0.6pt] (0,2.3) -- (2.3,0);%
    		\draw[line width=0.6pt] (0,2.3) -- (3.7,6);%
    		\draw[line width=0.6pt] (2.3,0) -- (6,3.7);%
    		\draw[dotted,line width=0.6pt] (0,0) -- (4.85,4.85);%
    		\draw[dashed,line width=0.6pt] (1.6,0) -- (1.6,6);%
    		\draw[dashed,line width=0.6pt] (0,1.6) -- (6,1.6);%
    		\draw[dashed,line width=0.6pt] (0,0.5) -- (6,0.5);%
		\node at (-0.5,0.5) {$-\tau_0^{-1}$};
    		%
    		\node at (2.3,-0.3) {$k$};
    		\node at (-0.3,2.3) {$k$};
    		\node at (6.3,-0.4) {$p$};
    		\node at (-0.3,6.3) {$q$};
    		\node at (1.5,-0.4) {$-\tau_i^{-1}$};
    		\node at (-0.5,1.5) {$-\tau_i^{-1}$};
    		%
		\node at (2.3,0.3) {D};
    		\node at (2.5,1.15) {E};
    		\node at (3.8,2.8) {F};
		\node at (0.9,1.1) {C};
		\node at (0.9,5.8) {\circle{15}};
		\node at (0.65,5.8) {7};
    		%
    		%
    		%
    		%
    		%
    		\fill[BrickRed] (10.6,0.5) - - (11.3,0.5) - - (10.8,0) - - (10.6,0.2) - - cycle;
    		\fill[LimeGreen](10.6,0.5) - - (10.6,2.1) - - (12.9,2.1) - - (11.3,0.5) - - cycle;
    		\fill[Orchid] (10.3,0.5) - - (10.6,0.5) - - (10.6,0.2) - - cycle;
    		\fill[Turquoise] (9.65,1.15) - - (10.6,2.1) - - (10.6,0.5) - - (10.3,0.5) - - cycle;
    		\fill[YellowOrange] (10.6,2.1) - - (13.35,4.85)  - - (14.5,3.7) - - (12.9,2.1) - - cycle;
    		\draw[->, line width=0.6pt] (8.5,0) -- (14.8,0);
    		\draw[->, line width=0.6pt] (8.5,0) -- (8.5,6.3);%
    		\draw[line width=0.6pt] (8.5,2.3) -- (10.8,0);%
    		\draw[line width=0.6pt] (8.5,2.3) -- (12.2,6);%
    		\draw[line width=0.6pt] (10.8,0) -- (14.5,3.7);%
    		\draw[dotted,line width=0.6pt] (8.5,0) -- (13.35,4.85);%
    		\draw[dashed,line width=0.6pt] (10.6,0) -- (10.6,6);%
    		\draw[dashed,line width=0.6pt] (8.5,2.1) -- (14.5,2.1);%
    		\draw[dashed,line width=0.6pt] (8.5,0.5) -- (14.5,0.5);%
    		\node at (8,0.5) {$-\tau_0^{-1}$};
    		%
    		\node at (11.0,-0.3) {$k$};
    		\node at (8.2,2.5) {$k$};
    		\node at (14.8,-0.4) {$p$};
    		\node at (8.2,6.3) {$q$};
    		\node at (10.2,-0.4) {$-\tau_i^{-1}$};
    		\node at (8,1.9) {$-\tau_i^{-1}$};
    		\node at (10.2,1.1) {C};
		\node at (10.2,0.25) {B};
		\node at (11.4,0.25) {D};
    		\node at (11.5,1.5) {E};
    		\node at (12.7,3.1) {F};
		\node at (9.4,5.8) {\circle{15}};
		\node at (9.15,5.8) {8};
    		%
    		%
  \end{tikzpicture}
  \begin{tikzpicture}%
    		\fill[Turquoise] (1.15,1.15) - - (2.5,2.5) - - (2.5,0.5) - - (1.8,0.5) - - cycle;
    		\fill[Orchid] (1.8,0.5) - - (2.5,0.5) - - (2.5,0.2) - - (2.3,0)- - cycle;
    		\fill[LimeGreen](2.5,2.5) - - (4.8,2.5) - - (2.5,0.2) - - (2.5,0.5) - - cycle;
    		\fill[YellowOrange](2.5,2.5) - - (4.85,4.85)  - - (6,3.7) - - (4.8,2.5) - - cycle;
		\fill[BrickRed] (2.5,0.5) - - (2.8,0.5) - - (2.5,0.2) - - cycle;
    		\draw[->, line width=0.6pt] (0,0) -- (6.3,0);%
    		\draw[->, line width=0.6pt] (0,0) -- (0,6.3);%
    		\draw[line width=0.6pt] (0,2.3) -- (2.3,0);%
    		\draw[line width=0.6pt] (0,2.3) -- (3.7,6);%
    		\draw[line width=0.6pt] (2.3,0) -- (6,3.7);%
    		\draw[dotted,line width=0.6pt] (0,0) -- (4.85,4.85);%
    		\draw[dashed,line width=0.6pt] (2.5,0) -- (2.5,6);%
    		\draw[dashed,line width=0.6pt] (0,2.5) -- (6,2.5);%
    		\draw[dashed,line width=0.6pt] (0,0.5) -- (6,0.5);%
    		\node at (-0.5,0.5) {$-\tau_0^{-1}$};
    		%
    		\node at (2.1,-0.3) {$k$};
    		\node at (-0.3,2.1) {$k$};
    		\node at (6.3,-0.4) {$p$};
    		\node at (-0.3,6.3) {$q$};
    		\node at (2.8,-0.4) {$-\tau_i^{-1}$};
    		\node at (-0.5,2.7) {$-\tau_i^{-1}$};
    		\node at (1.9,1.15) {C};
		\node at (1.8,0.25) {B};
		\node at (2.9,0.25) {D};
    		\node at (3.3,1.8) {E};
    		\node at (4.1,3.2) {F};
		%
		\node at (0.9,5.8) {\circle{15}};
		\node at (0.65,5.8) {9};
    		%
    		%
    		%
    		%
    		%
    		\fill[LimeGreen] (11.6,3.1) - - (13.9,3.1) - - (11.6,0.8) - - cycle;
    		\fill[Orchid] (10.3,0.5) - - (11.3,0.5) - - (10.8,0) - - cycle;
    		\fill[Turquoise] (9.65,1.15) - - (11.6,3.1) - - (11.6,0.8)  - - (11.3,0.5) - - (10.3,0.5) - - cycle;
    		\fill[YellowOrange] (11.6,3.1) - - (13.35,4.85)  - - (14.5,3.7) - - (13.9,3.1) - - cycle;
    		\draw[->, line width=0.6pt] (8.5,0) -- (14.8,0);
    		\draw[->, line width=0.6pt] (8.5,0) -- (8.5,6.3);%
    		\draw[line width=0.6pt] (8.5,2.3) -- (10.8,0);%
    		\draw[line width=0.6pt] (8.5,2.3) -- (12.2,6);%
    		\draw[line width=0.6pt] (10.8,0) -- (14.5,3.7);%
    		\draw[dotted,line width=0.6pt] (8.5,0) -- (13.35,4.85);%
    		\draw[dashed,line width=0.6pt] (11.6,0) -- (11.6,6);%
    		\draw[dashed,line width=0.6pt] (8.5,3.1) -- (14.5,3.1);%
    		\draw[dashed,line width=0.6pt] (8.5,0.5) -- (14.5,0.5);%
    		\node at (8,0.5) {$-\tau_0^{-1}$};
    		%
    		\node at (10.9,-0.3) {$k$};
    		\node at (8.2,2.3) {$k$};
    		\node at (14.8,-0.4) {$p$};
    		\node at (8.2,6.3) {$q$};
    		\node at (11.8,-0.4) {$-\tau_i^{-1}$};
    		\node at (8,3.1) {$-\tau_i^{-1}$};
    		%
    		\node at (10.8,0.3) {B};
    		\node at (10.7,1.23) {C};
    		\node at (12.4,2.3) {E};
    		\node at (13.2,3.7) {F};
		\node at (9.4,5.8) {\circle{15}};
		\node at (9.13,5.8) {10};
    		%
    		%
  \end{tikzpicture}
  \caption{Domains for the momentum integral (\ref{eq:pqint}). Labels coincide with those in (\ref{eq:XpXq}) and (\ref{eq:intcases}).} \label{fig:regions_irx2}
\end{figure}

To go further, it is convenient to break the integration region into subdomains depending on the values of $|k\tau_0|$ and $|k\tau_i|$ relative to 1, 2 and the ratio $\tau_0/\tau_i$. These correspond to the following cases,
\begin{alignat}{2} \displaybreak[0]  \notag
&\text{(1)}\qquad &&1 > |k\tau_0| > \tau_0/\tau_i-1\\ \displaybreak[0] \notag
&\text{(2)}\qquad &&1>|k\tau_0| \quad {\rm and}\quad  \tau_0/\tau_i-1 > |k\tau_0|\\ \displaybreak[0] \notag
&\text{(3)}\qquad && 2 > |k\tau_0| > 1 \quad {\rm and}\quad |k\tau_i|>1\\ \displaybreak[0] \notag
&\text{(4)}\qquad && 2 > |k\tau_0| > 1 \quad {\rm and}\quad 1>|k\tau_i|>1-\tau_i/\tau_0\\ \displaybreak[0] \label{eq:intcases}
&\text{(5)}\qquad && 2 > |k\tau_0| > 1 \quad {\rm and}\quad 1-\tau_i/\tau_0>|k\tau_i| \\ \displaybreak[0] \notag
&\text{(6)}\qquad && |k\tau_0|, |k\tau_i| > 2\\ \displaybreak[0] \notag
&\text{(7)}\qquad && |k\tau_0|>2> |k\tau_i|  \quad {\rm and}\quad  |k\tau_0|>\tau_0/\tau_i+1  \\ \displaybreak[0] \notag
&\text{(8)}\qquad && |k\tau_0|>2  \quad {\rm and}\quad   |k\tau_0|>\tau_0/\tau_i>|k\tau_0|-1\\ \displaybreak[0] \notag
&\text{(9)}\qquad && |k\tau_0|>2  \quad {\rm and}\quad   |k\tau_0|+1 >\tau_0/\tau_i>|k\tau_0| \\ \displaybreak[0] \notag
&\text{(10)}\qquad && |k\tau_0|>2 \quad {\rm and}\quad \tau_0/\tau_i>|k\tau_0|+1
\end{alignat}
These ten cases are illustrated in Figs.~\ref{fig:regions_irx1} and \ref{fig:regions_irx2}. Therein, within each diagram several integration domains are shown, coded by the letters A-F depending on the corresponding value of the ensemble average $\langle |X_p(\tau_i)|_{\rm AS}^2|X_q(\tau_i)|^2_{\rm AS}\rangle$, as found in (\ref{eq:XpXq}). For ease of visualization, these domains are also color coded.

In order to obtain manageable analytical expressions we will consider in full only two possibilities, $|k\tau_0|\gg 1$ (cases 10 and 6) and $|k\tau_0| \ll 1$ (case 2). Also, for the sake of simplicity, we will not assume the introduction of a cutoff scale at $k\gg k_f$, since it would multiply the number of cases to be considered. Nevertheless, even without this scale the stochastic contribution to the power spectrum will be shown to be negligible for sufficiently large wavenumbers. In what follows we denote the momentum integral over each of these subdomains as $\mathcal{I}_{n}$, with $n=\{1,\ldots,10\}$.

\subsubsection*{$\boldsymbol{k\gg k_0}$}

Let us consider here the case when the mode $k$ is inside the horizon before scatterings begin. The stronger condition $k\gg k_0$ simplifies the calculation, as it allows us to disregard cases 7, 8 and 9, since they are relevant only during a short period of time during scatterings. Note that case 6 cannot be outright neglected despite the fact that for it the $p$-mode is inside the horizon. For case 10 though, we can disregard any contributions from sub-horizon modes, and consider only the leading A, B and C integration domains, which lead to the following result:
\begin{align} \notag\displaybreak[0]
8\pi^2\mathcal{I}_{10} \;&\simeq \; \frac{1}{k^2}\Bigg\{ \left( \int_{k/2}^{k-k_0}dp \int_{k-p}^p dq + \int_{k-k_0}^{k+k_0}dp \int_{k_0}^p dq + \int_{k+k_0}^{-\tau_i^{-1}}dp \int_{p-k}^p dq \right) \\ \notag \displaybreak[0]
&\hspace{90pt} \times \left( |p\tau_i|^{-\alpha}|q\tau_i|^{-\beta}- |p\tau_i|^{-\beta} - |q\tau_i|^{-\beta} + 1 \right)\\ \notag \displaybreak[0]
&\hspace{60pt} + \int_0^{k_0}dq \int_{k-q}^{k+q}dp\, \left( |p\tau_i|^{-\alpha}|k_0\tau_i|^{-\beta}- |p\tau_i|^{-\beta} - |k_0\tau_i|^{-\beta} + 1 \right) \Bigg\}\\[5pt] \notag \displaybreak[0]
&=\; \frac{|k\tau_i|^{-\alpha-\beta}}{1-\beta} \Bigg\{ \frac{ 2^{\alpha+\beta-2} - |k\tau_i|^{\alpha+\beta-2}}{\alpha+\beta-2} - \frac{(k/k_0)^{\alpha+\beta-2}}{1-\alpha}\left[(1+k/k_0)^{1-\alpha}-(k/k_0-1)^{1-\alpha}\right]\\ \notag \displaybreak[0]
&\hspace{80pt} - B_{(1/2,1-k_0/k)}(1-\alpha,2-\beta) - B_{(|k\tau_i|,(1+k_0/k)^{-1})}(\alpha+\beta-2,2-\beta)\Bigg\}\\ \notag \displaybreak[0]
&\hspace{20pt} - \frac{|k\tau_i|^{-\beta}}{1-\beta} \Bigg\{ \frac{2(k/k_0)^{\beta-2}+|k\tau_i|^{\beta-2} - (|k\tau_i|/(1-|k\tau_i|))^{\beta-2} -2}{2-\beta}-2\left(\frac{k}{k_0}\right)^{\beta-2}   \Bigg\}\\ \notag \displaybreak[0]
&\hspace{20pt} + |k_0\tau_i|^{-\beta}|k\tau_i|^{-\alpha} \frac{(1+k_0/k)^{2-\alpha} + (1- k_0/k)^{2-\alpha}-2}{(2-\alpha)(1-\alpha)}\\  \displaybreak[0] \label{eq:I10}
&\hspace{20pt} - \frac{\beta |k\tau_i|^{-1} }{1-\beta}   -  |k_0\tau_i|^{-\beta}\left(\frac{k_0}{k}\right)^2 -\frac{3}{4}\,.
\end{align}
In the previous expression $B$ denotes the generalized incomplete beta function,
\beq\label{eq:betafunct}
B_{(z_1,z_2)}(a,b) \equiv \int_{z_1}^{z_2} t^{a-1}(1-t)^{b-1} dt\,.
\eeq
In the limit $|k\tau_i|\ll 1$ (\ref{eq:I10}) can be approximated as a power series in $|k\tau_i|$,
\begin{align} \notag
8\pi^2\mathcal{I}_{10} \;\simeq\; &|k\tau_i|^{-\alpha-\beta}C_{\alpha\beta}(k/k_0) + \frac{|k\tau_i|^{-\beta}}{2-\beta} \left[ \frac{2}{1-\beta} + \beta\left(\frac{k}{k_0}\right)
^{\beta-2}\right] -\frac{\alpha+3 \beta}{4 (\alpha +\beta)} \\
& -\left( \frac{\beta+1}{1-\beta} + \frac{1}{\alpha+\beta-1} \right)|k\tau_i|^{-1}\,.
\end{align}
where the function $C_{\alpha\beta}(k/k_0)$ has been defined in (\ref{eq:calphabeta}). \par\bigskip

Case 6 can be evaluated in a similar way, albeit in this case it is the F-domain only the one that we disregard. Here integration leads to the following expression,
\begin{align} \notag
8\pi^2\mathcal{I}_{6} \;&\simeq \; \frac{1}{8k^2}\mathcal{N}_s\left(\frac{\sigma}{H}\right)^2 \Bigg\{ \int_0^{k_0}dq\int_{k-q}^{k+q}dp\, |p\tau_i|^{-2}\left(|k_0\tau_i|^{-\beta}-1\right)\\ \notag \displaybreak[0] 
& \hspace{95pt}+ \int_{k_0}^{-\tau_i^{-1}}dq\int_{k-q}^{k+q}dp\, |p\tau_i|^{-2}\left( |q\tau_i|^{-\beta}-1\right) \Bigg\} \\ \notag
&= \; \frac{\S}{8|k\tau|^2} \Bigg\{ |k_0\tau_i|^{-\beta} \ln\left(\frac{1}{1-(k_0/k)^2}\right) - \ln\left(\frac{1}{1-|k\tau_i|^{2}}\right) \\ \displaybreak[0]
& \hspace{85pt}+ |k\tau_i|^{-\beta} B_{(k_0/k,|k\tau_i|^{-1})}\left(1-\frac{\beta}{2},0\right) \Bigg\} \\ \displaybreak[0]
&\simeq\; \frac{1}{8}\mathcal{N}_s\left(\frac{\sigma}{H}\right)^2 \Bigg\{ |k\tau_i|^{-\beta-2} \left(\frac{k}{k_0}\right)^{\beta/2-1} \left[\left(\frac{k}{k_0}\right)^{\beta/2-1}-\frac{2}{2-\beta}\right] + \frac{2}{2-\beta}|k\tau_i|^{-\beta/2-3} - |k\tau_i|^{-4}\Bigg\}\,.
\end{align}
In the last line we have approximated the transcendental functions by working in  the limit $|k\tau_i|\gg 1$.

\subsubsection*{$\boldsymbol{k\ll k_0}$}

We now consider those Goldstone modes that were already outside the horizon when scatterings started, $k\ll k_0$. For these, cases 1 and 2 in (\ref{eq:intcases}) are relevant, and it is case 2 the one that will provide the dominant contribution, given that case 1 applies only for a limited amount of time during scattering. Similarly to case 10 discussed above, the contribution to the momentum integral of case 2 can be evaluated as follows,
\begin{align} \notag \displaybreak[0]
8\pi^2 \mathcal{I}_2 \;&=\; \frac{1}{k^2}\Bigg\{ \int_{k/2}^{k_0} dp \int_{|p-k|}^p dq\, \left(|k_0\tau_i|^{-\alpha-\beta} - 2|k_0\tau_i|^{-\beta} +1\right)\\ \notag \displaybreak[0]
&\hspace{40pt} + \int_{k_0}^{k+k_0}dp\int_{p-k}^{k_0}dq\, \left(|p\tau_i|^{-\alpha}|k_0\tau_i|^{-\beta}- |p\tau_i|^{-\beta} - |k_0\tau_i|^{-\beta} + 1\right)\\ \notag \displaybreak[0]
&\hspace{40pt} + \left( \int_{k_0}^{k+k_0}dp \int_{k_0}^p dq + \int_{k+k_0}^{-\tau_i^{-1}}dp \int_{p-k}^p dq \right) \left( |p\tau_i|^{-\alpha}|q\tau_i|^{-\beta}- |p\tau_i|^{-\beta} - |q\tau_i|^{-\beta} + 1 \right) \Bigg\}   \\[5pt] \notag \displaybreak[0]
%
%
&=|k \tau_i|^{-\alpha-\beta} \Bigg\{ \left[\frac{\beta \left(1-(k/k_0+1)^{1-\alpha }\right)}{(1-\alpha ) (1-\beta )}+\frac{1-(k/k_0+1)^{2-\alpha }}{2-\alpha}+\frac{1}{(1-\beta) (\alpha+\beta-2)}\right] \left(\frac{k}{k_0}\right)^{\alpha+\beta-2}\\ \notag \displaybreak[0]
&\hspace{70pt} +\left[\frac{(k/k_0+1)^{1-\alpha}-1}{1-\alpha)}+1\right] \left(\frac{k}{k_0}\right)^{\alpha+\beta -1}-\frac{3}{4} \left(\frac{k}{k_0}\right)^{\alpha+\beta}\\ \notag\displaybreak[0]
&\hspace{70pt} -\frac{1}{1-\beta} B_{\left(|k \tau_i|,(1+k_0/k)^{-1}\right)}(\alpha +\beta-2,2-\beta) \Bigg\}\\ \notag \displaybreak[0]
&\hspace{15pt} - |k\tau_i|^{-\beta} \Bigg\{ \left[\frac{(k/k_0+1)^{1-\alpha}-1}{1-\alpha}-\frac{(k/k_0+1)^{2-\alpha}-1}{2-\alpha}\right] \left(\frac{k}{k_0}\right)^{\alpha -2}\\ \notag\displaybreak[0]
&\hspace{70pt} +\frac{\left((k/k_0+1)^{1-\alpha }-1\right) }{1-\alpha }\left(\frac{k}{k_0}\right)^{\alpha -1}-\frac{\left((k/k_0+1)^{2-\beta }-1\right) }{(1-\beta) (2-\beta )} \left(\frac{k}{k_0}\right)^{\beta -2}\\ \notag\displaybreak[0]
&\hspace{70pt}  +\frac{1-2 \beta }{1-\beta} \left(\frac{k}{k_0}\right)^{\beta-1} - \left(\frac{k}{k_0}\right)^{\beta}\Bigg\} -\frac{3}{4} -\left(\frac{\beta }{1-\beta}\right) |k\tau_i|^{-1}\\ \displaybreak[0]
&\hspace{20pt} - \left[ \frac{1-(1-|k\tau_i|)^{2-\beta}}{2-\beta }+\frac{1}{\alpha +\beta-2} \right] \frac{|k\tau_i|^{-2}}{ 1-\beta }\\[5pt] \notag \displaybreak[0]
&\simeq\; |k\tau_i|^{-\alpha -\beta } \left(\frac{k}{k_0}\right)^{\alpha +\beta } \left[\frac{\alpha +\beta }{\alpha +\beta -1} \left(\frac{k_0}{k}\right)+\frac{1}{4}\right] - |k\tau_i|^{-1}\left(\frac{1}{\alpha+\beta-1}+\frac{\beta+1}{1-\beta}\right) \\  \displaybreak[0]
&\hspace{20pt} + |k\tau_i|^{-\beta} \left(\frac{k}{k_0}\right)^{\beta }  \left[ \frac{2 \beta }{ 1-\beta  }\left(\frac{k_0}{k}\right) -\frac{1}{2} \left(\frac{k_0}{k}\right)^{\alpha -\beta }+\frac{3}{2}\right] - \frac{1}{4}\,.
\end{align}

\subsection{The Mean Curvature Power Spectrum}\label{sec:meanps}

Having computed the ensemble averaged momentum integral, we can now proceed to evaluate the sum (\ref{eq:deltadeltaav}). Here we must break the calculation into three cases, depending on the magnitude of the Goldstone wavenumber compared to $k_0$ and $k_f$. The most interesting regime for our purposes is evidently that in which $k_0\ll k\ll k_f$. In this case we can compute the sum in an analogous manner to the second term in (\ref{eq:howtosum}), which leads to the following result, 
\begin{align} \notag \displaybreak[0]
\langle \delta\Delta_{\zeta}^2(k_0\ll k \ll k_f) \rangle  \;&\simeq\; \frac{1}{2} (\Delta_{\zeta,0}^2)^2 \left(\frac{\sigma}{H}\right)^2 \Bigg\{\sum_{i=1}^{i_*} (k\tau_i)^4 \mathcal{G}_k^2(\tau,\tau_i)\,  \mathcal{I}_{6} + \sum_{i=i_*}^{N_s} (k\tau_i)^4 \mathcal{G}_k^2(\tau,\tau_i)\,  \mathcal{I}_{10} \Bigg\}\\ \notag
&\simeq\; \frac{1}{16} (\Delta_{\zeta,0}^2)^2 \mathcal{N}_s^{\,2}\left(\frac{\sigma}{H}\right)^4 \Bigg\{ \frac{1}{\beta}\left(\frac{k}{k_0}\right)^{\beta/2-1} \left[ \left(\frac{k}{k_0}\right)^{\beta/2-1}-\frac{2}{2-\beta}\right]\left[1-\left(\frac{k}{k_0}\right)^{-\beta}\right] \\ \notag \displaybreak[0]
&\hspace{30pt} + \frac{4}{4-\beta^2}\left[1-\left(\frac{k}{k_0}\right)^{-\beta/2-1}\right]-\frac{1}{2}\left[1-\left(\frac{k}{k_0}\right)^{-2}\right]\Bigg\}\\ \notag 
&\hspace{30pt} +  \frac{2}{9} (\Delta_{\zeta,0}^2)^2 \mathcal{N}_s\left(\frac{\sigma}{H}\right)^2 \Bigg\{ \frac{C_{\alpha\beta}(k/k_0)}{\alpha+\beta-4}\left[\left(\frac{k}{k_0}\right)^{4-\alpha-\beta}e^{(\alpha+\beta-4)N_{\rm tot}}-1\right]\\ \notag
&\hspace{30pt} + \frac{1}{(2-\beta)(\beta-4)} \left[ \frac{2}{1-\beta} + \beta\left(\frac{k}{k_0}\right)
^{\beta-2}\right] \left[\left(\frac{k}{k_0}\right)^{4-\beta}e^{(\beta-4)N_{\rm tot}}-1\right]\\ \notag
&\hspace{30pt}  +\frac{1}{3}\left( \frac{\beta+1}{1-\beta} + \frac{1}{\alpha+\beta-1} \right) \left[\left(\frac{k}{k_0}\right)^{3}e^{-3N_{\rm tot}}-1\right] \\ \displaybreak[0] \label{eq:fullintk}
&\hspace{30pt} +\frac{\alpha+3 \beta}{16 (\alpha +\beta)} \left[\left(\frac{k}{k_0}\right)^{4}e^{-4N_{\rm tot}}-1\right] \Bigg\}\\[5pt] \notag
&\simeq\; \frac{2}{9} (\Delta_{\zeta,0}^2)^2 \mathcal{N}_s\left(\frac{\sigma}{H}\right)^2  \times \begin{cases}
D_{\alpha\beta}\,, & \alpha+\beta<4\,,\\[10pt]
\dfrac{C_{\alpha\beta}(k/k_0)}{\alpha+\beta-4}\left(\dfrac{k}{k_0}\right)^{4-\alpha-\beta}e^{(\alpha+\beta-4)N_{\rm tot}} \,, & \alpha+\beta>4\,.
\end{cases}
\end{align}
The $k$-independent function of the scattering parameter $D_{\alpha\beta}$ is provided in Eq.~(\ref{eq:dab}). Besides the lack of scale invariance for strong scattering, further discussed in the main text, we note that the contribution to the power spectrum coming from sub-horizon Goldstone modes, which corresponds to the terms in (\ref{eq:fullintk}) proportional to the square of the scattering parameter, can be neglected with respect to the super-horizon contribution, as it would be expected from the discussion in Section~\ref{sec:brownian}. Note also the agreement with the schematic result~(\ref{eq:K1cont}) in the exponential regime.
\begin{figure}[!t]
\centering
    \includegraphics[width=\textwidth]{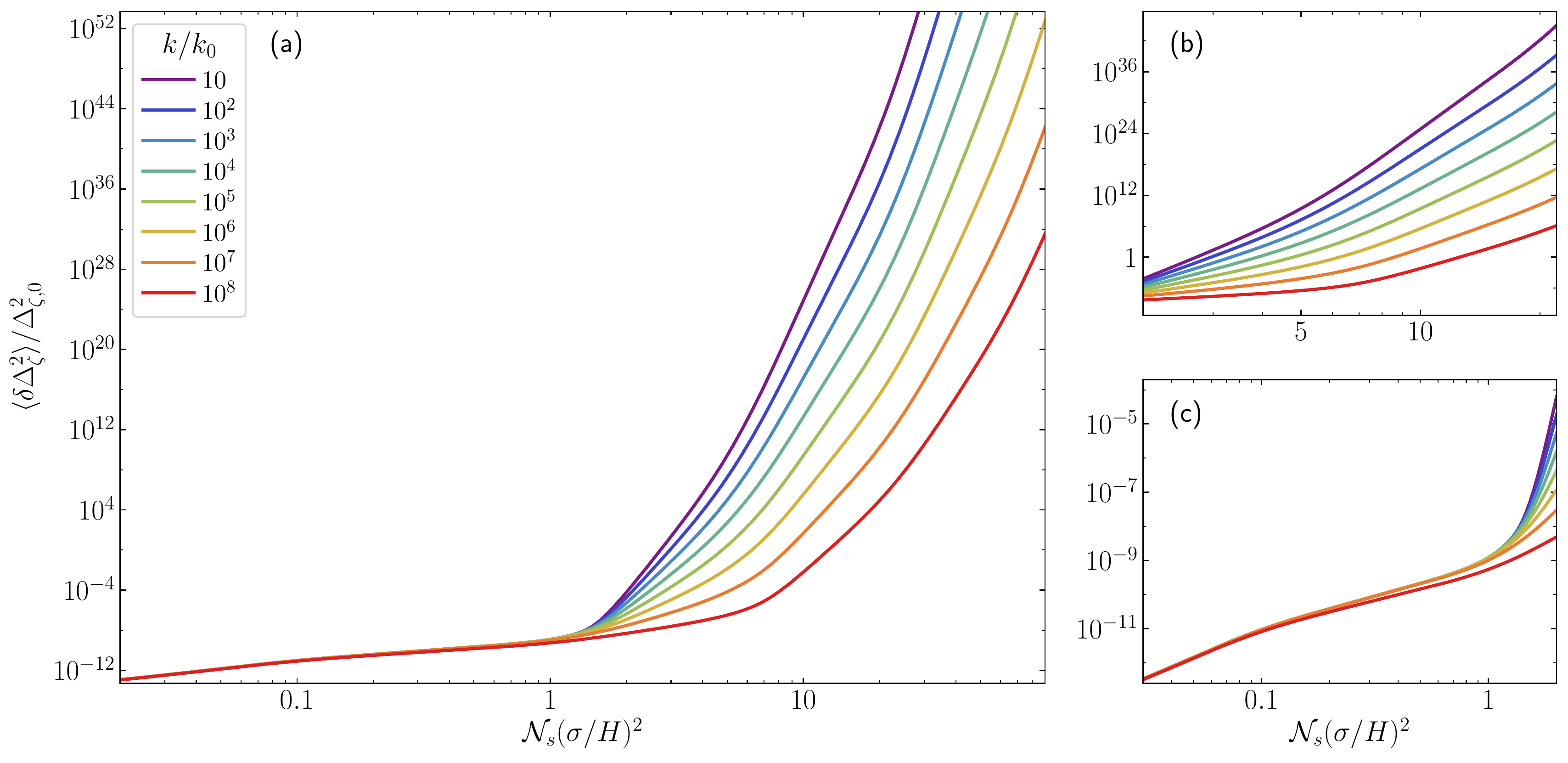}
    \caption{Ratio of the mean stochastic component of the curvature power spectrum to the adiabatic one, as a function of the scattering parameter, for $ k_0\ll k\ll k_f$, $N_{\rm tot}=20$ and $\Delta_{\zeta,0}^2=\Delta_{\zeta,\,{\rm Planck}}^2$. Panel (a) shows the form of the correction for $2\times 10^{-2}\lesssim \mathcal{N}_s(\sigma/H)^2\lesssim 90$. Panel (b) shows the detail of the correction in the regime where $\alpha+\beta>4$. Panel (c) shows the detail of the correction in the regime where $\alpha+\beta<4$. }
    \label{fig:nsplots1}
\end{figure}

Fig.~\ref{fig:nsplots1} shows the dependence on the scattering parameter of the ratio $\langle \delta\Delta_{\zeta}^2\rangle/\Delta_{\zeta,0}^2$ for a few decades in $\S$, for $k_0<k<k_f$. Panel (a) shows the full form of the correction to $\Delta_{\zeta}^2$. Note that for $\S\lesssim 1.5$, for which $\alpha+\beta<4$, all curves are approximately the same, manifesting the scale invariance of the spectrum. Only for the curve for which $k\sim k_f$ there is a visible deviation from the trend, consistent with the suppression of sub-horizon modes. Note that in this regime the stochastic component of the power spectrum is never dominant. This can be more clearly appreciated in panel (c).
For $\S\gtrsim 1.5$, scale invariance is lost, and the spectator field is in the regime of exponential excitation. Not only the spread of the curves for different values of $k$ is evident, but also the steep growth of the power spectrum as a function of the scattering parameter. \par\bigskip

In analogy with the previous case, the always-super-horizon scenario with $k\ll k_0$ can be evaluated in a straightforward way to give
\begin{align} \notag \displaybreak[0]
\langle \delta\Delta_{\zeta}^2(k\ll k_0) \rangle  \;&\simeq\; \frac{1}{2} (\Delta_{\zeta,0}^2)^2 \left(\frac{\sigma}{H}\right)^2 \sum_{i=0}^{N_s} (k\tau_i)^4 \mathcal{G}_k^2(\tau,\tau_i)\,  \mathcal{I}_{2}\\ \displaybreak[0] \notag
&\simeq\; \frac{2}{9} (\Delta_{\zeta,0}^2)^2 \mathcal{N}_s\left(\frac{\sigma}{H}\right)^2 \Bigg\{ \frac{1}{\alpha+\beta-4}\left[\frac{\alpha +\beta }{\alpha +\beta -1} \left(\frac{k_0}{k}\right)+\frac{1}{4}\right] \left(\frac{k}{k_0}\right)^{4} \left(e^{(\alpha+\beta-4)N_{\rm tot}}-1\right)\\ \displaybreak[0] \notag
&\hspace{30pt} + \frac{1}{\beta-4} \left[ \frac{2 \beta }{ 1-\beta  }\left(\frac{k_0}{k}\right) -\frac{1}{2} \left(\frac{k_0}{k}\right)^{\alpha -\beta }+\frac{3}{2}\right] \left(\frac{k}{k_0}\right)^{4} \left(e^{(\beta-4)N_{\rm tot}}-1\right)\\ \displaybreak[0] \notag 
&\hspace{30pt}  + \frac{1}{3} \left(\frac{1}{\alpha+\beta-1}+\frac{\beta+1}{1-\beta}\right)  \left(\frac{k}{k_0}\right)^{3} \left(e^{-3N_{\rm tot}}-1\right) \\ \displaybreak[0]
&\hspace{30pt} +\frac{1}{16} \left(\frac{k}{k_0}\right)^{4} \left(e^{-4N_{\rm tot}}-1\right) \Bigg\} \\[5pt]  \displaybreak[0] \notag
&\simeq\; \frac{2}{9} (\Delta_{\zeta,0}^2)^2 \mathcal{N}_s\left(\frac{\sigma}{H}\right)^2 \left(\dfrac{k}{k_0}\right)^3 \times \begin{cases}
\dfrac{1}{3} \left(1+\dfrac{8}{\beta-4}-\dfrac{4}{\alpha+\beta-4}\right)\,, & \alpha+\beta<4\,,\\[10pt]
\left(\dfrac{\alpha+\beta}{\alpha+\beta-1}\right)\dfrac{e^{(\alpha+\beta-4)N_{\rm tot}}}{\alpha+\beta-4}\,, & \alpha+\beta>4\,.
\end{cases}
\end{align}
Note the expected cubic scaling with momenta for any value of the scattering parameter, and the exponential dependence on $N_{\rm tot}$ for strong scattering (c.f.~Eq.~(\ref{eq:K1k0})). 
\begin{figure}[!t]
\centering
    \includegraphics[width=0.65\textwidth]{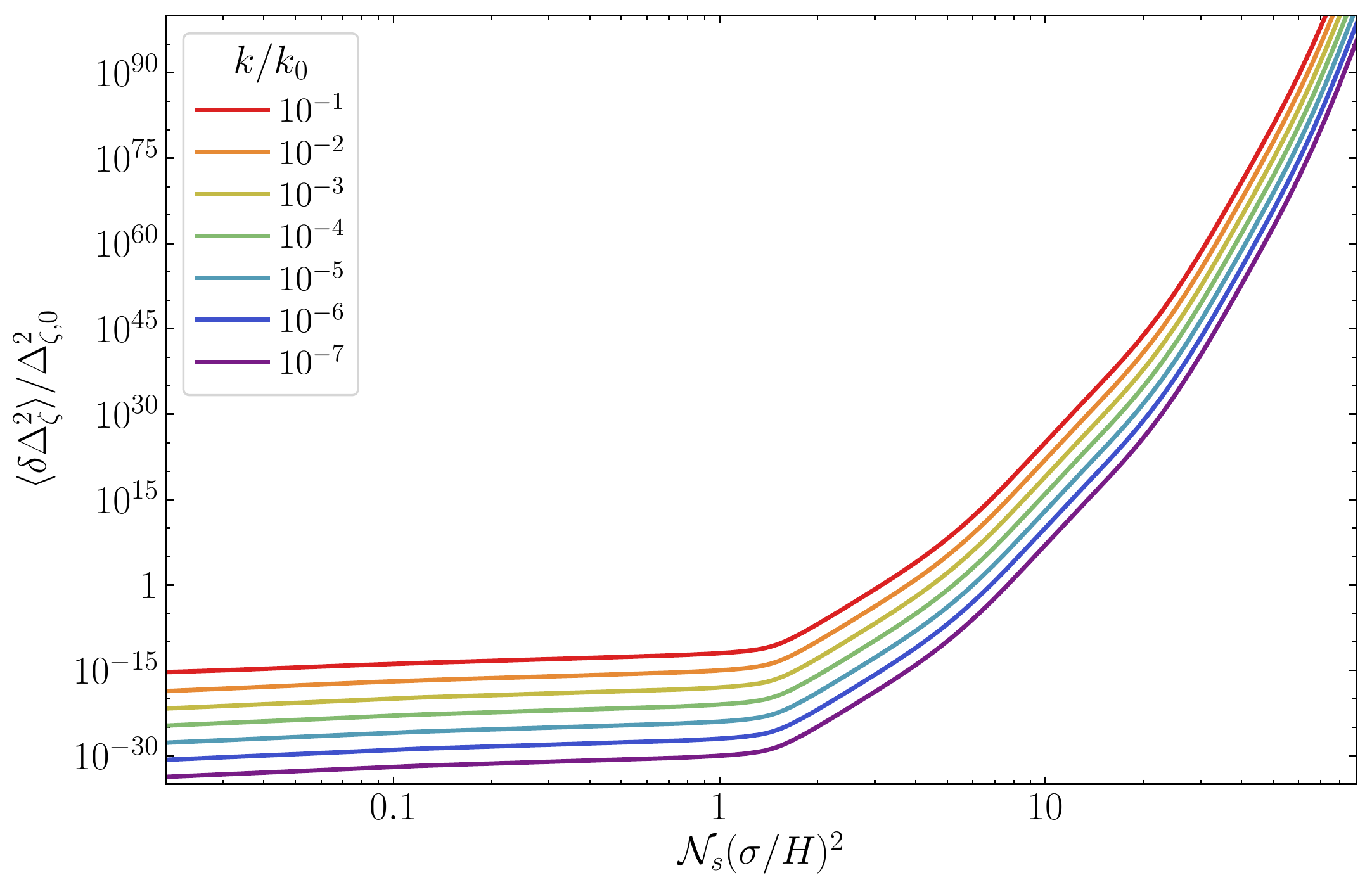}
    \caption{Ratio of the mean stochastic component of the curvature power spectrum to the adiabatic one, as a function of the scattering parameter, for $k\ll k_0$, $N_{\rm tot}=20$ and $\Delta_{\zeta,0}^2=\Delta_{\zeta,\,{\rm Planck}}^2$.}
    \label{fig:nsplots2}
\end{figure}

Fig.~\ref{fig:nsplots2} shows the dependence of the power spectrum on the scattering parameter for $k<k_0$. Although not immediately evident due to the compression of the curves due to the enormous amount of exponential enhancement for large scattering parameter, the cubic scaling is preserved at all scales, in agreement with our causality argument. The difference between weak and strong scattering is also marked for this case, with the stochastic component begin subdominant for $\S\lesssim 1$, and it being by far dominant for $\S\gtrsim 10$.\par\bigskip

Finally, let us consider the always-sub-horizon case with $k\gg k_f$. Computation of the sum (\ref{eq:deltadeltaav}) leads to the following expression,
\begin{align} \notag \displaybreak[0]
\langle \delta\Delta_{\zeta}^2(k_f\ll k) \rangle  \;&\simeq\; \frac{1}{2} (\Delta_{\zeta,0}^2)^2 \left(\frac{\sigma}{H}\right)^2 \sum_{i=0}^{N_s} (k\tau_i)^4 \mathcal{G}_k^2(\tau,\tau_i)\,  \mathcal{I}_{6}\\ \displaybreak[0] \notag
&\simeq\; \frac{1}{16} (\Delta_{\zeta,0}^2)^2 \mathcal{N}_s^{\,2}\left(\frac{\sigma}{H}\right)^4 \Bigg\{ \frac{1}{\beta}\left(\frac{k}{k_0}\right)^{-\beta/2-1} \left[\left(\frac{k}{k_0}\right)^{\beta/2-1}-\frac{2}{2-\beta}\right]\left(e^{\beta N_{\rm tot}}-1\right)\\ \displaybreak[0] \label{eq:fullintk3} 
&\hspace{30pt} + \frac{4}{4-\beta^2}\left(\frac{k}{k_0}\right)^{-\beta/2-1}\left(e^{(\beta/2+1) N_{\rm tot}}-1\right) - \frac{1}{2}\left(\frac{k}{k_0}\right)^{-2}\left(e^{2 N_{\rm tot}}-1\right) \Bigg\}\\[5pt] \displaybreak[0] \notag
&\simeq\; \frac{1}{16} (\Delta_{\zeta,0}^2)^2 \left[\mathcal{N}_s\left(\frac{\sigma}{H}\right)^2\right]^2  \times \begin{cases}
\dfrac{4 }{4-\beta^2} \left(\dfrac{k}{k_0}\right)^{-\beta/2-1} e^ {(\beta/2+1) N_{\rm tot} }\,, & \beta<2\,,\\[10pt]
\dfrac{1}{\beta} \left(\dfrac{k}{k_0}\right)^{-2} e^{\beta N_{\rm tot}}\,, & \beta>2\,.
\end{cases}
\end{align}
The proportionality to the square of the scattering parameter $\S$ in the previous result is expected due to the AS suppression of super-horizon modes. 
\begin{figure}[!t]
\centering
    \includegraphics[width=0.65\textwidth]{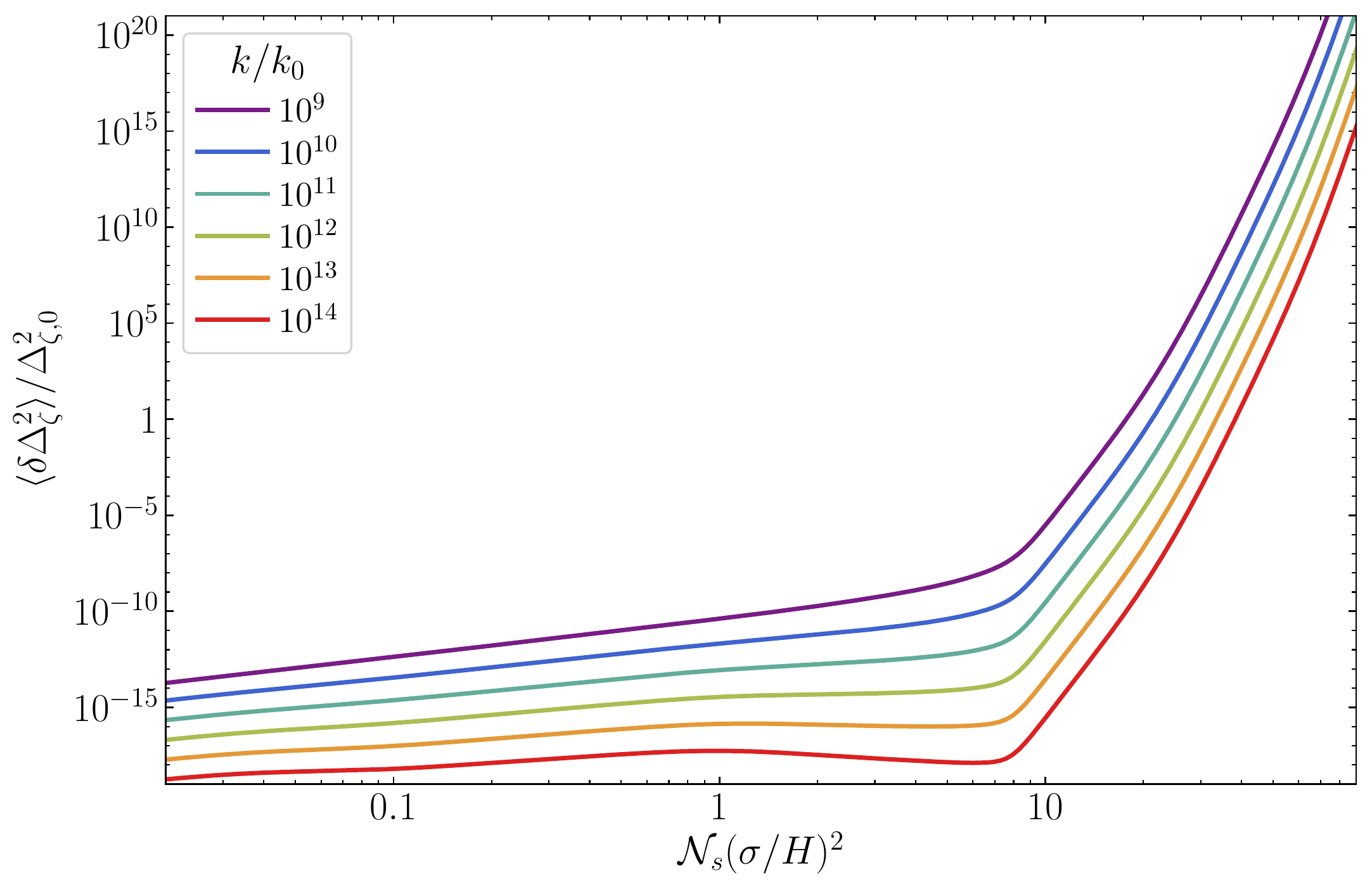}
    \caption{Ratio of the mean stochastic component of the curvature power spectrum to the adiabatic one, as a function of the scattering parameter, for $k_f< k$ , $N_{\rm tot}=20$ and $\Delta_{\zeta,0}^2=\Delta_{\zeta,\,{\rm Planck}}^2$.}
    \label{fig:nsplots3}
\end{figure}

Fig.~\ref{fig:nsplots3} shows the ratio $\langle \delta\Delta_{\zeta}^2\rangle/\Delta_{\zeta,0}^2$ as a function of the scattering parameter $\S$ for $k>k_f$. The decrease in power with increasing $k$ is evident for all scattering strengths. Also clear is the difference in the magnitude of the enhancement for weak and strong scattering, it being negligible for the former, and exponentially large for the later. Although not immediately clear in the figure, the difference in the scaling with $k$ must be noted. For weak scattering, $\langle \delta\Delta_{\zeta}^2\rangle \sim k^{-1}$, while for strong scattering, $\langle \delta\Delta_{\zeta}^2\rangle \sim k^{-2}$.\par\bigskip

\addcontentsline{toc}{section}{References}
\bibliographystyle{utphys}
\bibliography{Refs}

\end{document}